\documentclass[rmp,aps,floatfix,nofootinbib,preprint]{revtex4}

\usepackage{color}
\usepackage{epsfig}
\usepackage{graphics}
\usepackage{graphicx}
\usepackage{helvet}

\def\inbar{\,\vrule height1.5ex width.4pt depth0pt}
\def\IR{\relax{\rm I\kern-.18em R}}
\def\IC{\relax\hbox{$\inbar\kern-.3em{\rm C}$}}

\newcommand{\sfrac}[2]{\mbox{\footnotesize $\frac{#1}{#2}$}}

\definecolor{darkgreen}{rgb}{0,0.5,0}
\definecolor{purple}{rgb}{0.5,0,0.5}
\definecolor{nblue}{rgb}{0.0,0.0,0.50}
\definecolor{scarlet}{rgb}{1.0,0.2,0}
\usepackage[colorlinks=true, pdfstartview=FitV, linkcolor=purple, citecolor= purple, urlcolor=blue]{hyperref}

\begin{document}
\title{Distribution Functions of the Nucleon and Pion in the Valence Region}
\author{Roy J. Holt}
\email{holt@anl.gov}
\affiliation{Physics Division, Argonne National Laboratory, \\Argonne, Illinois 60439, USA}
\author{Craig D. Roberts}
\email{cdroberts@anl.gov}
\affiliation{Physics Division, Argonne National Laboratory, \\Argonne, Illinois 60439, USA}
\affiliation{Department of Physics, Peking University, Beijing 100871, China}

\renewcommand{\baselinestretch}{1.1}

\begin{abstract}
\begin{footnotesize}
We provide an experimental and theoretical perspective on the behavior of unpolarized distribution functions for the nucleon and pion on the valence-quark domain; namely, Bjorken-$x \gtrsim 0.4$.  This domain is key to much of hadron physics; e.g., a hadron is defined by its flavor content and that is a valence-quark property.  Furthermore, its accurate parametrization is crucial to the provision of reliable input for large collider experiments.
We focus on experimental extractions of distribution functions via electron and muon inelastic scattering, and from Drell-Yan interactions; and on theoretical treatments that emphasize an explanation of the distribution functions, providing an overview of major contemporary approaches and issues.
Valence-quark physics is a compelling subject, which probes at the heart of our understanding of the Standard Model.  There are numerous outstanding and unresolved challenges, which experiment and theory must confront.
In connection with experiment, we explain that an upgraded Jefferson Lab facility is well-suited to provide new data on the nucleon, while a future electron ion collider could provide essential new data for the mesons.  There is also great potential in using Drell-Yan interactions, at FNAL, J-PARC and GSI, to push into the large-$x$ domain for both mesons and nucleons.
We argue furthermore that explanation, in contrast to modeling and parametrization, requires a widespread acceptance of the need to adapt theory: to the lessons learnt already from the methods of nonperturbative quantum field theory; and a fuller exploitation of those methods.  %
\end{footnotesize}
\end{abstract}
%\date{July 2007}
%\date{November 2009}

\maketitle

\renewcommand{\baselinestretch}{1.6}

\tableofcontents
\newpage

\section{INTRODUCTION}
\label{sec:intro}
From the first deep inelastic scattering experiments \cite{Bloom:1969,Breidenbach:1969} and the advent of the parton model \cite{Bjorken:1967,Bjorken:1969,Feynman:1969} there has been a tremendous effort to deduce the parton distribution functions of the most stable hadrons -- the proton, neutron and pion.  The long sustained and thriving interest in these structure functions is motivated by the necessity to understand hadron structure at a truly fundamental level.  While it is anticipated that quantum chromodynamics (QCD) will provide the explanation of hadron structure, a quantitative description of hadrons within QCD is not yet at hand.  An explanation and prediction of the behavior of parton distribution functions in the valence-quark region;\footnote{As will become clear, for a given hadron in its infinite momentum frame, a parton's Bjorken-$x$ value specifies the fraction of the hadron's momentum carried by this parton.  There is no unambiguous beginning to the valence-quark domain.  We choose $x \gtrsim 0.4$ because thereafter the gluon distribution can be said to be much smaller than the valence $u$-quark distribution in the proton.}  viz., Bjorken-$x \gtrsim 0.4$, poses an important challenge for QCD and related models of hadron structure.  This is the domain on which the transition takes place from deep inelastic scattering, with incoherent elastic scattering from numerous loosely-correlated small-$x$ partons, to scattering from dressed-quarks that become increasingly well correlated as $x\to 1$.  %Elastic scattering from the target as a whole.
We will focus our attention herein on the behavior of unpolarized structure functions on the valence-quark domain.

The parton distributions are also essential to an understanding of QCD's role in nuclear structure.  It has been known \cite{Aubert:1983} for more than two decades that the parton distribution functions in a nucleus cannot simply be obtained by adding together the distributions within the constituent nucleons.  This mismatch is the so-called \emph{EMC effect} and, although concerted efforts have led to the identification of some of the ingredients necessary to an explanation, we still lack a completely satisfactory understanding of the nuclear dependence of parton distribution functions.  This hampers us enormously.  For example, an accounting for the EMC effect in light nuclei is key to extracting the neutron structure function.  That is a necessary precursor to a veracious determination of the differences between the parton distribution functions of the light-quarks.

An accurate determination of the pointwise behavior of distribution functions in the valence region is also important to very high-energy physics.  Particle discovery experiments and Standard Model tests with colliders are only possible if the QCD background is completely understood.  QCD evolution, apparent in the so-called scaling violations by parton distribution functions, entails that with increasing center-of-mass energy, $s$, the support at large-$x$ in the distributions evolves to small-$x$ and thereby contributes materially to the collider background.

Deep inelastic scattering (DIS) of electrons from protons and bound neutrons at the Stanford Linear Accelerator Center (SLAC) led to the discovery of quarks.  These experiments observed more electrons scattering with high energy at large angles than could be explained if protons and neutrons were uniform spheres of matter.\footnote{In both method and results this series of experiments, conducted from 1966-1978 and for which Taylor, Kendall and Friedman were awarded the 1990 Nobel Prize in Physics, was kindred to that which led Rutherford to discovery of the nucleus in 1911.}  In the approximately forty intervening years, electron DIS has played a central role in measuring structure functions, and SLAC has been especially effective in mapping the proton's structure functions in the valence-quark region.

More recently, muon scattering experiments -- performed by the European Muon Collaboration (EMC), the New Muon Collaboration (NMC) and the Bologna-CERN-Dubna-Munich-Saclay Collaboration (BCDMS) -- have contributed to our store of information.  There have also been a substantial number of neutrino scattering experiments.  However, they have generally used nuclear targets rather than pure hydrogen targets.  Drell-Yan experiments have been effective at measuring the anti-quark distributions in the proton and nuclei.  Putting all this together, it can be said that as a consequence the proton structure function is extremely well known, at least for $x\lesssim 0.7$.

Herein we will discuss the status of charged-lepton and Drell-Yan experiments, as well as prospects for new experiments at, e.g., the Thomas Jefferson National Accelerator Faclity (JLab) and FermiLab (FNAL).  On the other hand, although a number of experiments at the \emph{Conseil Europ\'{e}en pour la Recherche Nucl\'{e}aire} (CERN) and \emph{Deutsches Elektronen-Synchrotron} (DESY) have also provided measurements of the proton structure function, many of these efforts focused on the low-$x$ behavior and hence they will not be discussed.

An experimental determination of the neutron structure function at high Bjorken-$x$ has proved especially troublesome, the main reason being that most of our information about the neutron structure function is obtained from DIS experiments on a deuteron target, for which the nature of the EMC effect is simply unknown.  Consequently, inference of the neutron structure function from proton and deuteron measurements is model-dependent.  We will treat this topic in some depth and, in addition, canvass prospects for the future.

On the theoretical side the challenge is first to parametrize and ultimately to calculate the parton distribution functions.  We write \emph{ultimately} because the distribution functions are essentially nonperturbative and therefore cannot be calculated in perturbation theory.  Thus, absent a truly-accurate, quantitative and predictive nonperturbative tool, the QCD calculation of parton distribution functions remains an alluring but distant prospect.  Herein we will provide both an historical perspective on past attempts at a theoretical interpretation and an overview of recent progress towards this goal.  Notably, today, there are no computations of the pointwise behavior of the nucleon's valence-quark distribution function that agree with the predictions of the QCD parton model.

There has, on the other hand, been much success to date with parametrization.  A number of independent groups have analyzed and performed a global fit to the vast body of extant DIS and Drell-Yan nucleon structure function data.  In this approach the physical processes are factorized into the product of short- and long-distance contributions.  The so-called short-distance parts are calculable in perturbation theory, whereas the long-distance parts are determined by the parton distribution functions.  In this way, via the operator product expansion in QCD, scattering processes involving hadrons are connected to subprocesses involving partons.  The utility of this approach is grounded on the fact that so long as factorization is valid, the parton distribution functions are universal; namely, all hadron-level interactions for which factorization applies are described by the same small body of parton distribution functions.  This being true, the nonperturbative problem is reduced to that described above: instead of calculating all cross-sections from scratch, one has \emph{merely} to calculate the distribution functions.  Herein we will introduce the modern parametrizations.

At its simplest, the nucleon is a three-body problem.  As ostensibly a two-body problem, it would appear theoretically simpler to calculate properties of the pion and hence the parton distributions therein.  However, the pion is in fact both a bound state and the Goldstone mode associated with dynamical chiral symmetry breaking (DCSB) in QCD.  This amplifies the importance of understanding its properties.  However, it also significantly complicates the calculation of the pion's distribution functions and places additional constraints on any framework applied to the task: the pion simply cannot veraciously be described as a constituent quark-antiquark pair.  Instead, key features of nonperturbative quantum field theory must be brought to bear.

Although it is not presently possible to perform deep inelastic scattering on a free pion, the pion structure function has been measured in pionic Drell-Yan experiments at CERN and FNAL.  Owing to predictions from the QCD parton model, the form of the pion's valence-quark distribution function provides a stringent test of our understanding of both QCD and QCD-based approaches to hadron structure.  The FNAL Drell-Yan experiment \cite{Conway:1989fs} measured the pion structure function up to high Bjorken-$x$, where the QCD predictions were expected to be realized.  However, they were not manifest, so that a very puzzling discrepancy remains at this time.  The status of structure function measurements of the pion as well as prospects for future experiments at JLab and a possible future electron ion collider (EIC) will be presented, as will a full description of the theoretical perspective.

The purpose of this article is not to present a comprehensive review of all the issues involved in extracting structure functions from high energy data, as in \cite{Sterman:1995fz}, nor an exhaustive digest of theoretical developments in the field.  Rather the focus will be on structure functions in the valence region, which have chiefly been extracted from electron and muon inelastic scattering, as well as from Drell-Yan interactions.  Moreover, we will focus on the theoretical treatments that emphasize an explanation of these structure functions and summarize the major contemporary approaches and issues.  To end this Introduction, a brief remark on notation: unless denoted otherwise, when addressing a distribution function, it will be that associated with the proton; e.g., $u_v(x)$ will mean the proton's valence $u$-quark distribution. 
%\newpage

%\setcounter{figure}{0}
%\setcounter{table}{0}
\setcounter{equation}{0}

\section{Nucleon structure functions from electron and muon scattering}
\label{sec:kin}

\subsection{Kinematics}
\label{sec:kine}

\begin{figure}[t]
\includegraphics[clip,height=60em]{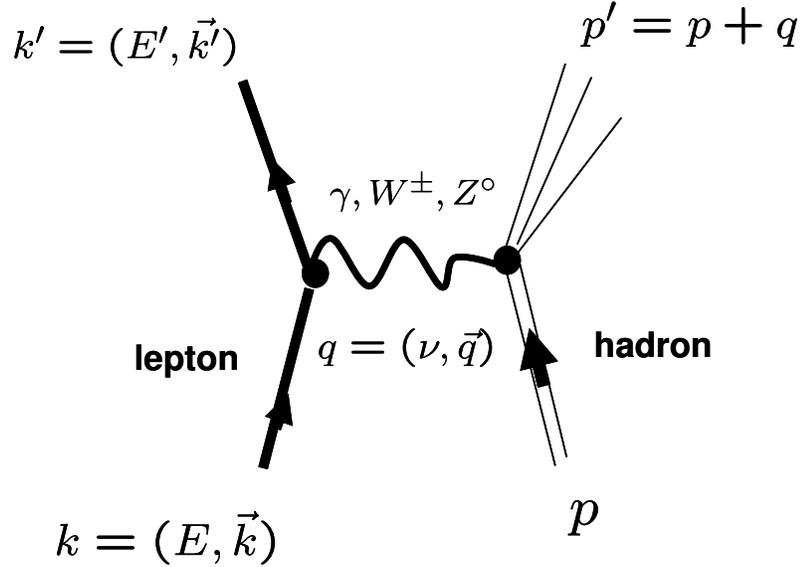}
\vspace*{-80ex}

\caption{Feynman diagram representing deep inelastic lepton scattering from a hadron. \label{feynm}}
\end{figure}

The Feynman diagram for lepton inelastic scattering from a hadron is given in Fig.~{\ref{feynm}}.  Practically, the leptons in a laboratory scattering process can be electrons, muons or neutrinos, as illustrated in the figure, and the current can be a photon, a charged weak current $W^\pm$, or a neutral weak current $Z^0$ of four momentum transfer, $q$.  The incident lepton has incoming and outgoing four momenta of $k$ and $k^\prime$, respectively, so that the momentum- and energy-transfers are given by
%We shall occasionally generalize Eq.~(\ref{eq:rnderivativestwo}) to
%\begin{equation}
%\label{eq:defPhi}
%[\partial_i,\partial_j] = -i\Phi_{ij} ,
%\end{equation}\
\begin{eqnarray}
\label{eq:qtransfer}
  q &= k - k^\prime\,, \\ ~
\label{eq:etransfer}
\nu &= E - E^\prime\,,
\end{eqnarray}
where the incident momentum and mass of the hadron are given by $p$ and $M$, while the outgoing debris after the inelastic event has a momentum $p^\prime$.

The invariants in the scattering process are: the square of the four momentum transfer; the square of the invariant mass; $p\cdot q$, which is related to the energy transfer in the target rest-frame; and the square of the total energy in the center-of-mass frame.  The square of the four-momentum transfer depends on the energy of the beam and scattered lepton as well as the lepton scattering angle: neglecting the lepton's mass,
\begin{equation}
Q^2:= -q^2  = 4EE'sin^2\frac{\theta}{2}\,.
\end{equation}
The square of the invariant mass, W, is given by
\begin{equation}
\label{eq:w2q2}
W^2 = (p + q)^2 = M^2 - Q^2 + 2 \, p\cdot q =  M^2 + \frac{1-x}{x}Q^2,
\end{equation}
where $M$ is the mass of the hadron and
\begin{equation}
\label{eq:xbj}
x = \frac{Q^2}{2 \, p\cdot q}
\end{equation}
is the Bjorken variable.  The square of the total energy in the center of mass is given by the Mandelstam variable, $s$:
\begin{equation}
  s = (k + p)^2
\end{equation}
A final, useful invariant is
\begin{equation}
y = \frac{p\cdot q}{p\cdot k}\,,
\end{equation}
which, in the target rest-frame, is a measure of the fractional energy loss by the incident lepton.

The Bjorken limit is theoretically defined as
\begin{equation}
Q^2\to \infty \,,\; 2\,p\cdot q \to \infty \,, x=\,{\rm fixed}\,.
\end{equation}
In this limit, $x$ can be shown to equal the fraction of the hadron's momentum carried by the struck quark.  Empirically, in order to avoid complications associated with the production of hadron resonances, the Bjorken scaling regime is explored via
\begin{equation}
\label{eq:dis}
 W \rightarrow \infty \,,\;  Q^2 \rightarrow \infty\,,\; x = \frac{Q^2}{Q^2+W^2-M^2} = {\rm fixed}\,.
\end{equation}

In principle, the deep inelastic approximation is only good when
\begin{equation}
\label{eq:discondition}
  \frac{\nu^2}{Q^2}    =  \frac{Q^2}{4M^2x^2} \gg  1.
\end{equation}
However, it is empirically well established that scaling in deep inelastic scattering (DIS) sets in at tractable values of $W$ and $Q^2$.  As a practical matter, evaluations of data typically impose the requirement that $Q^2 \ge  4\ GeV^2$ and $W \ge 3.5\ GeV$.  As discussed and explored, e.g., in \cite{Malace:2009kw,Accardi:2009br}, data at high Bjorken $x$ but not meeting these kinematic requirements would be subject to large target-mass corrections, which are kinematic and owe to binding of partons in the hadron.  Before discussing these topics in more detail, the cross section for inelastic electron scattering will be presented.

\subsection{The cross section for charged lepton scattering}
The cross section for inelastic charged-lepton scattering from a hadron via a one-photon exchange process, as indicated in Fig.~{\ref{feynm}}, is well known and given by
\begin{equation}
\label{eq:xsec}
  \frac{d\sigma  }{d\Omega dE' }(E,E',\theta) =
\left(\frac{\alpha_e^2}{Q^4 }\right)\left(\frac{E'}{E}\right) L_{\mu\nu}^eW^{\mu\nu},
\end{equation}
where $L_{\mu\nu}^e$ is the leptonic tensor and $W^{\mu\nu}$ is the hadronic tensor.  The leptonic tensor has the following straightforward dependence on the kinematic variables:
\begin{equation}
\label{eq:lepton}
L^{\mu\nu}_e = 2\left[k^\mu k'^\nu + k^\nu k'^\mu + g^{\mu\nu}\left(m^2-(k\cdot k')\right)\right],
\end{equation}
where $m$ is the lepton's mass.  Conserving current and parity, the hadronic tensor for a spin-$\frac{1}{2}$ charged particle depends on two structure functions, $W_1(\nu,Q^2)$ and $W_2(\nu,Q^2)$; viz.,
\begin{equation}
\label{eq:hadron}
W_{\mu\nu} = W_1\left(-g_{\mu\nu} + \frac{q_\mu q_\nu}{q^2}\right)+
\frac{W_2}{M^2}\left(p_\mu - \frac{p\cdot q}{q^2}q_\mu\right)\left(p_\nu - \frac{p\cdot q}{q^2}q_\nu\right).
\end{equation}
Contracting the leptonic tensor with the hadronic tensor yields the well-known cross section of the form:
\begin{equation}
\label{eq:xsec1}
  \frac{d\sigma  }{d\Omega dE' }(E,E',\theta) =
4\frac{\alpha_e^2 E'^2}{Q^4}\left[W_2(\nu,Q^2)\cos^2\frac{\theta}{2}
 + 2W_1(\nu,Q^2)\sin^2\frac{\theta}{2}\right].
\end{equation}

The quantities $W_1$ and  $W_2$ in Eq.\,({\ref{eq:xsec1}}) are often written in terms of dimensionless structure functions, $F_1$ and $F_2$, for the nucleon:
\begin{eqnarray}
\label{eq:f1}
  MW_1(\nu,Q^2) &= F_1(x,Q^2)\,,\\ ~
\label{eq:f2}
\nu W_2(\nu,Q^2) &= F_2(x,Q^2).
\end{eqnarray}
In turn, these structure functions are frequently written in terms of the transverse ($F_T$) and longitudinal ($F_L$) structure functions, which correspond to absorption of transverse and longitudinal virtual photons, respectively:
\begin{eqnarray}
\label{eq:transverse}
  F_T(x,Q^2) &=& 2xF_1(x,Q^2)\,,\\
\label{eq: long}
  F_L(x,Q^2) &= & F_2(x,Q^2) - 2xF_1(x,Q^2)\,.
\end{eqnarray}
They can, in principle, be measured separately by performing a Longitudinal/Transverse (L/T) separation, as will be discussed in Sec.~{\ref{sec:longitudinal}}.
%It was discovered in the late 1960s at SLAC that these quantities were nearly independent of $Q^2$ over a very large kinematic range.
The most intuitive physical picture of the interaction in the deep inelastic scattering regime is provided by the parton model.

\subsection{The parton model}
It was discovered at SLAC in the late 1960s that the structure functions appeared to scale; i.e., their evolution with $x$ is nearly independent of $Q^2$ over a very large kinematic range.  These important points are well documented, not only in the original papers \cite{Bloom:1969,Breidenbach:1969,Bjorken:1967, Bjorken:1969, Feynman:1969} but also in the Nobel lectures summarized in \cite{Friedman:1991, Kendall:1991, Taylor:1991}, and here will only be reviewed briefly.

The observed scaling feature means that the cross-section is not separately a function of the two kinematic variables, energy transfer, $\nu$, and $Q^2$ .  Instead, the cross-section's behavior can be expressed in a dependence on only one scaling variable; namely, Bjorken $x$.  Indeed, in the Bjorken-limit, Eq.\,(\ref{eq:dis}),
%; viz.,
%\begin{equation}
%\label{bjorkenlimit}
%Q^2\to\infty\,,\; \nu\to\infty\,,\; x={\rm constant},
%\end{equation}
the proton structure functions $W_1$ and $W_2$ assume a form consistent with elastic scattering from a point fermion; viz.,
\begin{eqnarray}
\label{eq:bjorken}
%  2MW_1(\nu,Q^2) &\rightarrow &\frac{1}{x}\delta\left(1 - \frac{Q^2}{2Mx\nu}\right) = 2F_1(x)\,,\\
%\nu W_2(\nu,Q^2) &\rightarrow & \delta\left(1 - \frac{Q^2}{2Mx\nu}\right) = F_2(x),
%
2 M W_1(\nu,Q^2) & \to & 2\,  F_1(x) \sim \frac{Q^2}{2 M \nu\,  x^2} \delta\left(1 - \frac{Q^2}{2 M \nu \, x}\right)\,, \\
\nu W_2(\nu,Q^2) &\to& F_2(x) \sim  \delta\left(1 - \frac{Q^2}{2 M \nu \, x}\right)\,,
\label{eq:bjorken1}
\end{eqnarray}
where $x$ is given by Eq.\,({\ref{eq:xbj}}).  This near-dependence on only a single dimensionless variable is referred to as Bjoken scaling.  Assuming this to be the case, then the expression for the cross section, Eq.\,({\ref{eq:xsec}}), becomes
\begin{equation}
\label{eq:pxsec}
  \frac{d\sigma  }{d\Omega dE' }(E,E',\theta) =
4\frac{\alpha_e^2 E'^2}{Q^4}\left[\frac{F_2(x)}{\nu} \cos^2\frac{\theta}{2}
 + \frac{2F_1(x)}{M} \sin^2\frac{\theta}{2}\right].
\end{equation}
Capitalizing on these results and assumptions, further analysis leads to the relation \cite{Callan:1969}:
\begin{equation}
\label{eq:callan}
F_2(x) = 2xF_1(x).
\end{equation}
This ``Callan-Gross relation'' is observed to be approximately valid, a feature that is often cited as evidence that the pointlike constituents of the proton; i.e., the quarks, are spin-$\frac{1}{2}$ degrees-of-freedom.

An intuitive understanding of the observed scaling behavior is provided by the quark-parton-model (QPM) or QCD parton model.  In this model the electron collides with a collection of partons within the hadron.  Owing to Lorentz contraction of the hadron at extremely high momentum, these partons are practically frozen in time during the extremely brief collision process.  Moreover, the probability of finding two partons near enough to interact with each other drops as $1/Q^2$.  Thus, initial and final state interactions are suppressed in the DIS regime.   Within this model, the photon interacts incoherently with individual partons and the cross section depends in a simple way on the probability of finding a quark of flavor ``$i$'' with fraction $x$ of the proton's momentum.  These properties are realized in the following simple expression for the electromagnetic structure function
\begin{equation}
\label{eq:parton}
 F_2(x) = 2xF_1(x) = \sum_i e_i^2x\left[q_i(x) + \bar{q}_i(x)\right],
\end{equation}
where the $e_i$ are the charges of the individual quarks and $q_i(x)$ are the probability densities for finding quark-$i$ with fraction-$x$ of the hadron's momentum.  Assuming that only the relatively light quarks ($u$, $d$, $s$, $c$) and their respective antiquarks contribute to the proton structure function, then the proton structure function can be written as
\begin{equation}
\label{eq:pstruc}
\frac{1}{x}F_2^p(x) = \frac{4}{9}\left[u(x) + \bar{u}(x) +c(x) + \bar{c}(x) \right]
+ \frac{1}{9}\left[d(x) + \bar{d}(x) + s(x) + \bar{s}(x)\right].
\end{equation}

%Fig.2
\begin{figure}[t]
\includegraphics[clip,angle=-90,width=0.8\textwidth]{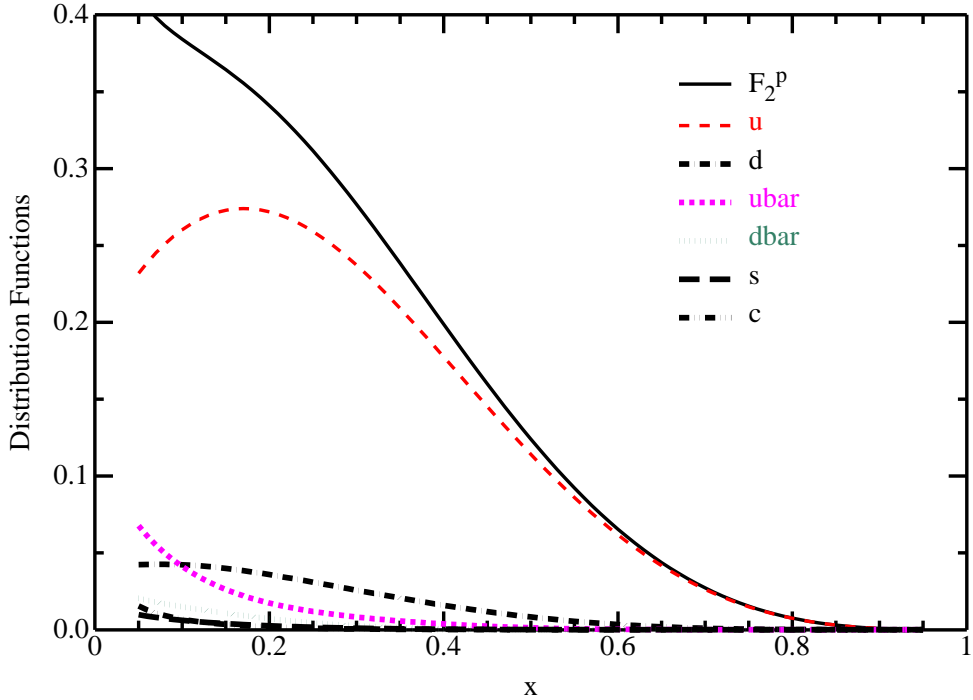}
\caption{\label{cteqquark} (Color online)
Quark flavor dependence of the proton structure function according to the CTEQ6L (leading order) evaluation at a scale $Q^2 = 10\,$GeV$^2$ \protect\cite{Stump:2003yu,Martin:2004}.  We plot the quantities $e_i^2\, xq_i(x)$ in order to emphasize the contribution to the structure function from  from the various quark flavors. }
\end{figure}

In Fig.~\ref{cteqquark} we depict the contributions of these quark flavors to the structure function in the valence region.  The CTEQ6L leading order (LO) evaluation \cite{Stump:2003yu,Martin:2004}, a modern global evaluation of existing data, was used for illustration.  The distribution of up and down quark flavors is fixed primarily by electron and muon scattering from the proton and neutron, while the distribution of the $u$ and $d$ quark sea is largely determined by Drell-Yan processes.  The strange and charm quark sea is determined by neutrino interactions.
The $u$-quark distribution is the dominant contribution to the proton structure function, while the $d$- and anti-$u$-quarks are significant.  The charm and strange quark distributions are small and are not easily distinguishable in the figure.
%These topics will be addressed in the following sections.  Nevertheless, we will see that
We note that semi-inclusive electron scattering from the proton and deuteron is emerging as an important tool for determining the strange quark distribution,
%thereby bringing into question the method of neutrino scattering from complex nuclei as a probe of proton structure functions.
thereby providing an important complement to the method of neutrino scattering from complex nuclei as a probe of non-valence proton structure functions.
Of course, the $u$ and $d$ quark distributions in the valence region are particularly important for testing descriptions of nucleon structure, and it is therefore vital to obtain accurate and precise data, and a reliable flavor separation in this region.

%A number of model calculations are aimed at the role of valence quarks in the hadron.  In contemporary parlance in lattice QCD, such calculations are known as quenched calculations.
A primary goal of modern theory is the computation of valence-quark distribution functions, using QCD-motivated or QCD-based models and nonperturbative methods in QCD.  Naturally, valence quarks have no strict empirical meaning because experiment cannot readily distinguish a valence- from a sea-quark.  However, valence-quark distributions for the $u$- and $d$-quarks in the nucleon can be defined as the following flavor-nonsinglet combinations:
\begin{eqnarray}
\label{eq:uvalence}
  u_v(x) &:= &u(x) - \bar{u}(x)\,, \\
\label{eq:dvalence}
 d_v(x) &:= &d(x) - \bar{d}(x)\,.
\end{eqnarray}
Conservation of charge leads to important normalization conditions:
%The integrals of $u_v(x)$ and $d_v(x)$ over all $x$ for the proton yield 2 and 1, respectively.
\begin{equation}
\label{uvdvnorm}
\int_0^1\, dx\, u_v(x) = 2\,,\;
\int_0^1\, dx\, d_v(x) = 1\,.
\end{equation}
Naturally,
\begin{equation}
\int_0^1\, dx\, s_v(x) = 0\,,
\end{equation}
with the same result for all heavier quarks.  However, it does not follow that $s_v(x)$ must be identically zero.

Consistent with nomenclature, the combinations in Eqs.\,(\ref{eq:uvalence}) and (\ref{eq:dvalence}) have nonzero values of various flavor quantum numbers, such as isospin and baryon number.  Singlet distributions also play a role and are constituted from a sum of quark distribution functions: \mbox{$\Sigma(x) = \Sigma_i [ q_i(x)+\bar q_i(x)]$}.  Under QCD evolution, discussed below, singlet distributions mix amongst themselves and with the gluon distribution, but the nonsinglet distributions do not.

\subsection{QCD and scaling violations}
\label{sec:QCDscalingV}
\begin{figure}[t]
\includegraphics[clip,angle = -90, width=0.8\textwidth]{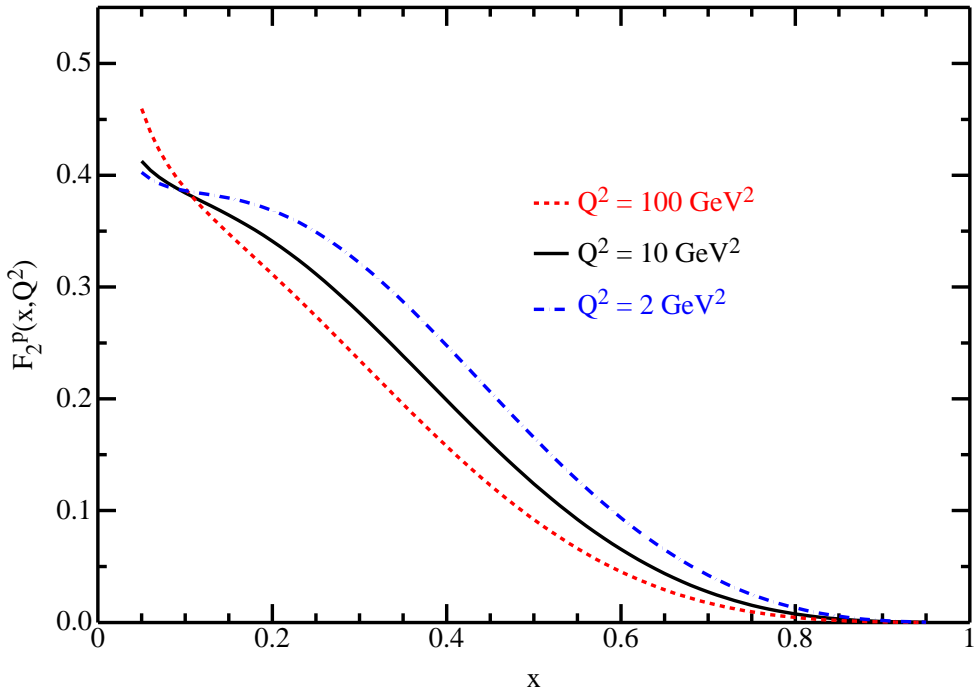}
\caption{(Color online) Structure function of the proton at three values of $Q^2$, computed from the CTEQ6L leading-order evaluation. \label{cteqq2}}
\end{figure}

The proton structure function in the valence region is shown in Fig.\,\ref{cteqq2} as a function of Bjorken $x$ and for three values of $Q^2$ for the CTEQ6L LO evaluation.  Although the $Q^2$ range illustrated is relatively large, from 2 GeV$^2$ to 100 GeV$^2$, the structure function does not exhibit a strong dependence on $Q^2$.  This result validates the parton model as an approximation to a QCD description of the structure function.  Of course, the structure functions have a dependence on $Q^2$, indicating the necessity of QCD rather than a parton model to describe the data.  This relatively small $Q^2$ dependence is known as a scaling violation.  %The very fact that the $Q^2$ dependence is relatively small indicates that there must be some underlying justification for the QPM.

The validity of the QPM led to the idea of factorization in deep inelastic scattering.  In factorization, the cross section is separated into two distinct contributions: a short-distance part, described by perturbative QCD (pQCD); and a long distance part.  The $Q^2$-dependence driving scaling violation originates in the pQCD part and owes to the same phenomena that generate running of the strong coupling constant, $\alpha_s$, and hence asymptotic freedom.  This factorization for the moments of the structure functions can be shown rigorously using the operator product expansion and the renormalization group equations.  Alternatively, a distribution function where the moments can be written as products can be expressed in terms of a convolution of two distributions.  Naturally, the long-distance part is only calculable using nonperturbative methods and presents a challenge for modern hadron theory.

%Fig.3

At small distances or, equivalently, large momentum transfers in QCD, the effective coupling constant becomes small.  This feature is known as asymptotic freedom \cite{Gross:1974,Politzer:1974}.  In terms of the renormalization scale, $\Lambda_{\rm QCD}$, the strong coupling constant has the perturbative form:
\begin{equation}
\label{eq:alpha}
\frac{\alpha_s(t)}{4\pi} = \frac{1}{\beta_0t} - \frac{\beta_1ln(t)}{\beta_0^3t^2} + \\
  \mathcal{O}\left(\frac{1}{t^3}\right),
\end{equation}
where: $t = ln(Q^2/\Lambda_{\rm QCD}^2)$; $\beta_0 = 11 -2n_f/3, \beta_1 = 102 - 38n_f/3$ and $n_f$ is the number of quark flavors involved in the process.  From Eq.\,({\ref{eq:alpha}}) it is apparent that $\alpha_s \rightarrow 0$ as $Q^2 \rightarrow \infty$, so that a parton-model limit is contained within QCD; e.g., as will become clear below, Eq.\,(\ref{eq:parton}) is obtained from Eq.\,({\ref{eq:nlo}}) in this limit.

Plainly, the definition of short- and long-distance phenomena is ambiguous, and associated with a separation of the cross-section into these two contributions is a factorization scale; namely, a mass-scale chosen by the practitioner as the boundary between hard and soft.
%Factorization is not unique and the functions can be different for the various factorization schemes.  This introduces a factorization scale into the problem.  The factorization scale is the scale at which the short and long distance phenomena can be factorized.
If that scale is large enough, then the hard part of the scattering cross-section can be calculated in pQCD.  This, however, introduces another scale, associated with renormalization.  For convenience and economy, the factorization and renormalization scales are usually chosen to be the same.
It follows, in addition, that factorization is scheme-dependent because the choice of renormalization scheme implicitly specifies a division of the finite pieces of the cross-section into those that are retained in the hard contribution and those understood to be contained in the soft piece.
The part identified as owing to long-distance effects is basically the parton density distribution function.

In fitting data to construct PDFs, it is usual to use the $\overline{MS}$ (modified minimal subtraction) renormalization scheme whereas, in regard of factorization, there are two commonly used schemes.
The DIS scheme \cite{Altarelli:1978id} was designed to ensure there are no higher-order corrections to the expression for the $F_2$ structure function in terms of the quark PDFs; i.e., all finite contributions are absorbed into the PDF.
On the other hand, in the more widely used $\overline{MS}$ scheme \cite{Bardeen:1978yd}, in addition to the divergent piece, only the usual $(\ln 4\pi - \gamma_E)$ combination is absorbed into $q(x)$ and hence the expression for $F_2$ exhibits explicit O$(\alpha_s)$ corrections.
%---Discussed in Ellis,Stirling,Webber, pp.103-104
[See \cite{Brock:1993sz} for more on these points.]

\begin{figure}[t]
\vspace*{-16ex}

\includegraphics[clip,width=1.0\textwidth]{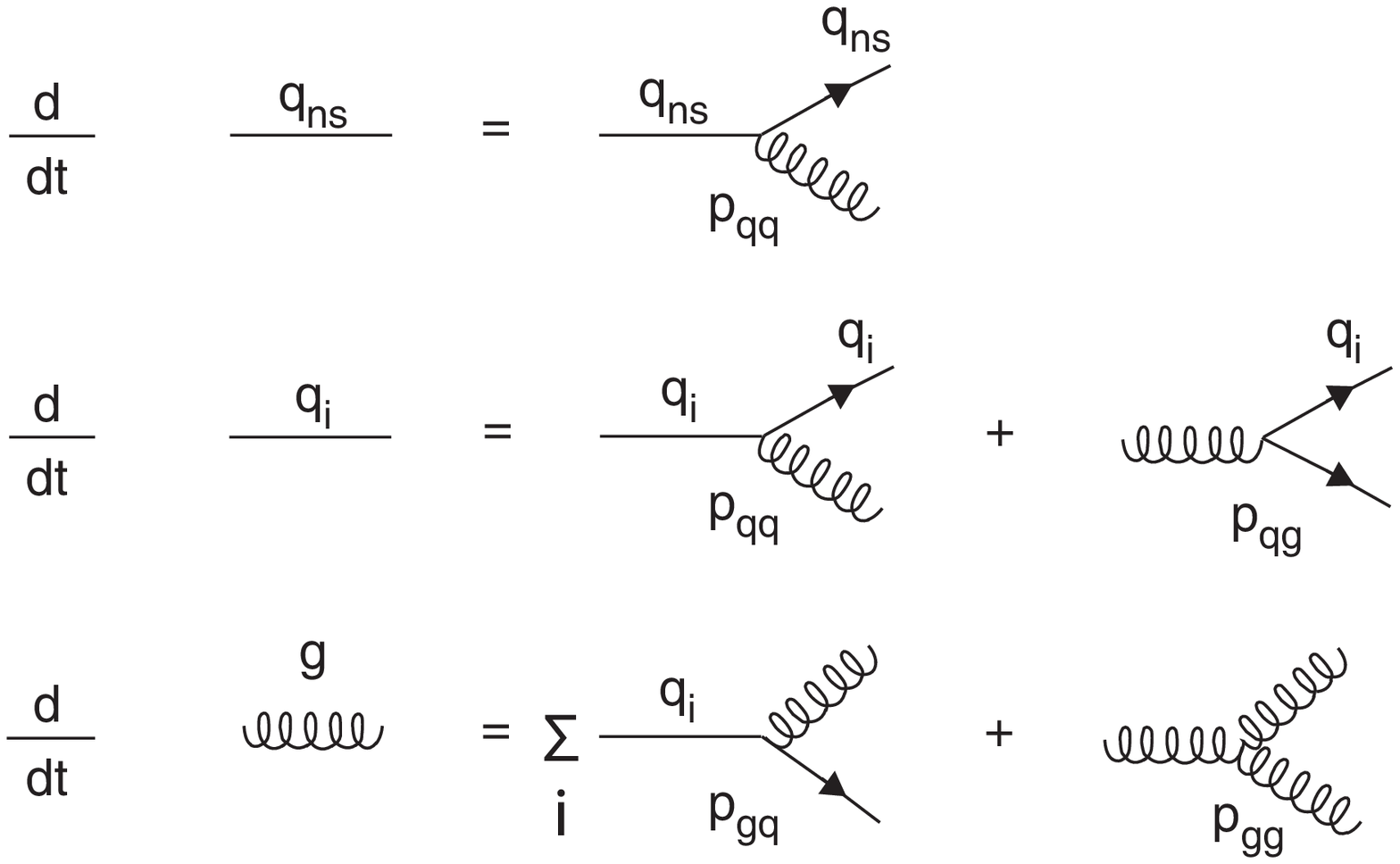}

\vspace*{-16ex}

\caption{Graphical representation of the leading order DGLAP equations. \label{dglap}}
\end{figure}

The CTEQ6L (leading-order) structure functions shown in Fig.\,\ref{cteqq2} were evaluated using the $\overline{MS}$-scheme (modified minimal subtraction).  At leading-order there is no difference between factorization schemes.  %This sets the renomalization scale.
As we will subsequently see (Fig.\,\ref{dglap}), radiated gluons give rise to the scaling violation or $Q^2$-dependence of the structure functions.  They, and the quark distribution functions defined therefrom, then depend on both $x$ and $Q^2$.  The $Q^2$ dependence is now routinely described within the framework of next-to-leading-order (NLO) QCD evolution.

An important and extremely useful feature of factorization is that a measurement of a structure function at relatively-low $Q^2$ permits the calculation, through the use of pQCD, of the structure function at high $Q^2$.  In leading order, a set of integro-differential equations, now known as the DGLAP evolution equations \cite{Dokshitzer:1977, Gribov:1972, Lipatov:1974, Altarelli:1977}, are used for this purpose.  Intuitively, one may think that as $Q^2$ increases, a parton can sometimes be resolved into two partons; e.g., a quark can split into a quark and a gluon, or a gluon into two gluons -- see Fig.~\ref{dglap} for a graphical representation of the leading order DGLAP equations.  The two resolved partons then share in the fraction of the nucleon momentum carried by the initial unresolved parton at lower $Q^2$.  The parton distribution thus becomes a function of $Q^2$.  Such evolution of the parton distribution function (PDF) can be seen in Fig.~\ref{cteqq2}.  At the highest values of $Q^2$ the structure function is shifted toward lower values of $x$.  This work has been generalized to other reaction processes \cite{Furmanski:1981}.

We note that in principle the domain of ``relatively-low $Q^2$'' means, nevertheless, \mbox{$Q^2\gg\Lambda_{\rm QCD}^2$}.  Notwithstanding this, in some empirical determinations of PDFs through data fitting [see Sec.\,\ref{sec:GRV}] and in the application of models to their calculation [see Sec.\,\protect\ref{sec:theory}], evolution is sometimes applied from low $Q^2\sim 4\Lambda_{\rm QCD}^2$.  In such cases an interpretation of the low-$Q^2$ distribution as a parton model density is questionable owing to the probable importance of nonperturbative corrections.  However, contributions from such corrections are suppressed by DGLAP evolution to larger-$Q^2$ and hence one anticipates that in these applications there is always a $Q^2$ whereafter the desired interpretation becomes valid.

As stated above, the evolution equations enable the accurate calculation of the PDFs at a general $Q^2$, provided they are known at another scale, $Q_0^2$, so long as pQCD is a valid tool at both scales.  At next-to-leading order the QCD evolution equations have the form \cite{Herrod:1980}
\begin{equation}
\label{eq:evol1}
\frac{dq_{NS \pm}(x,Q^2)}{d t} = P_{NS \pm}\otimes q_{NS\pm}(x',Q^2)\,,
\end{equation}
where the convolution is defined as
\begin{equation}
\label{defconv}
C \otimes f := \int_x^1\, \frac{dy}{y} \, C(y) \, f(\frac{x}{y})\,,
\end{equation}
and
\begin{equation}
\label{eq:evol2}
\frac{d}{dt}\left(\begin{array}{c}\Sigma(x,Q^2) \\  g(x,Q^2) \end{array}\right) =
P\otimes \left(\begin{array}{c} \Sigma(x',Q^2) \\ g(x',Q^2) \end{array}\right).
\end{equation}
In these formulae, $q_{NS-}= u-\bar{u}, d - \bar{d}$ and $q_{NS+} = (u + \bar{u}) - (d + \bar{d})$ or $(u + \bar{u}) + (d + \bar{d}) -2(s + \bar{s})$ are the non-singlet PDFs, while $\Sigma = \sum_i (q_i + \bar{q}_i)$ is the singlet combination.  The splitting functions are given by
\begin{eqnarray}
\label{eq:evol3}
P_{NS \pm} &= &\frac{\alpha_s}{2\pi}P_{qq}^0(x) + \left(\frac{\alpha_s}{2\pi}\right)^2P_{NS\pm}^{1}(x)\,,\\
\label{eq:evol4}
P &= &\frac{\alpha_s}{2\pi}P^0(x) + \left(\frac{\alpha_s}{2\pi}\right)^2P^{1}(x)\,,
\end{eqnarray}
where
\begin{equation}
\label{eq:evol5}
P^j =  \left(\begin{array}{cc} P_{qq}^j & 2n_fP_{qg}^j  \\
                                P_{gq}^j  &  P_{gg}^j
\end{array}\right).
\end{equation}
The leading order DGLAP equations are obtained simply by eliminating the $\alpha_s^2$ terms in the equations above.

It will be apparent from Eqs.\,(\ref{eq:evol1}) and (\ref{defconv}) that $x=1$ is a fixed point; namely, that the $x=1$ value of each distribution function is invariant under evolution -- it is the same at every value of the resolving scale $Q^2$.  This is because the right-hand-side of each evolution equation vanishes at $x=1$.  Naturally, when Bjorken-$x$ is unity, then $q^2+2 p\cdot q = 0$ and hence one is strictly dealing with the situation in which the invariant mass of the hadronic final state is precisely that of the target: $W=M$ in Fig.\,\ref{feynm}; viz., elastic scattering.  The structure functions inferred experimentally in the neighborhood of $x=1$ are therefore determined by the target's elastic form factors, which all vanish in the limit $Q^2\to\infty$.  This fact \emph{per se} is not interesting.  However, the rate at which distribution functions vanish with $x$ is, because it can lead to nonzero renormalization-scale-invariant distribution-function-ratios at $x=1$.  For example, the value of $\lim_{x\to 1} d_v(x)/u_v(x)$ in the nucleon is an unambiguous, scale invariant feature of QCD and hence a discriminator between models.  [For a concrete illustration of this point, see also Fig.\,\ref{uKpiPCT}, which depicts $u_v^K(x)/u_v^\pi(x)$.]  This is an important consequence, and one much discussed and explored in experiment and theory.  (See Secs.\,\ref{neutronstructurefunction} and \ref{sec:theory}.)
%http://www-spires.fnal.gov/spires/find/hep/www?eprint=arXiv:0911.2254  Melnitchouk

We reiterate, however, that the neighborhood of $x=1$ poses difficulties for experiment, and for theory, too, owing, e.g., to the importance on this domain of target-mass corrections and higher-twist contributions (explained below).  Whilst the equivalence between $x=1$ and elastic scattering cannot be avoided, one might choose to approach the limit obliquely.  For example, noting from Eq.\,(\ref{eq:dis}) that
\begin{equation}
1-x = 1 - \frac{Q^2}{Q^2+ W^2-M^2} \stackrel{Q^2\gg W^2-M^2}{\approx} \frac{(W-M) (W+M)}{Q^2}\,,
\end{equation}
one would ideally choose a path such that both $Q^2$ and $W-M$ are simultaneously kept as large as kinematically possible.  This is experimentally very difficult, as discussed in connection with Fig.\,\ref{wq2}.  Future experimental possibilities are outlined in Sec.\,\ref{EIC}.

A great triumph of QCD is the good agreement between the evolved structure function for the proton and experiment over many orders of magnitude in $Q^2$.  Of course, the structure function should be calculated at the same order as the PDFs.  The equation for the structure function at NLO in the DIS scheme is given by
\begin{eqnarray}
\nonumber F_2(x,Q^2) &=& x \sum_ie_i^2\left\{\rule{0em}{3.2ex} q_i(x,Q^2) + \bar{q}_i(x,Q^2)  \right. \\
&& \left. +
\frac{\alpha_s(Q^2)}{2\pi}\left[C_{i}\otimes\left(q_i(x,Q^2) + \bar{q}_i(x,Q^2)\right) + 2C_g\otimes g(x,Q^2)\right]\right\},
\label{eq:nlo}
\end{eqnarray}
where $g(x,Q^2)$ is the gluon distribution and the Wilson coefficients, $\{C_{i,g}\}$, are given in \cite{Furmanski:1981,Gluck:1994}.

\begin{figure}[t]
\includegraphics[clip,angle = -90, width=0.7\textwidth]{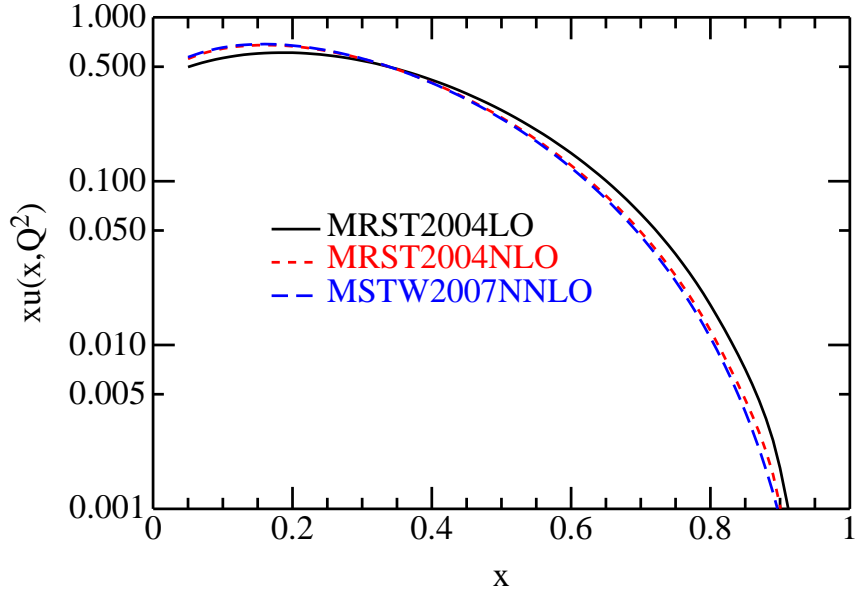}
\caption{(Color online) Comparison of the proton's $u(x)$ distributions in LO, NLO and NNLO at $Q^2 = 10\ GeV^2$ \label{nnlo}}
\end{figure}

During the past several years rather complete NNLO fits \cite{Martin:2007} in the $\overline{MS}$ scheme have been performed for the data.  These fits have made use of NNLO splitting functions \cite{Moch:2004pa, Vogt:2004mw}.  The LO, NLO and NNLO $u$-quark distribution functions are plotted at a scale of  $Q^2=10\ GeV^2 $ in Fig.~\ref{nnlo} for comparison.  Naturally, the structure functions themselves are measured quantities and do not depend on the order of the fit.  In the valence region, the largest difference occurs between LO and NLO as expected.  There is some evidence \cite{Yang:2000} from early NNLO analyses that evolution at NNLO largely offsets the effect of high twist in an NLO analysis on the large-$x$ domain.  However, a more recent analysis \cite{Blumlein:2006} performed up to N3LO contradicts this finding.

\subsection{High twist effects and target mass corrections}
As we have seen, the interactions between partons at short distances that give rise to scaling violations are reasonably well described in a pQCD approach through the $Q^2$-evolution equations.  Probes of a hadron at intermediate values of $Q^2$ might expose correlations among the partons.  Thus far, we have only considered the case in DIS where structure functions are governed by incoherent scattering from the partons.  As we move to lower $Q^2$ where nucleon resonances could dominate or to very high $x$ where the elastic scattering limit and complete coherence dominates, then correlations between the partons must be taken into account.  Ultimately a theoretical treatment of these correlations demands an understanding of the way that quarks and gluons bind to form hadrons.  Although perturbative QCD cannot explain binding effects, it does permit one to construct a model of power corrections to the perturbative result.  It was found \cite{Ellis:1982, Ellis:1983} that the first power correction, $1/Q^2$, to the structure function is governed by the intrinsic transverse components of the parton momentum.

The operator product expansion is normally used to order contributions to a DIS structure function according to their \textit{twist}, $\tau$, which is defined as the difference between the naive mass-dimension of an operator and its ``spin''.  Here ``spin'' means the number of vector indices contracted with the configuration-space four-vector, $z$.  For example \cite{Sterman:1995fz}, consider the expansion of a dimension-six current-current correlator, typical of DIS,
\begin{equation}
\label{twist}
J^\alpha(z) J_\alpha(0) \sim \sum_{\tau=2}^\infty \sum_{n=0}^\infty C_{\tau,n}(z^2,\mu^2)\, z^{\nu_1}\ldots\, z^{\nu_n} \, O^\tau_{\nu_1\ldots \nu_n}(z=0,\mu^2)\,,
\end{equation}
where $\mu$ is the factorization scale.
One is interested in the $z^2\sim 0$ behavior of the coefficient function $C_{\tau,n}(z^2,\mu^2) = (z^2)^{-d_C/2} \, \hat C(z^2 \mu^2)$ because the more highly singular is this function on the light-cone, the more important is the associated operator at large $Q^2$.  Now, suppose the operator $O^\tau_{\nu_1\ldots \nu_n}(\mu^2)$ has dimension $d_{O_n}$, then, in order to balance dimensions on both sides of Eq.\,(\ref{twist}), one must have $d_C = 3 - (d_{O_n}-n)/2$, since each factor of $z$ on the right-hand-side contributes mass dimension negative-one.  Plainly, the largest value of $d_C$ is obtained with the smallest value of twist $\tau=d_{O_n}-n$,  and the associated operators are those which dominate the large-$Q^2$ behavior of the correlator.
As Eq.\,(\ref{twist}) indicates, numerous operators with different dimensions have the same twist and all such operators are associated with the same degree of light-cone singularity.
%texkey Sterman:1995fz was useful and Roberts

As evident in the example above, the leading-twist contributions in DIS are twist-2.  In unpolarized DIS, the higher twist components; i.e., twist $= 4, 6,\,$\ldots, are suppressed by $1/Q^2$, $1/Q^4\,$, \ldots, respectively.  Thus, in general a structure function should be expressed in the form \cite{Alekhin:2003}
\begin{equation}
 \label{eq:twist}
    F_2(x,Q^2) = F_2^{LT}(x,Q^2) + \frac{H_2(x,Q^2)}{Q^2} + \mathcal{O}(1/Q^4), \\
\end{equation}
where $F_2^{LT}$ refers to the leading-twist part.  As a practical matter, the magnitudes of higher-twist terms are generally unknown and the estimates are therefore somewhat controversial.  Before one can claim discovery of higher twist components, it is essential that these be distinguished from target mass corrections and $Q^2$-evolution effects.

\begin{figure}[t]
\leftline{\includegraphics[clip,width=0.45\textwidth]{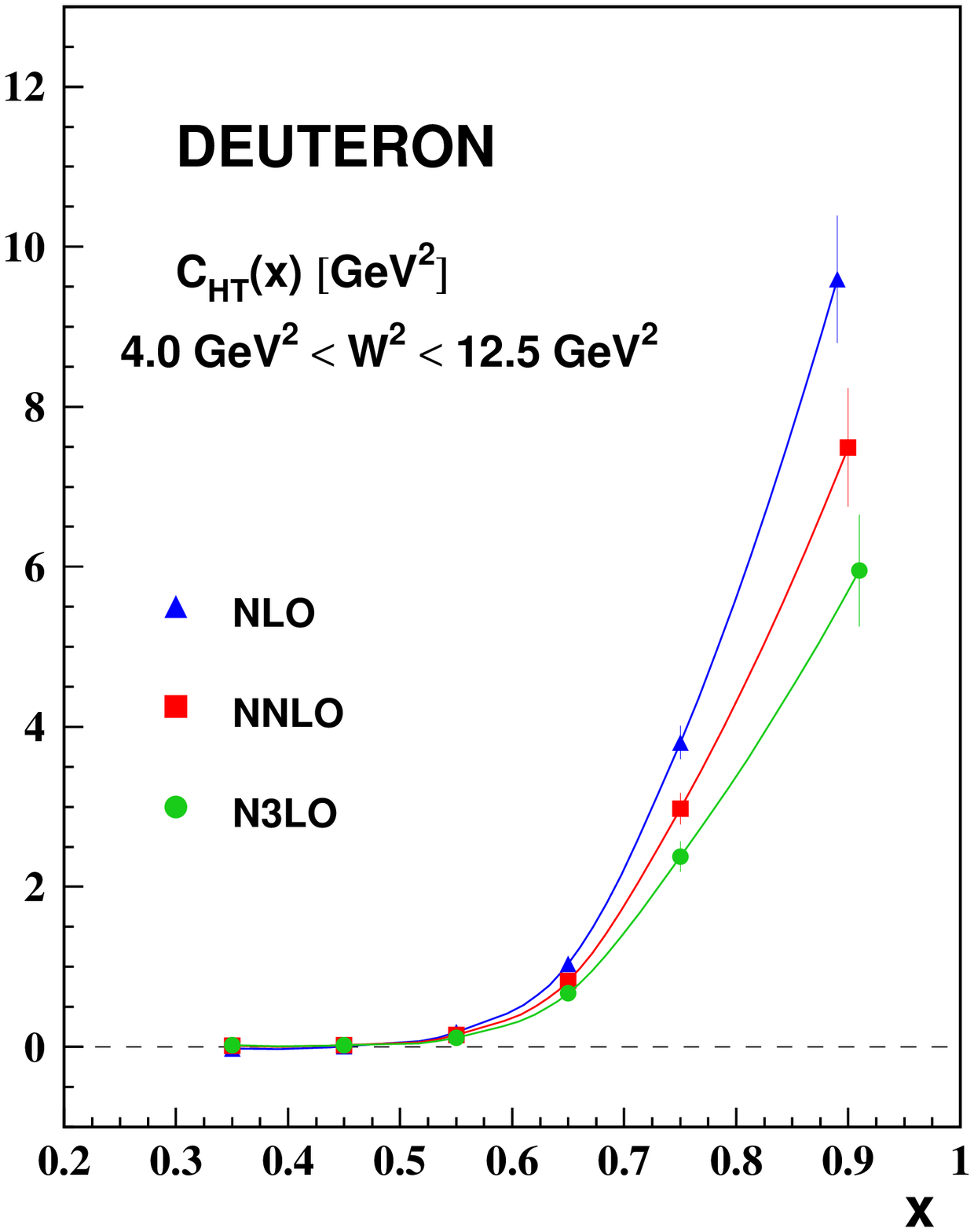}}
\vspace*{-57ex}

\rightline{\includegraphics[clip,width=0.45\textwidth]{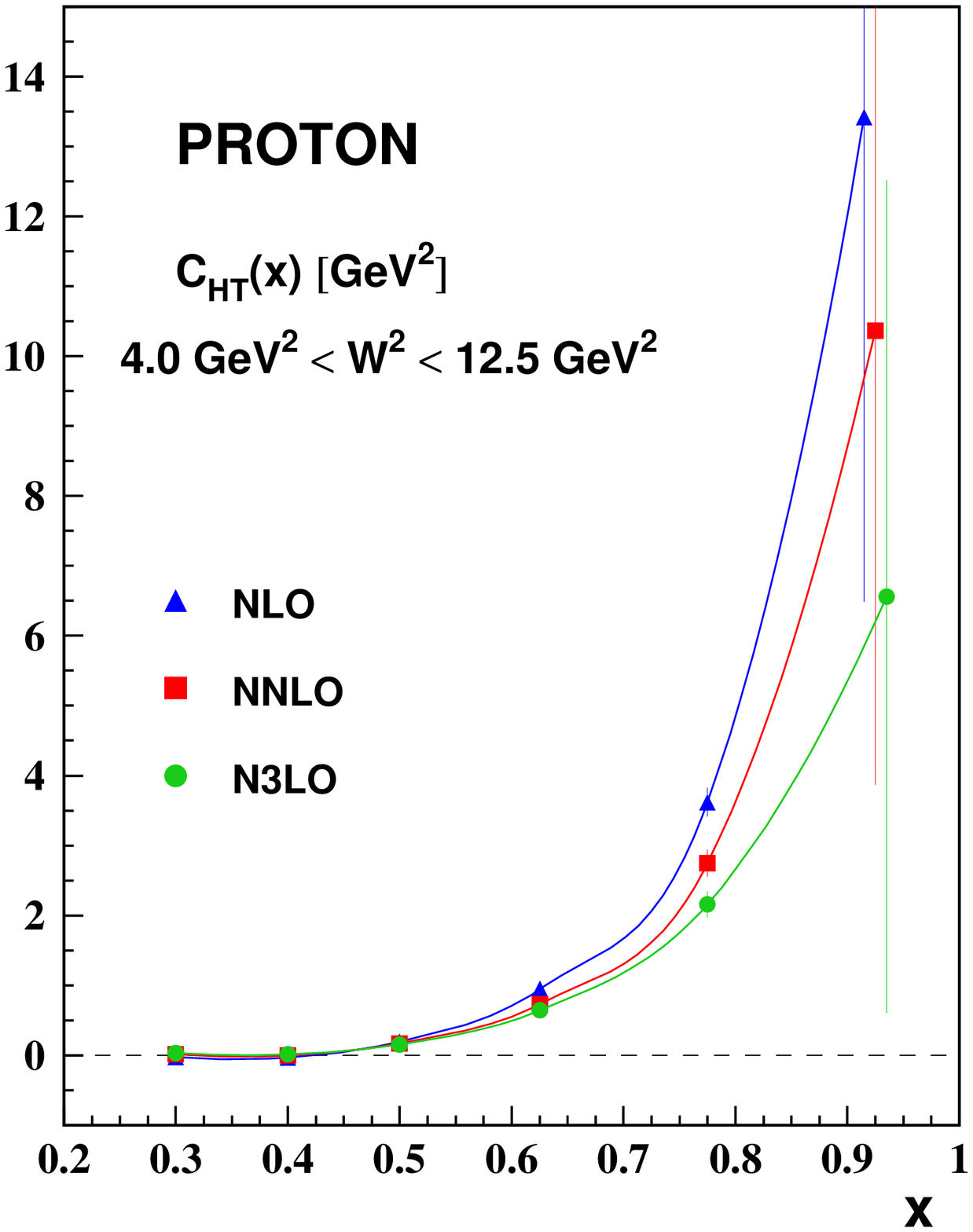}}
\caption{\label{fig:HTwist} (Color online)
Empirical nonsinglet higher-twist coefficient, denoted $C_{HT}$, as a function of Bjorken-$x$, for the proton (left panel) and deuteron (right panel), as obtained when the twist-2 contributions are treated at NLO, NNLO and N3LO.  [Figure adapted from \protect\cite{Blumlein:2006}.]}
\end{figure}

Several analyses have shown that higher-twist coefficients can become quite large at high-$x$ and relatively low $Q^2$.  An analysis \cite{Virchaux:1991} of BCDMS and SLAC data indicated that the twist-4 coefficient is sizable for the proton and deuteron.  Here, however, the inclusion of target mass corrections reduced the magnitude of the twist-4 coefficient.  This finding was supported in an analysis that included NNLO \cite{Yang:2000}. It was found that the NNLO contribution can partially compensate for a large higher-twist effect.  More recently, an analysis \cite{Blumlein:2006} of world data indicated that the twist-4 coefficient is sizable for NLO, NNLO and N3LO for the proton and deuteron.  While increasing the order of the QCD analysis of the data reduces the twist-4 coefficient, the coefficient is significant at high $x$.  This is displayed in Fig.\,\ref{fig:HTwist}, which plots an empirically determined correction defined via
\begin{equation}
F_2(x,Q^2) = O_{TMC}[ F_2^{{\rm twist}-2}(x,Q^2)] \left( 1 + \frac{C_{HT}(x,Q^2)}{Q^2} \right)\,,
\end{equation}
where $O_{TMC}$ describes inclusion of target mass corrections of the twist-2 contributions to the structure function.
At present, there exists no statistically significant determination of the magnitude of the twist-6 coefficient.
%EPRINT HEP-PH/0607200 Figs. 14&15

In connection with the size of higher-twist contributions, it is relevant to consider their impact on Bloom-Gilman duality \cite{Bloom:1971}.  This is an expectation that the nucleon structure function, when measured in the region dominated by low-lying nucleon resonances, should follow a global scaling curve that is determined by high energy data.
NB.\ Whilst it has been observed that the standard application of DGLAP evolution to deep inelastic structure functions can appear to be inconsistent with Bloom-Gilman duality at fixed $W$, this apparent conflict is resolved if one takes into account the fact that the struck quark is far-off shell in the $x\to 1$ domain \protect\cite{lepage80,Brodsky:2005wx}.

Underlying Bloom-Gilman duality is the assumption of quark-hadron duality; namely, a belief that one obtains equivalent descriptions of physical phenomena irrespective of whether one uses partonic or hadronic degrees-of-freedom \cite{Shifman:2000jv,Melnitchouk:2005zr}.  While this assumption is probably correct in principle, in comparisons between computations that use finite-order truncations, it can and is violated in practice.
%; e.g., \cite{GonzalezAlonso:2010rn}.
The presence of large higher-twist contributions can therefore entail an apparently strong violation of Bloom-Gilman duality.

Experiments, particularly at the highest values of $x$, might not be performed at sufficiently large momentum transfer to avoid high-twist effects and target-mass corrections.  A pQCD treatment is truly valid only in the limit that the mass of the hadron is negligible in comparison with the dominant scale, $Q^2$.  The target-mass corrections are kinematic and owe to binding of partons in the hadron.  Nachtmann first pointed out \cite{Nachtmann:1973} that at finite $Q^2$ and non-vanishing target mass, the scaling variable should be the fraction, $\xi$, of the nucleon's light front momentum carried by the parton; viz.,
\begin{equation}
 \label{eq:nachtmann}
    \xi(x, Q^2) = \frac{2x}{1+\kappa}\,,\; \kappa=\sqrt{1 + 4x^2M^2/Q^2}. \\
 \end{equation}
Naturally, this Nachtmann variable reduces to Bjorken-$x$ for $M^2/Q^2 \rightarrow 0$.

It was subsequently shown \cite{Georgi:1976} that at leading-order, if the quantity $x^2M^2/Q^2$ is reasonably small, then it is straightforward to apply corrections for finite mass to the structure function.  If the leading-twist structure function is denoted by $F_i^{LT}$ and the target-mass corrected structure function by $F_i^{TM}$, then
one has the expressions:
\begin{eqnarray}
\label{eq:F_1}
  F_1^{TM}(x,Q^2) &=& \frac{x}{2\kappa}F_1^{LT}(\xi) + \frac{M^2 x^2}{Q^2 \kappa^3}\int_\xi^1 dx'F_1^{LT}(x') +
2 \frac{M^4x^3}{Q^2\kappa^3}\int_{x'}^1dx''F_1^{LT}(x'')\,, \\
%\label{eq:F_2}
  F_2^{TM}(x,Q^2) &=& \frac{x^2}{\kappa^3}F_2^{LT}(\xi) + 6\frac{M^2 x^3}{Q^2 \kappa^4}\int_\xi^1 dx'F_2^{LT}(x') +
12 \frac{M^4x^4}{Q^4\kappa^5}\int_{x'}^1dx''F_2^{LT}(x'')\,.
\end{eqnarray}
%where $\kappa = \left(1 + \frac{4M^2x^2}{Q^2}\right)^\frac{1}{2}$.
%The effect of target mass corrections, which makes use of the Georgi-Politzer approach for LO, are demonstrated in Fig.~\ref{targmass} for the proton at a $Q^2 = 9\ GeV^2$.
%  As expected the effect is largest at highest $x$.

%\begin{figure}
%\epsfysize=3.0in
%\includegraphics[angle = -90, width=4.0in]{targmass.ps}
%\caption{Structure function of the proton with and without LO target mass corrections at a $Q^2 = 10\ GeV^2$.  The largest effect occurs at the highest values of $x$. \label{targmass}}
%\end{figure}

A concern with the Nachtmann or the Georgi-Politzer approach is the so-called ``threshold problem'' It results from the fact that the maximum kinematic value of $\xi$ is less than unity, which means that the corrected leading twist structure function does not vanish at $x=1$.

Target-mass corrections have been extended to NLO QCD \cite{Derujula:1977}, and generalized to include charge and neutral weak current deep inelastic scattering \cite{Kretzer:2003}. De R\'{u}jula {\it et al} argued that the threshold problem could be solved by considering higher-twist effects.

A recent interesting approach \cite{Steffens:2006} to the ``threshold problem'' followed the original Georgi-Politzer approach but with one critical difference.  Namely, the upper limit of the integrals in  Eq.\,(\ref{eq:F_1}) becomes the upper limits of $x'$ and $x''$ that are permitted by kinematics at $x=1$ rather than merely unity.  In terms of existing data, this solution does not result in a large effect for $Q^2 > 2\,{\rm GeV}^2$, but can have a significant effect below $1\,{\rm GeV}^2$ where the underlying assumption of factorization becomes questionable anyway.

In addition to target-mass corrections, there are also effects from nonzero current-quark mass \cite {Barbieri:1976}.  Such corrections will not be discussed herein because of our focus on high-$x$, whereat heavy quarks have little effect in nucleons and Goldstone bosons.  These effects, in addition to those arising from target-mass corrections, are discussed in a recent review \cite {Schienbein:2007}.  Furthermore, fresh work \cite {Accardi:2008ne} has pointed out that changing the upper limit of the integrals in the Georgi-Politzer equations leads to an abrupt cutoff at $x=1$, and proposed a ``quark jet'' mechanism to give a more reasonable approach to the very large $x$ behavior.

\subsection{The proton structure function}
\begin{figure}[t]
\includegraphics[clip,angle = -90, width=0.8\textwidth]{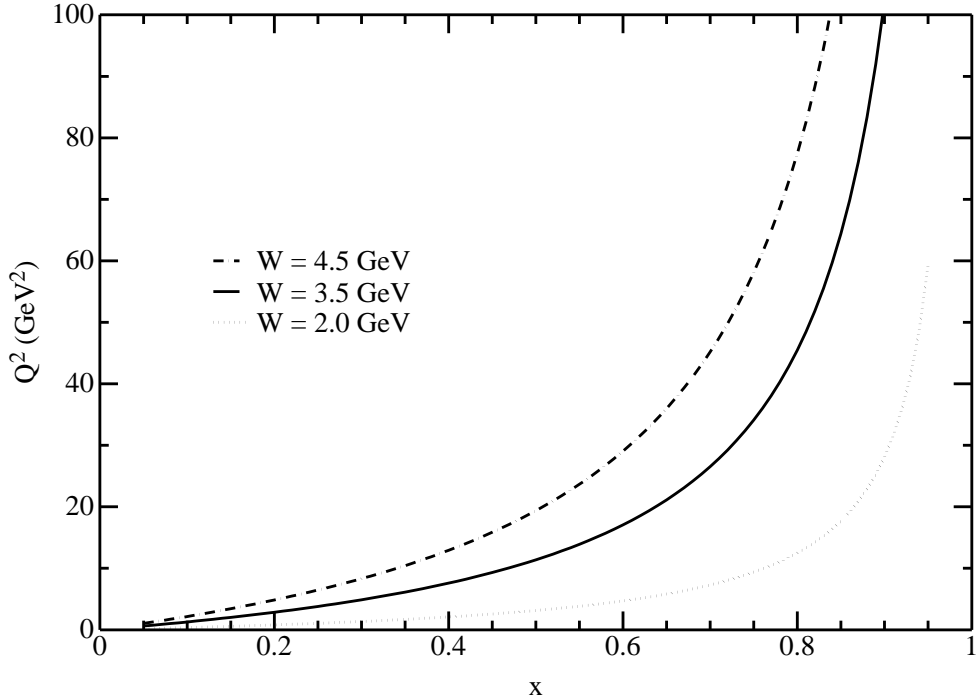}
\caption{$Q^2$ vs.\ $x$ for three values of $W$.  Measurements at very high $x$ demand very high $Q^2$ to remain in the DIS region. \label{wq2}}
\end{figure}

Measurements of structure functions at very high values of $x$ are extremely challenging. The main problems can be seen at a glance from Fig.~\ref{wq2}.  Here the $Q^2$ is plotted from  Eq.\,(\ref{eq:w2q2}) as a function of $x$.  Typical evaluations of the parton distribution functions require that $W \ge 3.5\,{\rm GeV}$.  Clearly, the $Q^2$ necessary to meet this demand is extremely high and impractical, at present, for $x \ge 0.7$.

\begin{figure}[t]
\includegraphics[clip,angle = -90, width=0.8\textwidth]{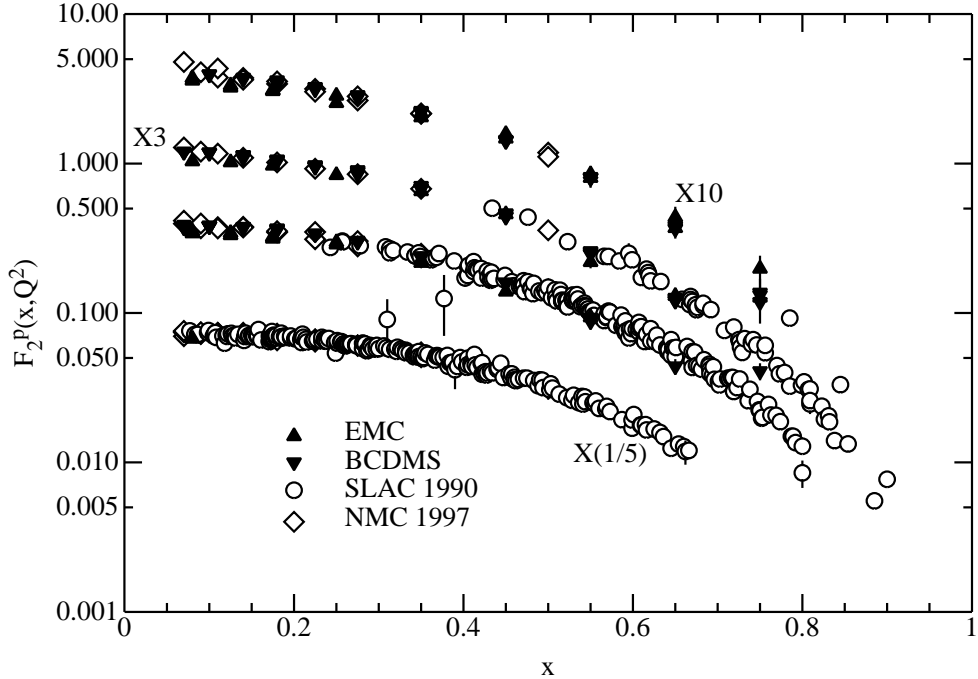}
\caption{Proton structure function vs.\ Bjorken-$x$ with $W \ge 2.0 \ GeV$.  For ease of representation: the upper band of data were multiplied by a factor of 10 and correspond to the domain $25\,{\rm GeV}^2 \leq Q^2 \leq 40\,{\rm GeV}^2$; the next band are scaled by a factor of 3 and cover the domain:  $15\,{\rm GeV}^2 \leq Q^2 \leq 25\,{\rm GeV}^2$; the third-lowest band are unscaled and cover: $7\,{\rm GeV}^2 \leq Q^2 \leq 15\,{\rm GeV}^2$; and the lowest band are scaled by a factor of 1/5 and cover:  $2\,{\rm GeV}^2 \leq Q^2 \leq 7\,{\rm GeV}^2$. \label{f2pq2lg}}
\end{figure}

The first DIS measurements, performed at SLAC, were limited to a beam energy of 24.5\,GeV.  Although this beam energy limit is a serious problem for very high $x$ measurements, the SLAC experiments had good control of systematic errors.  Relatively high luminosities were possible, the incident electron energy was very well known and the spectrometer properties were relatively well understood.  A consistent re-analysis of all the early SLAC experiments was performed in \cite{Whitlow:1992}, and the result is displayed in Fig.~\ref{f2pq2lg}.

After the importance of the early SLAC experiments was realized, the quest for additional data, particularly at high $Q^2$, led to muon scattering experiments at CERN and FNAL.  At CERN a series of experiments was performed at muon energies between 90 and 280\,GeV by the NMC collaboration \cite{Amaudruz:1992}, between 120 and 280\,GeV by the BCDMS collaboration \cite{Benvenuti:1990} and at 280\,GeV by the EMC collaboration \cite{Aubert:1987}.  In this case, the combination of very high beam energies and relatively low luminosities makes it difficult to obtain structure function data at very high $x$.  The practical limit is about $x=0.75$ as shown in  Fig.~\ref{f2pq2lg}.  The FNAL E665 experiment provides data at even lower $x$ than the CERN experiments since the beam energy was 470\,GeV.

Data for various $Q^2$ ranges from the BCDMS \cite{Benvenuti:1990}, EMC \cite{Aubert:1983, Aubert:1987}, NMC \cite{Amaudruz:1992,Arneodo:1996kd} and SLAC \cite{Whitlow:1990} experiments are plotted in Fig.~\ref{f2pq2lg}.  The effect of the kinematic constraints on the reach in $x$ is evident throughout the relatively wide range of values for $Q^2$.  The SLAC data persist out to the highest values of $x$, but only for relatively low values of $W$.

\begin{figure}[t]
\includegraphics[clip,angle = -90, width=0.8\textwidth]{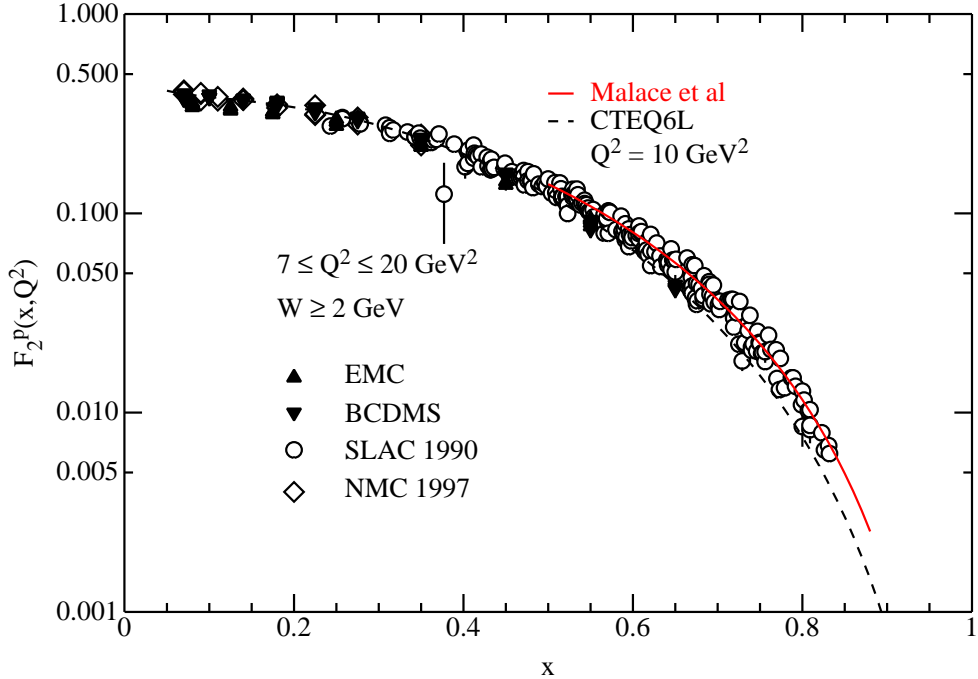}
\caption{(Color online) The proton structure functions vs.\  Bjorken-$x$ with $W \ge 2 \,$GeV.  The dashed black curve represents the CTEQ6L evaluation with $Q^2 = 10\ GeV^2$, while the solid red curve indicates the result obtained when target mass corrections are incorporated \cite{Malace:2009kw}.\label{f2ptgt}}
\end{figure}

If the condition on $W$ is relaxed to $W \ge 2\ GeV$, then data from SLAC exist up to $x = 0.85$ as indicated in Fig.~\ref{f2ptgt}.  Clearly, the CTEQ6L evaluation at the very highest values of $x$ deviate from the data.  However, if target mass corrections are taken into account, as shown by the dashed curve in the figure, then good agreement is restored.  This illustrates that target mass corrections are indispensable in order to fully understand and extract parton distributions at very high $x$.

%Fig.8

%\newpage
\subsection{The neutron structure function}
\label{neutronstructurefunction}
Interest in the neutron structure function at very high $x$ has burgeoned during the past decade.  The neutron structure function, or at least the ratio of the neutron structure function to that of the proton, is believed to be one of the best methods to determine the $d(x)/u(x)$ ratio.  A knowledge of the this ratio in the valence region would provide an important constraint on models for the nucleon \cite{Isgur:1999, Brodsky:1995}.  For example, if one were to assume that a simple $SU(6)$ flavor symmetric model for the nucleon were valid, then the $d/u$ ratio would just be 1/2 for the valence quark distributions.  In this model, the nucleon and $\Delta$ would be degenerate, so this is clearly not a correct description of the nucleon.  Other interesting limits occur, for example, where the $d$-quark in the proton is ``sequestered'' in a pointlike scalar diquark.  In this model, $\left. d/u\right|_{x\to 1}=0$ and $\left.F_2^n/F_2^p\right|_{x\to 1}=1/4$.  Furthermore, in a world where pQCD could be applied naively and scattering at large-$x$ involves only quarks with the same helicity as the target hadron, then $\left. d/u\right|_{x\to 1}=1/5$ and $\left.F_2^n/F_2^p\right|_{x\to 1}=3/7=0.43$.  [An extensive discussion of these issues is provided in Sec.\,\ref{sec:theory}, where it is argued, e.g., that $\left.F_2^n/F_2^p\right|_{x\to 1}=0.36$ and only with reliable data at $x\ge 0.8$ will one be empirically able to determine the $x\to 1$ behavior.]

The parton model is a good starting point for introducing the neutron structure function. Under the assumption of isospin symmetry and recalling Eq.\,(\ref{uvdvnorm}), the neutron structure function is simply obtained from Eq.\,(\ref{eq:pstruc}) by the operation $u\leftrightarrow d$; viz.,
\begin{equation}
\label{eq:nstruc}
 \frac{1}{x} F_2^n(x) = \frac{4}{9}\left[d(x) + \bar{d}(x) +c(x) +  \bar{c}(x) \right] + \frac{1}{9}\left[u(x) + \bar{u(x)} + s(x) + \bar{s(x)}\right]\\
\end{equation}
Neglecting the $c$-quarks, the expression for the neutron-to-proton structure function ratio is
\begin{equation}
 \label{eq:npratio}
    \frac{F_2^n(x)}{F_2^p(x)} = \frac{ u(x) + \bar{u}(x)
 + 4 \left[d(x) + \bar{d}(x)\right] + s(x) + \bar{s}(x)}
    {4\left[u(x) + \bar{u}(x)\right]
 + d(x) + \bar{d(x)} + s(x) + \bar{s}(x)}\,.\\
\end{equation}
{}From this equation it is readily apparent that the ratio is: $1/4$ when u quarks dominate; 4 when d quarks dominate; and, if the sea dominates, the ratio approaches unity.  Within the parton model, Eq.\,(\ref{eq:npratio}) leads to the Nachtmann limit \cite{Nachtmann:1973},
\begin{equation}
\label{eq:nacht}
\frac{1}{4} \leq \frac{F_2^n(x)}{F_2^p(x)} \leq 4\,.
\end{equation}

As interest has grown, a number of new methods for measuring the $F_2^n/F_2^p$ ratio have been proposed.  One approach \cite{Fenker:2003} is to tag very low momentum protons emerging from the deuteron when deep inelastic scattering from the neutron is performed.  This technique minimizes off-shell effects since at very low momentum the neutron is practically a spectator in the deuteron.  Another method \cite{Afnan:2000, Afnan:2003, Bissey:2000, Petratos:2006} is to perform deep inelastic scattering from $^3$He and $^3$H at very high $x$.  In forming the ratio of the scattering rates from these two nuclei, the nuclear effects cancel to a high degree in extracting the $F_2^n/F_2^p$ ratio.

Two proposed methods avoid nuclear effects altogether.  Both a ratio of charge-current neutrino to antineutrino scattering from the proton and parity-violating deep inelastic electron scattering at high $x$ from the proton are sensitive to the $d/u$ ratio.  Nevertheless, both of these methods are fraught with technical difficulties, particularly at high $x$.

The primary problem with measuring the neutron structure function is that there exists no practical free neutron target.  Typically, a deuteron target is employed in experiments.  The neutron structure function is then extracted from a measurement of the proton and
deuteron structure functions by employing a model for the deuteron wave function.  As an example, we consider the covariant approach of \cite{Thomas:1998}.  In this case, the proton and neutron structure functions are convoluted with a nucleon density function in the
deuteron.  The neutron is also off shell, so that a model for the off-shell behavior is necessary.  Then the expression for the deuteron structure function is given by
\begin{equation}
\label{eq:convul}
F_2^D(x) = \frac{1}{2}\sum_N\int_x^{M_D/M}dy'\rho_{N/D}(y')
F_2^N\left(\frac{x}{y'},Q^2\right) + \delta^{off}F_2^D \,,
\end{equation}
where $\rho_{N/D}(y')$ is the probability of finding a nucleon of momentum $y'$ in the deuteron and $\delta^{off}F_2^D$ is the off shell correction.  The extraction process is iterative.  First, the off-shell effect is subtracted from the measured deuteron structure function.  Then the proton structure function, convoluted as indicated by the first term in Eq.\,(\ref{eq:convul}), is subtracted from the remainder, leaving the convoluted neutron structure function.  The neutron structure function is then deconvoluted to give the neutron structure function.  The last three steps are then repeated until convergence is achieved.

\begin{figure}
\includegraphics[clip,angle = -90, width=0.8\textwidth]{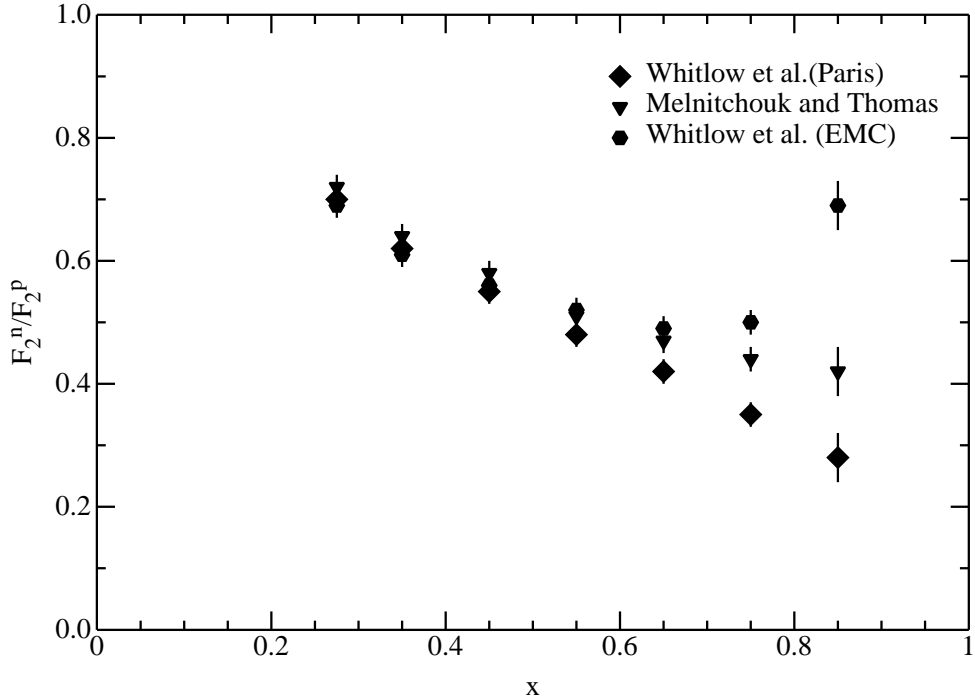}
%\epsfbox{mrstq2.ps}
\caption{\label{f2nratio} The ratio of the neutron to proton structure functions as a function of  $x$. The results of three extraction methods are shown.}
\end{figure}

The results \cite{Thomas:1998} of this procedure for the ratio of the neutron to proton structure functions are shown in Fig.~\ref{f2nratio}.  In this work a covariant deuteron wave function and a consistent off-shell correction were used \cite{Melnitchouk:1994}.  Another application \cite{Burov:2004} of the covariant approach indicates that data for the deuteron structure function at very high $x$ are essential for constraining the high $x$ neutron structure function.  Although the convolution approach has been used by a number of authors as a step toward explaining the EMC effect, the convolution method has no firm theoretical basis.

Two other extractions, also shown in the figure, yield very different results at high $x$.  The extraction, labeled ``Whitlow {\it et al} (Paris)'' \cite{Whitlow:1992} made use of light-front kinematics with the null vector along the incident beam direction and an early deuteron wave function.  The extraction labeled ``Whitlow {\it et al} (EMC)'' \cite{Whitlow:1992} made a density-dependent extrapolation of the EMC effect in heavy nuclei to that in the deuteron.  There is significant controversy \cite{Yang:1999, Melnitchouk:2000, Yang:2000} surrounding the density-dependent extrapolation method.

A recent approach \cite{Arrington:2008zh} made use of the light-front, with null-plane dynamics and modern deuteron wavefunctions.  This framework, which can be applied even when the the constraints of DIS kinematics are not met, greatly simplifies the convolution equation in terms of the nucleon structure functions.  The analysis established that it is important to account properly for the $Q^2$-dependence of the data and incorrect to use a simple convolution formula to analyze data, if that data does not satisfy the DIS kinematic constraints.  When the CD-Bonn potential is used to determine the deuteron wave function, the method of \cite{Arrington:2008zh} produces a ratio similar to that labeled ``Whitlow {\it et al} (Paris)'' in Fig.\,\ref{f2nratio}, which was obtained using a light-front approach with the null vector aligned along the direction of the incident electron beam \cite{Frankfurt:1981mk}.  Whilst this particular choice of null-vector orientation is consistent with DIS kinematics, the very high $x$ SLAC data, included in the analysis, are not.  They are usually excluded when fitting PDFs.

%\subsection{Moments of structure functions}

\subsection{The longitudinal structure function}
\label{sec:longitudinal}
Before the $F_2$ structure function can be extracted from data, one must either know or measure the $R$ structure function.  By combining Eqs.~(\ref{eq:xsec1}), (\ref{eq:f1}) and (\ref{eq:f2}), the expression for the cross section in terms of $F_1(x,Q^2)$ and $F_2(x,Q^2)$ is given by
\begin{equation}
\label{eq:xsec2}
  \frac{d\sigma  }{d\Omega dE' }(E,E',\theta) =
4\frac{\alpha_e^2 E'^2}{Q^4}\left[\frac{F_2(x, Q^2)}{\nu} \cos^2\frac{\theta}{2}
 + \frac{2F_1(x,Q^2)}{M} \sin^2\frac{\theta}{2}\right].
\end{equation}

This cross section is often expressed in terms of the $F_2$ and $R$ structure functions in the following way:
\begin{equation}
\label{eq:xsec3}
  \frac{d\sigma}{d\Omega dE' }(E,E',\theta) =
4\frac{\alpha_e^2 E'^2}{Q^4}\frac{F_2(x,Q^2)}{\nu} \cos^2\frac{\theta}{2}\left[ 1 + \frac{1-\epsilon}{\epsilon}\frac{1}{1 + R(x,Q^2)}\right],
\end{equation}
where
\begin{equation}
\label{eq:epsilon}
  \epsilon = \left[ 1 + 2 ( 1 +\nu^2/Q^2)\tan^2\frac{\theta}{2}\right]^{-1},
\end{equation}
and
\begin{equation}
\label{eq:R}
R(x,Q^2) \equiv \frac{\sigma_L}{\sigma_T} =
\frac{F_2(x,Q^2)}{2xF_1(x,Q^2)}\left(1+\frac{2Mx}{\nu}\right) - 1 \,.\\
\end{equation}

Here, $\sigma_L$ and $\sigma_T$ refer to the longitudinal and transverse cross sections, respectively.  Since the quarks are pointlike spin-$1/2$ objects, the cross-section for absorption of a longitudinal photon is small in comparison with that for a transverse photon.  Furthermore, a vector-vector interaction preserves chirality, so the longitudinal cross-section will depend on violations of this chirality.  Violations of order $m_q^2/Q^2$ would be expected, where $m_q$ is the struck quark's current-mass.  Intrinsic transverse quark momentum, $k_T$, as well as higher orders in $\alpha_s$ from the gluon distribution also can give rise to an increase in the longitudinal cross section. For example, the Bjorken-Feynman model gives the result \cite{Feynman:1972} that
\begin{equation}
\label{eq:RFeynman}
R = \frac{4(m_q^2 + <k_T^2>)}{Q^2}\,.
\end{equation}

The QCD contribution to $F_L$, and consequently to $R$, of order $\alpha_s$ is given by \cite{Reya:1981, Altarelli:1982}
\begin{equation}
\label{eq:FL}
F_L(x,Q^2) = \frac{\alpha_s}{2\pi}x^2 \int_x^1 \frac{dy}{y^3}
\left[\frac{8}{3}F_2(y,Q^2) + \frac{40}{9}\left(1 - \frac{x}{y}\right)yG(y,Q^2)\right] + \mathcal{O}(\alpha_s^2)\,, \\
\end{equation}
where $G$ is the gluon distribution function, and only four flavors of quarks are considered in the second term.
%The longitudinal--transverse separation is performed by the conventional Rosenbluth technique.
An excellent extraction \cite{Whitlow:1990} of $F_2$ and $R$ was performed from a global analysis of the SLAC data.% and is shown for some kinematics in Fig. -- Missing figure}

It has long been believed that a tight constraint on the gluon structure function can be built from good measurements of $F_L$.  The best extractions for $R$ in the valence region are from the lower energy SLAC and JLab data \cite{Whitlow:1990,Yang:2000, Tvaskis:2007}.  The best extractions of $F_L$ are from HERA data, but at lower $x$ than the subject of this review.  Nevertheless, the deduced gluon distribution is very sensitive to the order of $\alpha_s$ used in the analysis \cite{Martin:2006}.

\begin{figure}[t]

\centerline{%\hspace*{2em}%
\includegraphics[clip,width=0.8\textwidth]{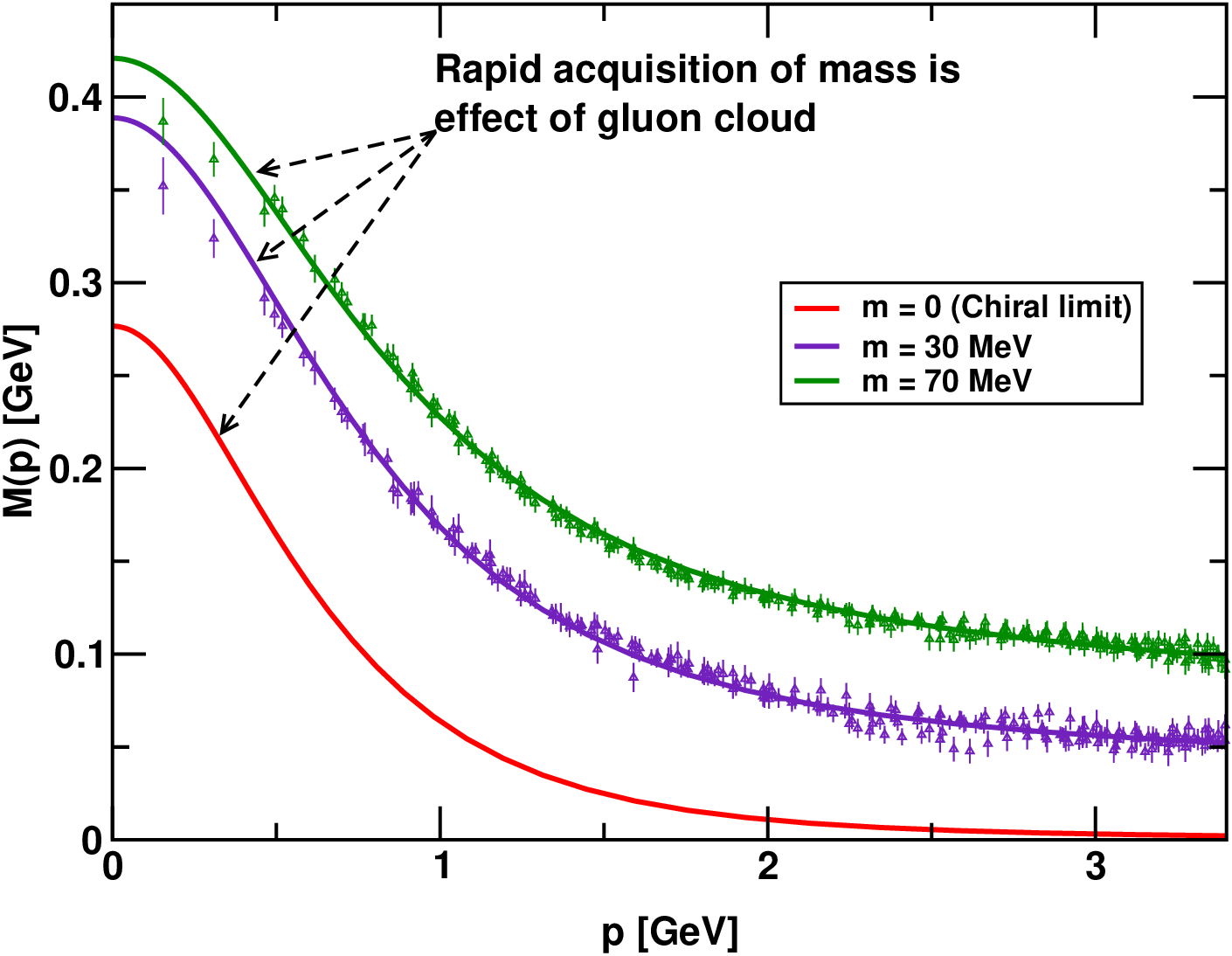}}

\caption{\label{gluoncloud} (Color online)
Dressed-quark mass function, $M(p)$: solid curves -- Dyson-Schwinger equation (DSE) results, obtained as explained in \protect\cite{Bhagwat:2003vw,Bhagwat:2006tu}, ``data'' -- numerical simulations of unquenched lattice-QCD \protect\cite{Bowman:2005vx}.  (NB.\ $m=70\,$MeV is the uppermost curve and current-quark mass decreases from top to bottom.)  One observes the current-quark of perturbative QCD evolving into a constituent-quark as its momentum becomes smaller.  The constituent-quark mass arises from a cloud of low-momentum gluons attaching themselves to the current-quark.  This is dynamical chiral symmetry breaking (DCSB): an essentially nonperturbative effect that generates a quark mass \emph{from nothing}; namely, it occurs even in the chiral limit.  %In QCD, the vast bulk of the light-quark constituent mass arises from gluons through DCSB.
[Adapted from \protect\cite{Bhagwat:2007vx}.]}
\end{figure}

%\begin{figure}
%\epsfysize=3.0in
%\includegraphics[angle = 0, width=4.0in]{quarkmass.eps}
%\caption{Numerical simulations of lattice QCD (data, at two different bare masses) have confirmed model predictions (solid curves) that the vast bulk of the light-quark constituent mass arises from the gluons\label{massfromnothing}}
%\end{figure}

\subsection{The gluon structure function}
Interest in the gluon structure function has grown markedly during the past decade.  Part of the interest resides in the fact that gluons comprise more than 98\% of the rest-mass of the nucleon as well as more than half of it's light-front momentum.  This interesting aspect of the role of glue in the nucleon can be seen from Fig.~\ref{gluoncloud} .  Here the mass of a light-quark is plotted as a function of the quark's momentum.  The ``data'' in the plot are results from lattice simulations, while the curves are from Dyson-Schwinger equation (DSE) calculations.  The curve labeled ``chiral limit'' is the DSE result obtained with the current-quark mass set to zero.  Clearly, even if the current-quark mass vanishes, the quark mass rises rapidly to near 0.3\,GeV at low quark momentum.  This is dynamical chiral symmetry breaking (DCSB) and a clear demonstration in QCD of the effect on the quark mass produced by the presence of gluons with strong self-interactions.

Sensitivity to the gluon structure function via lepton beams can be produced by a careful measurement of scaling violations; i.e., measurements of the longitudinal structure function or the partial derivative of $F_2(x, Q^2)$ with respect to $\ln{Q^2}$.  This is particularly effective at low values of $x$ where this derivative is directly proportional to the gluon distribution function at first order.

%\subsection{Future experiments at JLab and EIC}
%\label{EIC}
%
%The 12\,GeV upgrade at JLab permits a new opportunity to study structure functions at extremely high $x$.  Although the new energy that will be available at JLab will not be sufficient to avoid target-mass corrections for the nucleon at the higher values of $x$, it does represent a new opportunity to obtain precise data on the proton and neutron structure functions.  It might also represent a new opportunity \cite{Wijesooriya:2001} to study the pion structure function, although assumptions concerning virtual pions would have to be invoked.

%The electron ion collider (EIC) is aimed at polarization studies of the nucleon structure function at lower values of $x$.  Nevertheless, the EIC may permit new studies of the ratio of the neutron to proton structure functions at high $x$.  The EIC will permit access to the virtual pion and kaon structure function through measurements \cite{Holt:2000cv} of the forward nucleon structure functions. 
%\newpage

%\section{Nucleon distribution functions from neutrino scattering}
%\label{sec:neutrino}
%\subsection{The hadronic current}
%\subsection{The parity odd distribution function}
%\subsection{Strange quark distributions}
%\subsection{Future plans at FNAL}
%\newpage

%\setcounter{figure}{0}
%\setcounter{table}{0}
\setcounter{equation}{0}
\section{Distribution functions from Drell-Yan interactions}
%\section{Structure Functions from Drell-Yan Interactions}
\label{sec:drellyan}
Informative reviews on the Drell-Yan process exist \cite{Kenyon:1982,Reimer:2007}.  In this section we emphasize that the Drell-Yan interaction provides: (i) the cleanest access to the antiquark distributions in the valence region in hadrons; (ii) a means to determine the quark distributions in the proton at very high $x$; and (iii) the most information on the parton distributions in mesons.  Naturally, in order to isolate the valence quark distributions in the proton, it is imperative to measure the antiquark distributions.

\subsection{The Drell-Yan interaction}
\label{sec:dy}
The Drell-Yan process was devised \cite{Drell:1970a, Drell:1971} to explain hadron-hadron collisions where an anti-lepton lepton pair is produced.  For example, two hadrons, $A$ and $B$, collide and produce the lepton pair $l^+ l^-$:
\begin{equation}
\label{eq:dy}
 A + B \longrightarrow l^+ + l^- + X \,.
\end{equation}
At relatively large values of momentum transfer, say greater than a few GeV$^2$, the underlying process is believed to be dominated by antiquark-quark annihilation:
\begin{equation}
\label{eq:dyq}
 \bar{q} + q \longrightarrow l^+ + l^-\,.
\end{equation}

\begin{figure}[t]
\vspace*{-16ex}

\includegraphics[clip,width=0.80\textwidth]{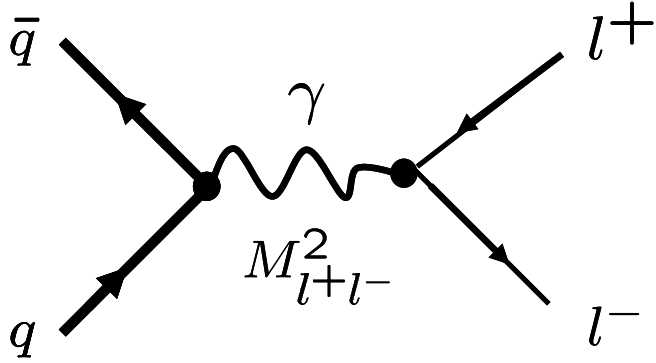}
\vspace*{-16ex}

\caption{Feynman diagram representing the Drell-Yan interaction \label{dyfeyn}}
\end{figure}

The Feynman diagram for this underlying Drell-Yan interaction is given in Fig.~{\ref{dyfeyn}}.  Here, two hadrons collide and a time-like photon of mass $M_{l^+l^-}$ is formed from the annihilation of a quark in one of the hadrons with an antiquark in the other.  The photon then decays by emitting an antilepton-lepton pair.  In a typical experiment, the lepton pair is often an antimuon-muon pair, as a matter of technical convenience.

In this process the square of the four momentum transfer is just given by the square of the dilepton mass, $M_{l^+l^-}^2$.  At leading-order, the Drell-Yan cross-section can be determined in a straightforward manner to yield:
\begin{equation}
\label{eq:xsecdy}
  \frac{d\sigma  }{dx_b dx_t } =
\frac{4\pi\alpha_e^2}{9M_{l^+l^-}^2}\sum_i e_i^2\left[q_{bi}(x_b, Q^2)\bar{q}_{ti}(x_t,Q^2)
 + \bar{q}_{bi}(x_b,Q^2)q_{ti}(x_t,Q^2)\right],
\end{equation}
where $q_{b(t)i}(x_{b(t)i})$ refers to the structure function of the quark of flavor $i$ in the beam (b) or target (t) hadron, respectively.  It has been shown \cite{Altarelli:1979} that if the structure functions have the same definition as those in deep inelastic scattering; namely, that if one employs the DIS factorization scheme, then the NLO part is calculable and becomes a multiplicative factor to the expression above, as discussed in the next subsection.

\begin{figure}[t]

\includegraphics[width=0.50\textwidth]{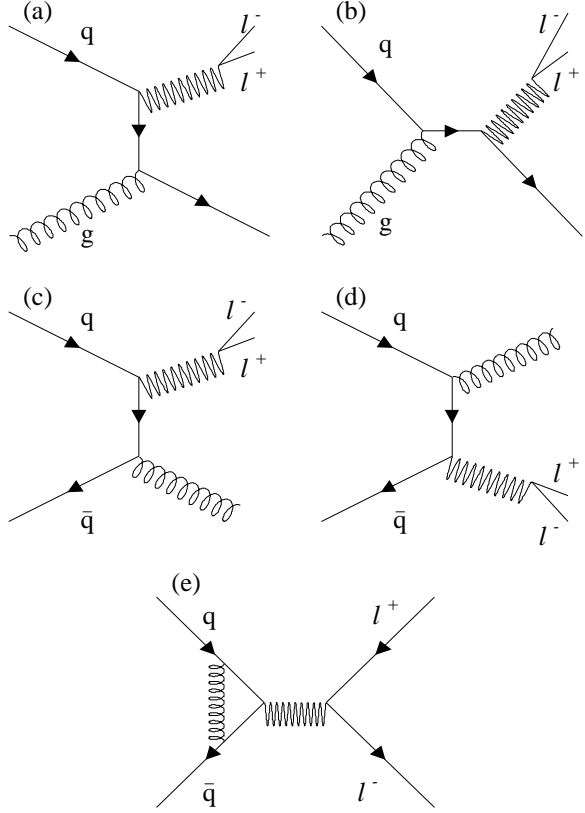}

\caption{Feynman graphs of NLO corrections to the Drell-Yan process.  \emph{Top row} -- Compton-like scattering in QCD; viz., $gq \rightarrow \gamma^*$ (and also $g\bar{q} \rightarrow \gamma^*$, not shown explicitly).  \emph{Middle row} -- QCD radiative correction; viz., $q\bar{q} \rightarrow \gamma^*$.  \emph{Bottom diagram} -- QCD vertex correction.
[Figure adapted from \protect\cite{Reimer:2007}.] \label{dynlo}}
\end{figure}

\subsection{QCD and higher order corrections}
The NLO QCD processes that contribute to Drell-Yan scattering are depicted in Fig.~\ref{dynlo}. These processes lead to a modification of the Drell-Yan cross section by introducing the so-called $K$ factor:
\begin{equation}
\label{eq:kxsecdy}
\frac{d\sigma }{dx_b dx_t }\left(NLO\right) = K_{NLO}\frac{d\sigma  }{dx_b dx_t }\left(LO\right) .
\end{equation}
With PDFs defined in the DIS factorization scheme, the $K_{NLO}$-factor is given approximately by \cite{Altarelli:1979}
\begin{equation}
\label{eq:kfacdy}
  K_{NLO} \approx 1 + \frac{\alpha_s}{2\pi}\left(1+\frac{4}{3}\pi^2\right)
\end{equation}
and assumes a value between 1.5 and 2.  The consideration of NNLO, as well as NLO diagrams, also leads to a simple factorization of the cross-section and an approximate factor of two for $K$.  The factorization scheme dependence of the K-factor is described at length in \cite{vanNeerven:1991gh}.
We note in addition that the $K$ factor depends on kinematics, a fact shown in \cite{Wijesooriya:2005ir} to be important at very high $x$ for pionic Drell-Yan studies.

\subsection{High-$x$ quark distribution functions}
The Drell-Yan process presents a valuable method for measuring parton distribution functions in hadrons at very high $x$.
For example, it can be used to probe the quark distribution in the beam proton.  To see how, consider that if $s$- and $c$-quarks are neglected and the beam-target kinematics are chosen such that $x_F:=x_b-x_t$ is large, then for proton$+$proton collisions, Eq.\,(\ref{eq:kxsecdy}) can be rewritten \cite{Webb:2003bj}
\begin{equation}
\label{eq:xsdy}
  \frac{d\sigma  }{dx_b dx_t } \approx
\frac{4\pi\alpha_e^2 K}{81s}\left[4 \, u_{b}(x_b, Q^2)\bar{u}_{t}(x_t,Q^2)
 + d_{b}(x_b,Q^2)\bar{d}_{t}(x_t,Q^2)\right]
\end{equation}
because $\bar q_b(x_b)\ll q_b(x_b)$ and $q_t(x_t)\ll \bar q_t(x_t)$ for large-$x_F$.
%If one assumes further that $|\bar u_t(x)-\bar d_t(x)| \ll |\bar u_t(x)+\bar d_t(x)|$ on this kinematic domain, then
%\begin{equation}
%\label{eq:xsdy1}
%  \frac{d\sigma  }{dx_b dx_t } \approx
%\frac{4\pi\alpha_e^2 K}{81 s} \left[ 4u_{b}(x_b, Q^2) + d_{b}(x_b,Q^2) \right]
%\frac{1}{2} \left[ \bar{u}_{t}(x_t,Q^2) + \bar{d}_{t}(x_t,Q^2) \right].
%\end{equation}
%If one chooses to average over $x_t$ when analyzing data, then Eq.\,(\ref{eq:xsdy1}) can be understood in a weaker sense; namely, so long as
%\begin{equation}
%\int_0^1 dx_t \, [\bar u(x_t) - \bar d(x_t)] \ll \int_0^1 dx_t \, [\bar u(x_t) + \bar d(x_t)]\,,
%\end{equation}
%then
%\begin{equation}
%\label{eq:xsdy2}
%\frac{d\sigma  }{dx_b} \approx \frac{4\pi\alpha_e^2 K}{81 s} \left[ 4u_{b}(x_b, Q^2) + d_{b}(x_b,Q^2) \right] \frac{1}{2} \int_0^1 dx_t
%\left[ \bar{u}_{t}(x_t,Q^2) + \bar{d}_{t}(x_t,Q^2) \right].
%\end{equation}
%\begin{equation}
%\label{eq:xsdy2}
%  \frac{d\sigma  }{dx_b dx_t } \approx
%\frac{4\pi\alpha_e^2 K}{81 s} \left[ 4u_{b}(x_b, Q^2) + d_{b}(x_b,Q^2) \right]
%\frac{1}{2} \left[ \bar{u}_{t}(x_t,Q^2) + \bar{d}_{t}(x_t,Q^2) \right].
%\end{equation}
%
Now suppose that the target is a deutron with a similar kinematic setup, then Eq.\,(\ref{eq:kxsecdy}) can be written \cite{Webb:2003bj}
\begin{equation}
\label{eq:xsbdy}
  \frac{d\sigma  }{dx_b dx_t } \approx
\frac{4\pi\alpha_e^2 K}{81s} \left[ 4u_{b}(x_b, Q^2) + d_{b}(x_b,Q^2) \right]
 \left[ \bar{u}_{t}(x_t,Q^2) + \bar{d}_{t}(x_t,Q^2) \right],
\end{equation}
where, as usual herein, $q(x)$ means the distribution of flavor-$q$-quarks in the proton; and one has assumed isospin symmetry and neglected nuclear binding effects in the weakly-bound deuteron.  %Following this reasoning, the $p\,d$ differential cross-section is approximately twice that for $p \,p$, a statement which should be most accurate when the data for both reactions are integrated over $x_t$.
It is thus apparent that this kinematic setup produces Drell-Yan cross sections that are primarily
sensitive to the valence distributions in the proton beam, and the antiquark distributions at small $x_t$ in the proton and deuteron targets.

\begin{figure}[t]
\includegraphics[clip,width=0.80\textwidth]{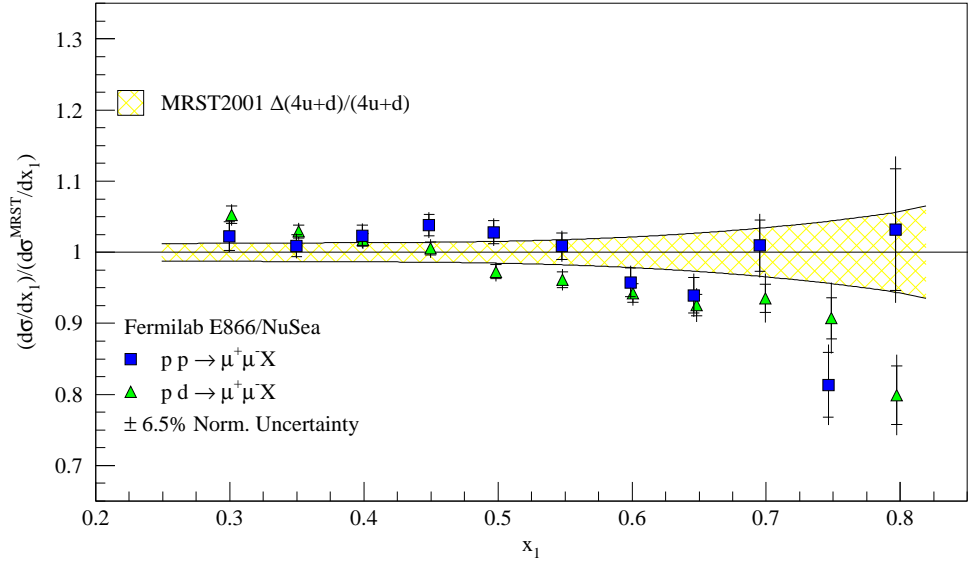}
\caption{\label{highxdy}
(Color online)
Ratios of the measured Drell-Yan $p\,p$ (squares) and $p\,d$ (triangles) cross-sections to NLO calculations based on the MRST\,2001 PDF fit \protect\cite{Martin:2001es}, plotted as a function of beam momentum fraction, here labeled $x_1$, with averaging over the target momentum fraction.
%
%As described in connection with Eqs.\,(\protect\ref{eq:xsdy}) and (\protect\ref{eq:xsbdy}), these cross-sections are primarily sensitive to the valence approximately proportional to $4u(x)+d(x)$ in the beam proton.
%
The shaded area between the solid lines represents the experimental uncertainty ranges \protect\cite{Martin:2002aw} on $4u(x)+d(x)$ in the MRST\,2001 PDF fit.
The data are from the FNAL E866 experiment. [Figure adapted from \cite{Webb:2003bj,Webb:2003ps}.]}
\end{figure}

Some results from analysis of the FNAL E866 Drell-Yan data are reproduced in Fig.\,\ref{highxdy}.  In this figure the E866 results for $p\,p$ and $p\,d$ collisions are divided by the appropriate differential cross-section computed using the MRST\,2001 PDFs \cite{Martin:2001es}.  %The deviation from unity at high-$x$ indicates that the MRST\,2001 values for the valence-parton distributions are overestimated.  Although it is probable the data are indicating that the $u(x)$ distribution has been overestimated, a stronger conclusion cannot be drawn without a separation of the flavor dependence of the parton distribution function.  It appears that the assumptions leading to Eqs.\,(\ref{eq:xsdy2}) and (\ref{eq:xsbdy}) are valid within the precision of the data.
Analogous ratios plotted as a function of $x_t$ \cite{Webb:2003ps} indicate that the MRST\,2001 PDFs provide a very good description of the $p\,p$ cross-section's $x_t$-dependence on the complete $x_t$-domain, and a good description of the $p\,d$ cross-section for $x_t < 0.15$.
In this light, consider first the $p\,p$ data, which hints that the plotted ratio is smaller than one at $x_b\gtrsim 0.6$.  Given that the $u$-quark is responsible for roughly 80\% of the cross-section, this observation can be interpreted as an indication that the MRST\,2001 PDFs overestimate the proton's valence $u$-quark distribution.  There is a strong signal from the $p\,d$ cross-section that the plotted ratio is less than unity for $x_b \gtrsim 0.5$.  Given that this cross-section is proportional to $4u(x)+d(x)$ and the greater suppression, one can argue that the PDFs overestimate the proton's valence $d$-quark distribution.
A consideration of the impact of this and other recent data on PDF fits is presented in \cite{Owens:2007kp}.

\subsection{$\bar{d}/\bar{u}$ ratio and the Gottfried sum rule}
One of the most celebrated applications of the Drell-Yan process is the measurement of the flavor dependence of the antiquark sea in the proton.  This was first suggested more than twenty years ago \cite{Bickerstaff:1985ax,Bickerstaff:1985da,Garvey:1986,Ellis:1990ti}, at which time it was usually assumed that the light-quark sea was flavor symmetric.  Such experiments led subsequently to a demonstration that this is not true; namely, the light-quark sea is flavor asymmetric.

One of the first indications that the light-quark sea might be flavor asymmetric was observation of the violation of the Gottfried sum rule.  The Gottfried sum rule is defined by \cite{Gottfried:1967}
\begin{equation}
\label{eq:gottfried}
S_{G} := \int_0^1 \frac{dx}{x}\, (F_2^p(x) - F_2^n(x))\,.
\end{equation}
Using Eqs.~(\ref{eq:uvalence}) -- (\ref{uvdvnorm}) and assuming isospin invariance, $S_G$ becomes:
\begin{equation}
\label{eq:SG}
S_{G} =
\frac{1}{3} - \frac{2}{3}\int_0^1\! dx\, \left[(\bar{d}(x) - \bar{u}(x)\right]=: \frac{1}{3} - \frac{2}{3} {\cal I}_{\bar d -\bar u} \,.
\end{equation}
If the sea is flavor symmetric, then the Gottfried sum just evaluates to $1/3$.  Any other value is termed a ``violation'' of the sum rule.

The earliest hints of a possible violation of the sum rule can be found in data from SLAC \cite{Bodek:1973} and Fermilab \cite{Ito:1980ev}.  The NMC experiment at CERN gave the result \cite{Amaudruz:1991at,Arneodo:1994sh,Arneodo:1996kd}
\begin{equation}
S_G = 0.235 \pm 0.026\,,
\end{equation}
significantly different from $1/3$.  A more recent re-evaluation of the Gottfried sum, using a neural network parametrization of all then available data on the nonsinglet structure function $F_2^{\rm NS}=F_2^p-F_2^n$ \cite{Benvenuti:1989fm,Benvenuti:1989rh,Arneodo:1996qe}, yielded  $S_G = 0.244 \pm 0.045 \Rightarrow {\cal I}_{\bar d -\bar u} = 0.134 \mp 0.068 $ \cite{Abbate:2005}, in agreement with the earlier analyses.

Of course, it was pointed out long ago that Pauli blocking would give some enhancement of the ratio $\bar{d}/\bar{u}$ \cite{Field:1977}.  This mechanism might have been sufficient to explain the early SLAC data but Drell-Yan experiments have since given far more information about the magnitude of the effect.  Pion cloud models also generate such an effect \cite{Thomas:1983fh}.

\begin{table}[t]
\caption{\label{gottfriedt} Values of ${\cal I}_{\bar d -\bar u}$, which appears in the Gottfried sum rule, Eq.\,(\protect\ref{eq:SG}), compiled in \protect\textcite{Peng:2003}.  The original sources are: NMC (DIS) \cite{Amaudruz:1991at,Arneodo:1994sh}; HERMES (semi-inclusive DIS) \protect\cite{Ackerstaff:1998sr}; and E866 (Drell-Yan) \protect\cite{Hawker:1998ty,Peng:1998pa,Towell:2001nh}.
NB.\ At NNLO, ${\cal I}_{\bar d -\bar u}$ is almost scale-independent for $Q^2\in [1,100]\,$GeV$^2$, as illustrated in Fig.\,3 of \protect\cite{Abbate:2005}.}
\begin{tabular}{|l|c|}
\hline
~Experiment~ & ~${\cal I}_{\bar d -\bar u} =\int_0^1\!dx\, \left[(\bar{d}(x) - \bar{u}(x)\right]$~ \\
\hline
~NMC         &  0.147 $\pm$ 0.039     \\
~HERMES    &    0.16~ $\pm$ 0.03~~      \\
~E866    &    0.118 $\pm$ 0.011      \\ \hline
\end{tabular}
\end{table}

A compilation \cite{Peng:2003} of values for ${\cal I}_{\bar d -\bar u}$, which characterizes the second term in Eq.\,(\ref{eq:SG}), are given in Table\,\ref{gottfriedt}.  The relative agreement is reasonable among these determinations.  The most accurate result quoted is from the Drell-Yan experiment, FNAL E866, and the comparison is meaningful because ${\cal I}_{\bar d -\bar u}$ is almost scale-independent for $Q^2\in [1,100]\,$GeV$^2$.

The power of the Drell-Yan technique can be illustrated by making some simple assumptions.  Consider the yield per nucleon from a proton beam incident on a target nucleus of atomic mass A, charge Z and neutron number N.  Assuming that the yield, $Y_{pA}$, from the process factorizes into a simple sum of $p\,p$ and $p\,n$ interactions, then
\begin{equation}
\label{eq:xA}
Y_{pA} \sim \left(Z\sigma_{pp} + N\sigma_{pn}\right)/A\,,
\end{equation}
where $\sigma_{pp}$ and $\sigma_{pn}$ are the proton-proton and proton-neutron Drell-Yan cross-sections, respectively.  If it is further assumed that $p\,p$ Drell-Yan interactions are dominated by the target proton's $\bar u_p\,$-quark distribution and those for $p\,n$ are dominated by the neutron's $\bar u_n=\bar d_p\,$-quark distribution, then
\begin{equation}
\label{eq:xsecratio}
\frac{\sigma_{pp}}{\sigma_{pn}} \approx \frac{\bar{u}_p}{\rule{0em}{2.5ex}\bar{d}_p}.
\end{equation}

With Eqs.\,(\ref{eq:xA}) and (\ref{eq:xsecratio}), it is readily seen that the ratio of yields for proton-nucleus to that of proton-deuteron interactions becomes
\begin{equation}
\label{eq:yratio}
\frac{Y_{pA}}{Y_{pd}} \approx 1 + \frac{(N - Z)(\bar{u} - \bar{d})}{A(\bar{u} + \bar{d})}\,,
\end{equation}
where we now return to our convention of omitting the subscript ``$p$'' when distribution functions in the proton are meant.  Clearly, if the flavor of the light-quark sea is symmetric; i.e., $\bar{u} = \bar{d}$, then this ratio of yields becomes unity.  By considering the ratio of $p\,p$ to $p\,d$ Drell-Yan scattering, then the ratio becomes:
\begin{equation}
\label{eq:pppdratio}
\frac{Y_{pp}}{Y_{pd}} \approx 1 + \frac{\bar{u} - \bar{d}}{\bar{u} + \bar{d}}
\end{equation}
and the ratio $\bar{d}/\bar{u}$ becomes accessible experimentally \cite{Bickerstaff:1985ax,Bickerstaff:1985da,Garvey:1986}.
Of course, the experiments are analyzed in a more sophisticated manner without these simplifying assumptions.

\begin{figure}[t]
\includegraphics[clip,width=0.80\textwidth]{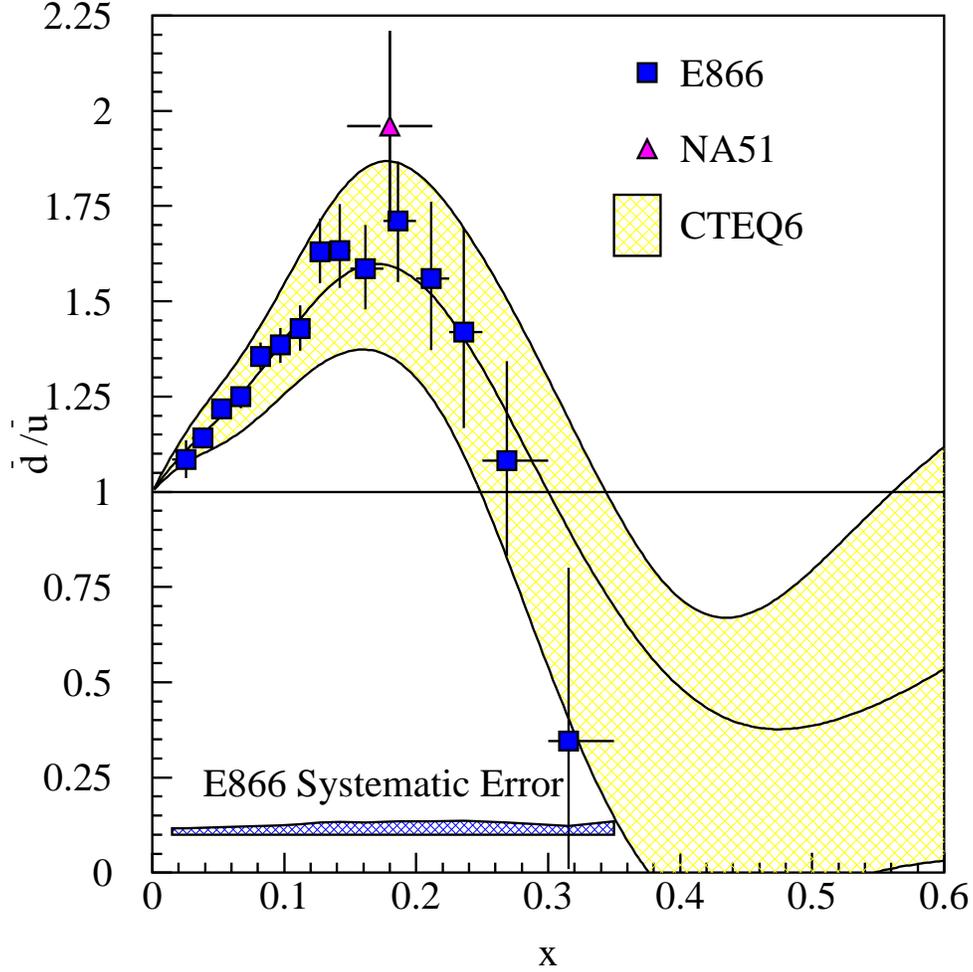}
\caption{(Color online)
$x$-dependence of $\bar d/\bar u$.  The solid blue squares represent the measurements from FNAL E866 \cite{Towell:2001nh,Hawker:1998ty}, while the red triangle indicates the CERN NA51 measurement \cite{Baldit:1994jk}.  The yellow shaded region indicates the uncertainty from the CTEQ6 evaluation of the world data. \label{dbarubar}}
\end{figure}

The first Drell-Yan experiment that indicated that the sea was not flavor symmetric was experiment NA51 at CERN \cite{Baldit:1994jk}.  This experiment found that $\bar{u}/\bar{d} = 0.51 \pm 0.04 \pm 0.05$ at $x = 0.18$.  Experiment E866 made a more comprehensive study by extending the x-range of the experiment \cite{Towell:2001nh,Hawker:1998ty}.  These results are shown in Fig.\,\ref{dbarubar}.

As argued, e.g., in \cite{Steffens:1996bc}, Pauli blocking is not sufficient to explain the large effect observed in modern experiments.  A number of theoretical explanations have been advanced to explain this effect in the context of models: pion cloud; chiral quark; chiral soliton; and instanton.  %These models will be discussed in section  XXX.
As real information about this phenomenon is only available for $x \lesssim 0.3$ and our focus is $x\gtrsim 0.4$, we do not discuss it further herein but refer the interested reader to \cite{Speth:1996pz,Kumano:1997cy,Garvey:2001yq}, and references therein and thereto.

From Eq.~(\ref{eq:yratio}), it should be apparent that the protonic Drell-Yan interaction with nuclei is an especially valuable means to measure anti-quark distributions in nuclei.  In fact, this method has been used \cite{McGaughey:1992kz} and proposed \cite{Reimer:2007} to search for an antiquark or, equivalently, a pion excess in nuclei.  It is believed \cite{Friman:1983,Thomas:1983fh} that observation of the pion excess in nuclei would provide a stringent confirmation of our understanding of conventional nuclear theory, where the nuclear binding is produced by pion exchange.  Thus far, no evidence for a pion or antiquark excess in nuclei has been discovered \cite{Gaskell:2001fn,McGaughey:1992kz}.
% Gaskell:2001fn=charged pion electroproduction from 1H, 2H, and 3He
% McGaughey:1992kz=DY

\subsection{The pion structure function}
\label{exptpionstructure}

\begin{figure}[t]
\includegraphics[clip,width=0.80\textwidth]{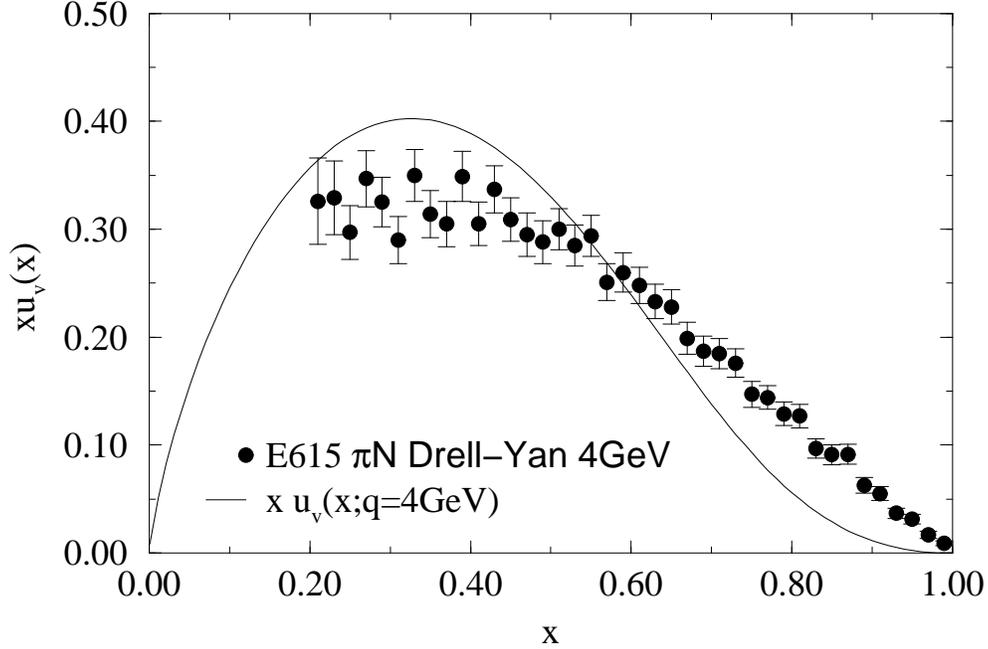}
\caption{The $\bar{u}$ quark distribution function in the $\pi^-$.  The data are from a LO analysis of the FNAL E615 experiment \cite{Conway:1989fs}.  The curve reproduces a Dyson-Schwinger equation calculation \cite{Hecht:2000xa}, described in Sec.\,\protect\ref{sec:DSE}, which is consistent with QCD expectations.  The discrepancy is disturbing and hitherto unresolved.}
\label{pionP}
\end{figure}

The pion plays a key role in nucleon and nuclear structure.  It has not only been used to explain the long-range nucleon-nucleon interaction, forming a basic part of the Standard Model of Nuclear Physics \cite{Pieper:2001mp,Wiringa:2006ih}, but also, e.g., to explain the flavor asymmetry observed in the quark sea in the nucleon.  However, compared to that of other hadrons, the pion mass is anomalously small.  This owes to dynamical chiral symmetry breaking and any veracious description of the pion must properly account for its dual role as a quark-antiquark bound-state and the Nambu-Goldstone boson associated with DCSB \cite{Maris:1997hd}.  It is this dichotomy and its consequences that makes an experimental and theoretical elucidation of pion properties so essential to understanding the strong interaction.

\begin{figure}
\includegraphics[angle=0, width=0.50\textwidth]{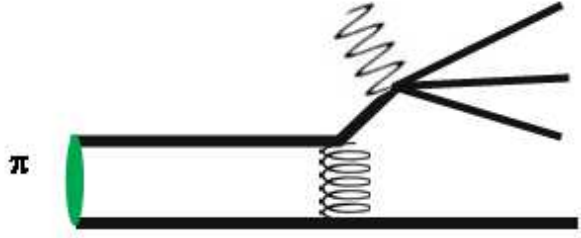}
\caption{(Color online)
Feynman diagram of the parton picture of DIS on the pion, where one hard gluon exchanges prior to absorption of the highly energetic photon, so that the struck parton carries most of the momentum of the pion.  Evaluation of this diagram leads to a $(1-x)^2$ dependence of the pion structure function at very high $x$ \protect\cite{Ezawa:1974wm,Farrar:1975yb}. \label{pionparton}}
\end{figure}

Experimental knowledge of the parton structure of the pion arises primarily from pionic Drell-Yan
scattering from nucleons in heavy nuclei \cite{Conway:1989fs, Conway:1987hv,Heinrich:1991, Falciano:1986wk, Guanziroli:1987rp, Badier:1983mj,Betev:1985pf}.  A LO analysis of results from FNAL\,E615 \cite{Conway:1989fs} are shown in Fig.\,\ref{pionP} but the shape of the empirically extracted pion distribution function at high-$x$ is contentious.
%As we explain in Sec.\,\ref{sec:DSE}, Here we note that theoretical descriptions of pionic parton structure at high-$x$ disagree.
%\cite{Hecht:2000xa}.
%At this point we note that theoretical descriptions of pionic parton structure at high-$x$ disagree.

\begin{figure}[t]
\includegraphics[clip,width=0.80\textwidth]{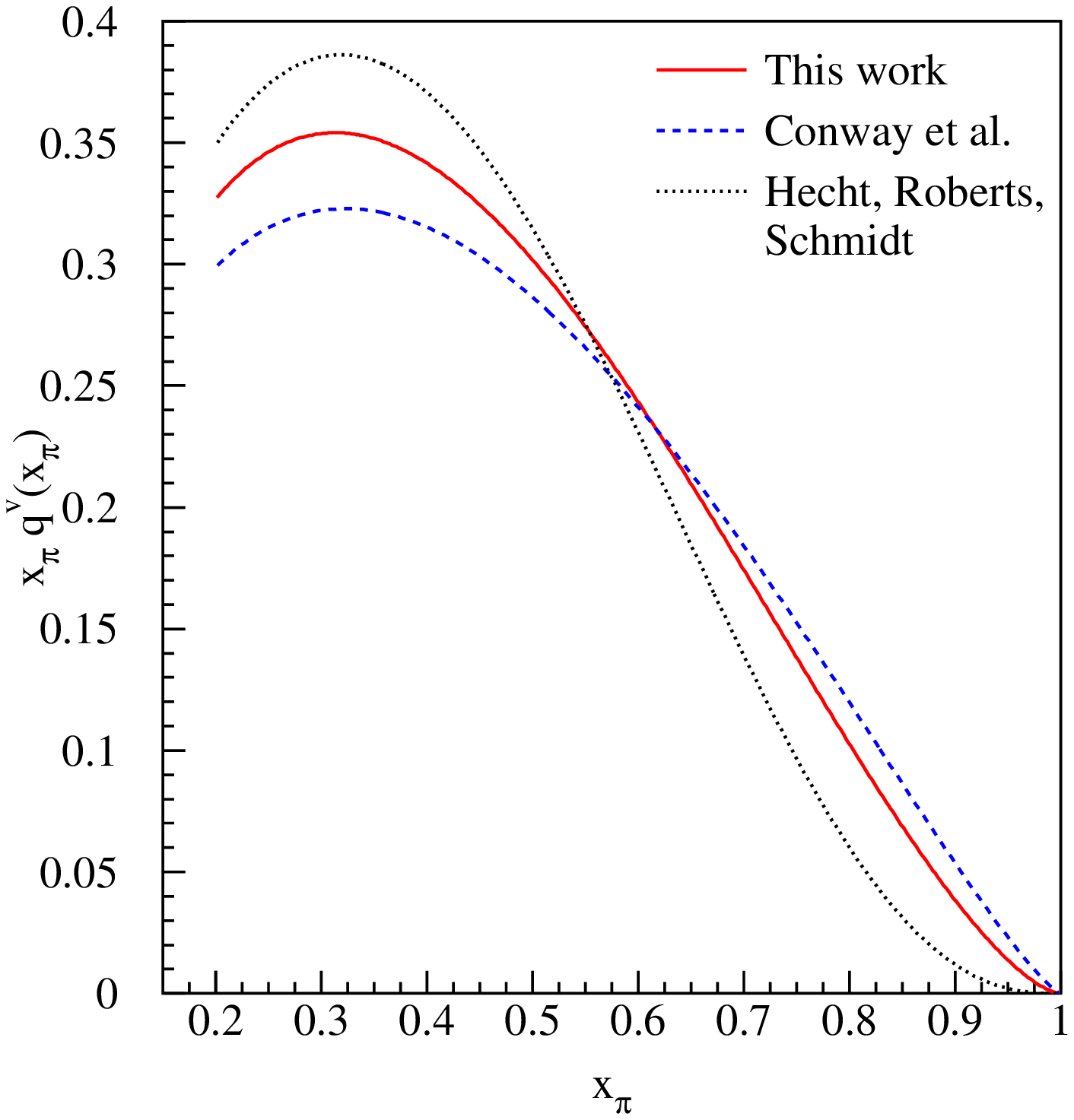}
\caption{(Color online)
The pion distribution function.  \emph{Dashed curve} -- fit from the LO analysis of the FNAL E615 data; \emph{solid curve} -- NLO fit to the E615 data \protect\cite{Wijesooriya:2005ir}; and \emph{dotted curve} -- calculation of the distribution function using a Dyson-Schwinger equation approach \protect\cite{Hecht:2000xa}, which manifestly incorporates the momentum-dependence of the dressed-quark mass function that is depicted in Fig.\,\protect\ref{gluoncloud}. \label{highxdypion}}
\end{figure}

Indeed, theoretical descriptions disagree.  The QCD parton model \cite{Ezawa:1974wm,Farrar:1975yb}, which determines the pion distribution function from the process depicted in Fig.~\ref{pionparton}, indicates that at very high $x$ the distribution should behave as $(1-x)^2$.  Perturbative quantum chromodynamics (pQCD) \cite{Ji:2004hz,Brodsky:1995} and continuum non-perturbative calculations, such as Dyson-Schwinger Equation (DSE) studies \cite{Hecht:2000xa, Maris:2003vk, Bloch:1999ke, Bloch:1999rm,Hecht:1999cr}, which express the momentum-dependence of the dressed-quark mass function that is evident in Fig.\,\ref{gluoncloud}, indicate that the high-$x$ behavior should be $(1-x)^{2+\gamma}$, with an anomalous dimension $\gamma>0$.

In contrast,
AdS/QCD models using light-front holography \cite{Brodsky:2007hb} yield $(1-x)^a$ with $a=0$,
as do Nambu-Jona-Lasinio models when a translationally invariant regularization is used \cite{Davidson:1995uv, Weigel:1999pc, Bentz:1999gx}.  
On the other hand, NJL models yield $a=1$ with a hard cutoff \cite{Shigetani:1993dx}, 
as do duality arguments \cite{Melnitchouk:2002gh}.
Relativistic constituent quark models \cite{Frederico:1994dx, Szczepaniak:1993uq} give $(1-x)^a$ with $0<a<2$ depending on the form of model wave function; and instanton-based models produce $(1-x)^a$ with $1<a<2$ \cite{Dorokhov:2000gu}.  A full discussion is presented in Sec.\,\ref{sec:theory} and, in particular, Sec.\,\ref{sec:DSE}.
% Frederico < 1  Szczepaniak ... looks like 2

Given the importance of the shape of the pion distribution function at high $x$ as a test of QCD, a NLO analysis of the FNAL\,E615 data was performed \cite{Wijesooriya:2005ir}.  The results of the this analysis are compared with those of the original LO analysis in Fig.\,\ref{highxdypion}.  The solid curve is from the NLO analysis and in comparison with the LO analysis (dashed curve) has some strength shifted from the very high $x$ region to the lower $x$ region, as one should expect from the gluon radiation involved in the NLO processes depicted in Fig.~\ref{dynlo}.  Nevertheless, the amount of additional depletion thus uncovered is not yet sufficient to agree with the non-perturbative DSE calculation, given by the black dotted curve in the figure, or the pQCD prediction, which both give a very-high-$x$ dependence of the form $(1-x)^{2+\gamma}$, where $\gamma >0$.

This discrepancy remains a crucial mystery for a QCD description of the lightest and subtlest hadron, and a number of explanations have been advanced to explain it.  These range from simple experimental resolution-in-$x$ problems \cite{Wijesooriya:2005ir}; to a higher-twist effect that, at fixed $Q^2 (1-x)$ with $Q^2\to \infty$, accentuates the structure function associated with longitudinal photon polarization \cite{Berger:1979}; and theoretical factorization issues \cite{Kopeliovich:2005} in Drell-Yan at high $x$.  Given this and some questions regarding the Boer-Mulders effect, which are discussed in the next section, a new pionic Drell-Yan experiment with better resolution is warranted.

% Lattice calculations yield only the moments of the distributions and not the PDFs themselves \cite{Best:1997qp, Detmold:2003tm}. (Either develop this more or eliminate this. Perhaps move this to the theory section.)

\subsection{Azimuthal asymmetries}
\begin{figure}[t]
\includegraphics[angle=0,width=0.8\textwidth]{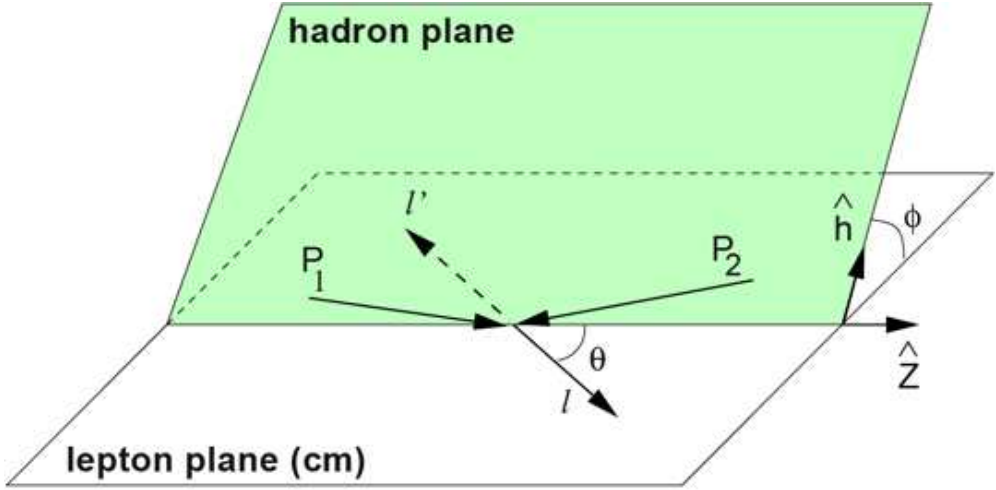}
\caption{(Color online) The decay angular distribution of the Drell-Yan process in the Collins-Soper frame. \label{angd}}
\end{figure}

The decay angular distribution of the lepton pair in Drell-Yan interactions provides interesting additional insight into the valence structure of the hadron. In the simplest case, the decay angular distribution for a purely transversely polarized Drell-Yan photon is given by
\begin{equation}
\label{eq:transverseangd}
\left(\frac{1}{\sigma}\right)\left(\frac{d\sigma}{d\Omega}\right) =
\left[\frac{3}{16\pi}\right]\left[1 + \cos^2 \theta  \right],
\end{equation}
where the angles are defined in Fig.~\ref{angd}.  For the more general case where the Drell-Yan photon also has a longitudinal component, the decay angular distribution with angles defined in  Fig.~\ref{angd} can be written as \cite{Soper:1977}
\begin{equation}
\label{eq:decayangd}
\left(\frac{1}{\sigma}\right)\left(\frac{d\sigma}{d\Omega}\right) =
\left[\frac{3}{4\pi}\frac{1}{\lambda + 3}\right]\left[1 + \lambda \cos^2 \theta + \mu \sin2\theta \cos\phi
 +\frac{\nu}{2}\sin^2\theta \cos 2\phi \right].
\end{equation}
This expression is valid in all reference frames.  Commonly used reference frames are the:
u-channel, in which the $z$-axis is chosen antiparallel to the target beam direction;
Gottfried-Jackson \cite{Gottfried:1964nx} (t-channel) -- $z$-axis is chosen parallel to the beam nucleon;
and Collins-Soper \cite{Soper:1977} -- $z$-axis bisects the angle between the $z$-axes in the other two frames.
The quantities $\lambda$, $\mu$, and $\nu$ in one frame can readily be related to their forms in another \cite{Conway:1989fs}.

\begin{figure}[t]
\includegraphics[width=0.70\textwidth]{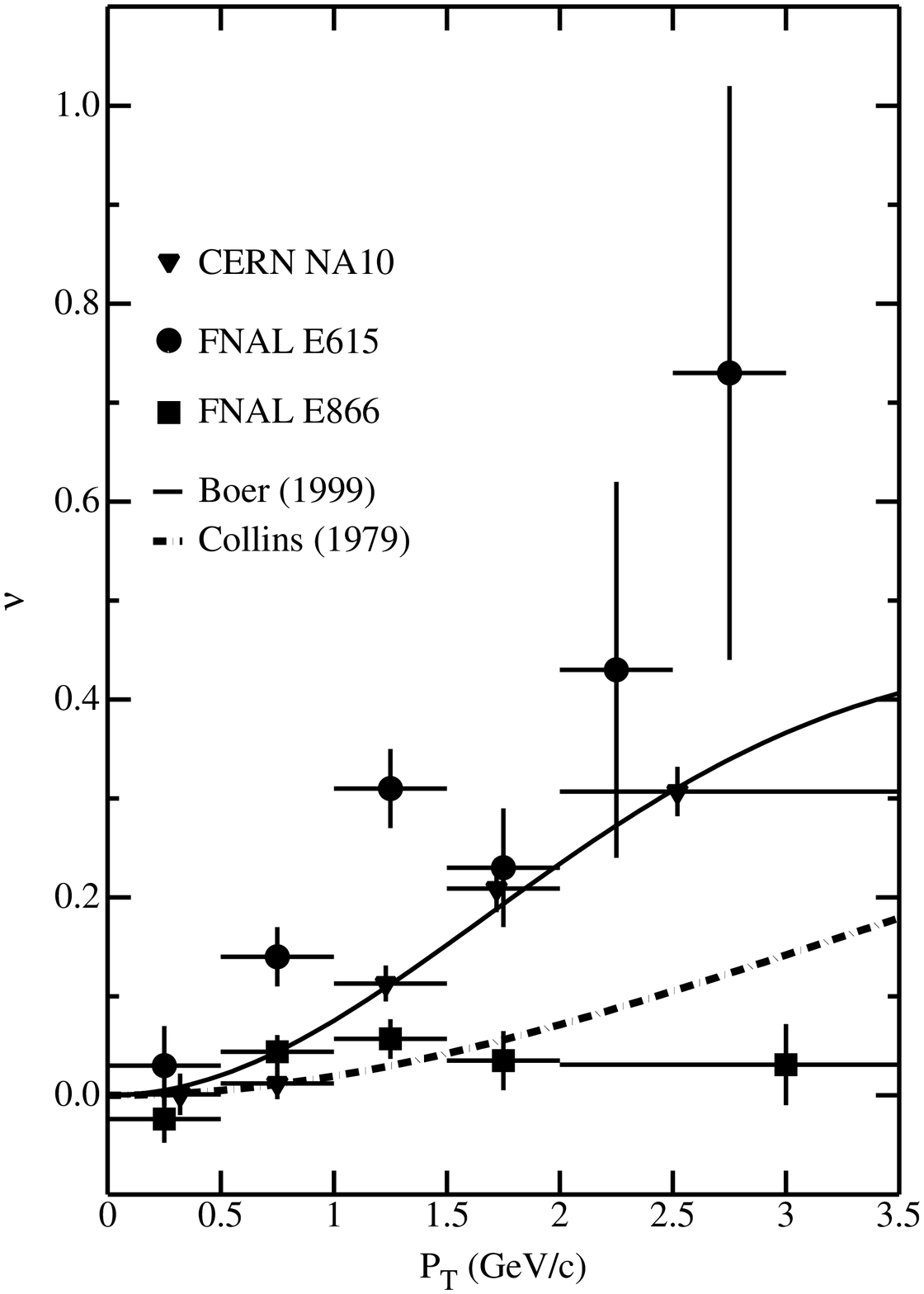}
\caption{\label{BoerMulders}
The $\nu$ or $\cos2\phi$ component of the Drell-Yan angular distributions obtained with a pion (NA10 and E615) or proton (E866) beam plotted as a function of the transverse momentum of the lepton pair.
The dotted curve depicts a pQCD prediction \protect\cite{Collins:1978}. [Eq.\,(\protect\ref{eq:collinsangd}) and associated discussion.]
The solid curve, which represents the Boer-Mulders effect \protect\cite{Boer:1999}, was adjusted to describe the NA10 data.  [Eq.\,(\protect\ref{eq:boerA}) and associated discussion.]}
\end{figure}

For pionic Drell-Yan interactions, a very interesting result is that the parameter $\nu$ in the expression above was found  \cite{Conway:1989fs, Conway:1987hv,Heinrich:1991, Falciano:1986wk, Guanziroli:1987rp, Badier:1983mj,Betev:1985pf} to be large and dependent on the transverse momentum of the lepton pair, as shown in Fig.\,\ref{BoerMulders}.

In the late 1970's, two processes were investigated that could produce an azimuthal asymmetry in Drell-Yan processes.  The first was a higher twist effect \cite{Berger:1979}, while the second was a single gluon radiation process \cite{Collins:1978}.  In the former, the high-twist diagrams considered gave rise to a $\cos\phi$-dependence in the Drell-Yan process.  This is considered to be important at very high $x$ and relatively low $Q^2$.  It is now believed \cite{Boer:1999} that high-twist effects, such as those explored in \cite{Brandenburg:1994}, cannot simultaneously describe the observed $\mu$ and $\nu$ in the Drell-Yan experiments.

In connection with the latter process, diagrams (c) and (d) in Fig.\,\ref{dynlo} were considered.  The gluon radiation gives rise to the transverse quark momentum.  This simple process produces a $\cos2\phi$ dependence in the angular distribution of the Drell-Yan lepton pair.  For example, it was found that \cite{Collins:1978}
\begin{equation}
\label{eq:collinsangd}
\nu = \frac{P_T^2}{Q^2 + \frac{3}{2}P_T^2}\,,
\end{equation}
where $P_T$ is the total transverse momentum observed in the process.  It appears that this simple pQCD process could explain a significant fraction of the large $\nu$ observed in the pionic Drell-Yan experiments, as shown by the dotted curve in Fig.\,\ref{BoerMulders}.

It is notable that for some time it was believed that if soft gluon resummation is included \cite{Chiappetta:1986}, then this process predicts very small values of $\nu$ and restores in large part the naive Drell-Yan cross section given by  Eq.\,(\ref{eq:transverseangd}).  However, it was recently shown \cite{Boer:2006eq} that the gluon resummation in that work was not applied to both the numerator and denominator in the $\cos2\phi$ component of the cross section.  When the gluon resummation is applied correctly, then the effect partially cancels out, leaving the simple pQCD process as dominant.  This result is verified in \cite{Berger:2007}.  It remains mysterious why the simple pQCD process overestimates $\nu$ for the proton Drell-Yan experiment, depicted by the filled-squares in Fig.\,(\ref{BoerMulders}).

The mystery increases when the Lam-Tung relation is considered.  In the context of the parton model, a relationship exists between two of the decay angular distribution parameters in Eq.~{\ref{eq:decayangd}} \cite{Lam:1980}:
\begin{equation}
\label{eq:lamtung}
 1 - \lambda = 2\nu \, .
\end{equation}
The Lam-Tung relation is a consequence of the spin-$1/2$ nature of quarks.  It is an analogue of the Callan-Gross relation in DIS, Eq.\,(\ref{eq:callan}), but is less sensitive to QCD corrections.  Nevertheless, a violation of the Lam-Tung relation would suggest a rather significant non-perturbative process.  The pionic Drell-Yan data indicate a large violation of the Lam-Tung relation, whereas the proton data do not indicate a violation.  A more recent treatment \cite{Berger:2007} demonstrated that a resummed part of the helicity structure functions preserves the Lam-Tung relation as a function of $P_T$ to all orders in $\alpha_s$.

The observed violation of the Lam-Tung relation led to the suggestion of a new non-perturbative process \cite{Brandenburg:1993}.  Therein, the violation of the Lam-Tung relation was parametrized in terms of $\kappa$
\begin{equation}
\label{eq:brandenburg1}
     1 - \lambda -2 \nu = -4 \kappa\,,
\end{equation}
where an \emph{Ansatz} was proposed for the $P_T$ dependence of $\kappa$.  After that work, a new structure function was advocated \cite{BoerMulders:1998}, known now as the Boer-Mulders structure function.  This quantity, $h_1^\perp(x,k_T)$, represents the T-odd, chiral-odd structure function that describes quarks with Bjorken-$x$ and intrinsic transverse momentum $k_T$ in one hadron, while  $\bar{h}_1^\perp(\bar{x},p_T)$ represents the antiquarks in the second hadron.   The $\cos2\phi$ term arises from a double helicity flip process \cite{Brandenburg:1993}, and is generally expressed by the product of two single chiral-odd, time-reveral odd (T-odd) helicity flip amplitudes, one for each of the hadrons involved in the process.  Then the expression for $\nu$ is proportional to the product of the two T-odd structure functions:
\begin{equation}
\label{eq:boermulder}
\nu \propto h_1^\perp(x,k_T) \otimes \bar{h}_1^\perp(\bar{x},p_T).
\end{equation}
[See also \cite{Bodwin:1988fs,Boer:2002ju,Collins:2007nk}.]
It is notable that a similar product gives rise to the $\cos2\phi$ asymmetry in semi-inclusive DIS, where the antiquark distribution is replaced by the Collins fragmentation function.  A nonzero Boer-Mulders function would signal a correlation between the transverse spin and the transverse momentum of quarks inside an unpolarized hadron.

We note that in contrast to the parton distribution functions with whose properties we are primarily concerned, the Boer-Mulders and Collins functions are examples of ``dynamic'' structure functions, which do not have a probabilistic interpretation.  As we have indicated here, if nonzero, these functions can lead to a wide range of effects that are not usually apparent in the parton model \cite{Brodsky:2009dv}.

An \emph{Ansatz} similar to that in \cite{Brandenburg:1993} was used in later work \cite{Boer:1999}, wherein it was advocated that the parametrization of the Boer-Mulders structure function should assume the same form as that for the Collins fragmentation function \cite{Collins:1992kk}.  This parametrization uses a quark mass scale, $M_c$ in a modified fermion propagator:
\begin{equation}
\label{eq:boer}
h_1^\perp(x,p_T) = \frac{\alpha_T}{\pi}C_H\frac{M_cM_H}{p_T^2 + M_c^2}\exp(-\alpha_Tp_T^2) f(x) \, ,
\end{equation}
where $M_H$ is the hadron mass, the parameter $C_H$ is chosen to be unity, $\alpha_T = 1\ GeV^{-2}$, and $f(x)$ is a parton distribution function.  In assuming that the Boer-Mulders structure function and the Collins fragmentation function have exactly the same form, the result for $\nu$ reduces to
\begin{equation}
\label{eq:boerA}
\nu = 2\kappa = \kappa _1 \frac{P_T^2 M_c^2}{(P_T^2 + 4 M_c^2)^2}\,.
\end{equation}
The solid curve in  Fig.\,\ref{BoerMulders} results from the above equation and gives a reasonable description of the NA10 (pion) data.  However, to be consistent with this data, the quark mass scale parameter must be unnaturally large [$\sim 2.3\,$GeV in \cite{Boer:1999}] in comparison with the chiral symmetry breaking scale, typically $0.3\,$GeV, as evident in Fig.\,\ref{gluoncloud}.  There is no sound basis for this and so, clearly, an improved theoretical description of this structure function is necessary.   In order to accept this simple model as the explanation for the large asymmetry in pionic Drell-Yan data, one would also require an understanding of how it may correctly be combined with the apparently large pQCD component identified in \cite{Collins:1978}.

The Boer-Mulders asymmetry was found \cite{Zhu:2006gx} to be extremely small in the proton Drell-Yan data of FNAL\,E866 depicted in Fig.\,\ref{BoerMulders}.  This small asymmetry is not understood, but it might be our first indication that both the Boer-Mulders structure function is small for sea quarks and the pQCD part is suppressed.  The pionic Drell-Yan data involve valence quarks in both the pion and the nucleon target, while these proton Drell-Yan data were taken with kinematics that focused attention mainly on the valence domain in the beam proton and the sea region in the target.
A very interesting test of this idea would be a Drell-Yan experiment with an anti-proton beam and a proton target where the kinematics were chosen such that the annihilating quarks could both be valence quarks.  Experiments of this type are planned for the Facility for Antiproton and Ion Research (FAIR) at Darmstadt.
New studies of pionic Drell-Yan will be initiated in the near future in the COMPASS experiment at CERN.

A new suggestion \cite{Lu:2006, Lu:2007} is to make use of the $\cos2\phi$ dependence in unpolarized Drell-Yan scattering to measure the flavor dependence of the Boer-Mulders structure function for both quarks and antiquarks.  In order to measure the quark (antiquark) distribution, a pion (proton) beam on both hydrogen and deuteron targets is proposed to perform the flavor separation.

%\begin{table*}
%\caption{Lam-Tung test for two different pionic Drell-Yan experiments, taken from
%\protect\textcite{Zhu:2006gx}.}
%\label{lamtungt}
%\begin{tabular}{|c|c|c|}
%\hline
%Lam-Tung test & CERN NA10 & FNAL E615 \\
%\hline
%$2\nu + (\lambda -1)$         &  0.01 $\pm$ 0.04  &  0.51 $\pm$ 0.07    \\
%\hline
%\end{tabular}
%\end{table*}

The Boer-Mulders asymmetry can also be measured in unpolarized semi-inclusive DIS (SIDIS).  The general expression for this cross section \cite{Ahmed:1999}, written to emphasize the azimuthal dependence, is given by:
\begin{equation}
\label{eq:sidis}
\frac{d\sigma}{dxdQ^2dzdP_T^2d\phi} = \frac{\alpha_e^2\pi}{2Q^6z}(A + B\cos\phi + C\cos2\phi),
\end{equation}
where $z$ is the fraction of the energy transfer imparted to the produced hadron.  In this case, the Boer-Mulders structure function also gives rise to an azimuthal asymmetry $\cos2\phi$.  In fact, the parameter $C$ in Eq.\,(\ref{eq:sidis}) has a term that is proportional to the product of the Boer-Mulders structure function and the Collins fragmentation function.

This asymmetry must, however, be disentangled from the Cahn effect \cite{Cahn:1978, Cahn:1989}, which also gives rise to $\cos\phi$ and $\cos2\phi$ dependence.  The Cahn effect can arise from the transverse momentum of the quark in a manner similar to that for $R$ given by Eq.\,(\ref{eq:RFeynman}).  At relatively large values of $P_T$, say $P_T \ge 1\ GeV/c$,  the parameters $B$ and $C$ can also arise from gluon radiation effects \cite{Georgi:1977}.  The azimuthal asymmetries arising from these effects have been very well established in SIDIS from: the CERN EMC experiment \cite{Arneodo:1986}; FNAL E665 \cite{Adams:1993}; and HERA \cite{Derrick:1995}.  At Jefferson Lab the azimuthal asymmetries were found \cite{Mkrtchyan:2007} to be small, and consistent with the Cahn effect and the Boer-Mulders effect at very low values of $P_T$.  Future results from HERMES and COMPASS, and an upgraded JLab facility \cite{P12-06-112} will probe a more comprehensive kinematic space, so as to pin down these effects in the valence region and provide new information about the Boer-Mulders asymmetry.
%\textcolor[rgb]{0.98,0.00,0.00}{One might also expect that studies of the azimuthal asymmetry in semi-inclusive DIS (SIDIS) at an upgraded Jefferson Lab could provide new information about the Boer-Mulders asymmetry.}

Given the importance of DCSB in QCD, it is also worth mentioning in connection with SIDIS that a leading-twist mechanism within QCD has been identified which can generate a transverse spin asymmetry that directly probes partonic structure associated with chiral-symmetry breaking \cite{Brodsky:2002cx}.  This is the so-called ``Sivers asymmetry'' \cite{PhysRevD.41.83,PhysRevD.43.261}.   In hadron-induced hard processes; e.g., Drell-Yan, this asymmetry exists and is reversed in sign, and thereby violates naive universality of parton densities \cite{Collins:2002kn,Brodsky:2002rv}.  Indeed, in the Drell-Yan process, even when both the beam and target are unpolarized, the annihilating quark and antiquark have a transverse-momentum-dependent transversity.  Extensive discussions of distributions associated with quark spin asymmetries are presented in \cite{Barone:2001sp,D'Alesio:2007}.

We close this subsection by reiterating that anti-proton--proton Drell-Yan interactions, of the type planned for FAIR, are a particularly powerful method for isolating the Boer-Mulders structure function.  In this case, the $\cos2\phi$ term is again given by  Eq.\,(\ref{eq:boermulder}), where now the process involves annihilation of the quarks in the valence region of the proton with the valence antiquarks in the antiproton.  Similarly, the Collins fragmentation function can be determined from $e^+e^- \rightarrow h\bar{h}$ where $h$ and $\bar{h}$ refer to the outgoing hadron and anti-hadron.  Some work \cite{Ogawa:2007} has been performed at Belle in this regard.  A recent review \cite{D'Alesio:2007} of azimuthal asymmetries and single spin asymmetries in hard scattering processes covers these topics in more detail.

\subsection{The kaon distribution functions}

\begin{figure}[t]
\includegraphics[width=0.80\textwidth]{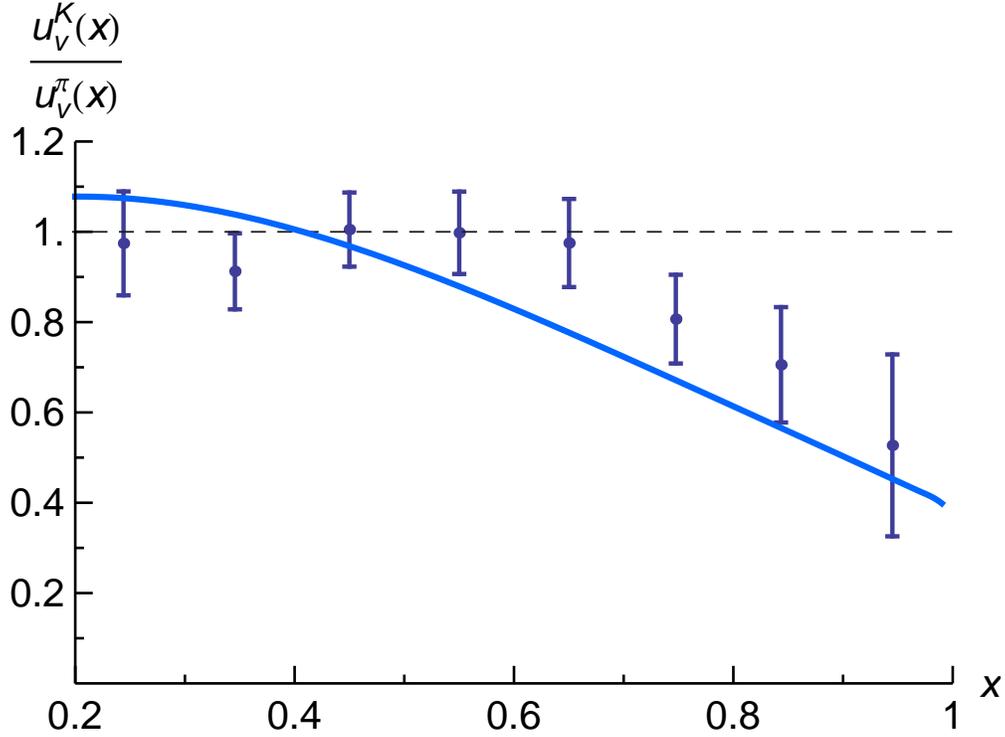}
\caption{Ratio of kaonic to pionic Drell-Yan cross sections, which is usually interpreted as the ratio of the $\bar{u}$ quark distribution function in the $K^-$ to that in the $\pi^-$.  Data from \cite{Badier:1980}, which is obtained from a sample of dimuon events with invariant mass $4.1<M<8.5\,$GeV; and \emph{solid-curve} from \protect\cite{Shigetani:1993dx}, an NJL-model result evolved from to $Q_0=0.5\,$GeV to $4.5\,$GeV.  (See also Fig.\,\protect\ref{uKpiPCT} and associated discussion.) \label{kaon}}
\end{figure}

The valence structure of the kaon is comprised of a light $u$ or $d$ quark/antiquark and a strange quark/antiquark.  If our understanding of meson structure is correct, then the large difference between the current-masses of the $s$-quark and the $u$- and $d$-quarks should give rise to some very interesting effects in the kaon structure function.  For example, owing to its larger mass, the $s$-quark should carry more of the charged-kaon's momentum than the $u$ quark.  Then, the $u_v$ quark distribution in the kaon should be weighted to lower values in $x$ than that in the pion.  A Nambu--Jona-Lasinio model calculation \cite{Shigetani:1993dx} exhibits this behavior, as shown in Fig.~\ref{kaon}.  A modern DSE prediction, discussed in Sec.\,\ref{sec:DSE} and depicted in Fig.\,\ref{uKpiPCT}, confirms this trend and provides an improved understanding of its origin.

Experimentally, a Drell-Yan measurement \cite{Badier:1980} of the ratio of $K^-$ to $\pi^-$ is consistent with unity over most of the $x$ region, with a suggestion that the ratio is dropping at high $x$.  However, the data are not of sufficient quality to test and verify our understanding of pion and kaon structure.
To explain, we note that one might claim from the data in Fig.\,\ref{kaon} that for $x>0.7$ there is a deviation from $u_v^K(x)=u_v^\pi(x)$ at the $3.8\,\sigma$-level.  However, following \cite{Badier:1980}, which relied on simple models for motivation, we observe that the entire data set is fitted by (see Fig.\,\ref{uKpiPCT})
\begin{equation}
\label{RDYKpi}
R_{K^-/\pi^-}(x) = 1.1 (1-x)^{0.22}
\end{equation}
with $\chi^2/8=0.45$, and this drops to just $\chi^2/3=0.032$ for data at $x>0.7$.  Such values are unreasonably small.
Hence, it is essential to make a high accuracy measurement of the structure function of the kaon at $x>0.6$ before the behavior in Fig.\,\ref{kaon} can be viewed as anything more than a hint.

\pagebreak

\section{Future experiments at JLab, EIC, FNAL, J-PARC and FAIR}
\label{EIC}
\begin{figure}[t]
\includegraphics[clip,width=0.75\textwidth]{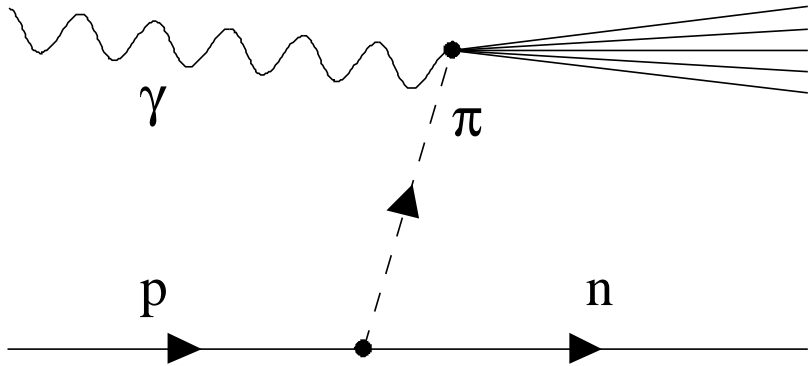}
\caption{Feynman diagram for a possible DIS interaction with a virtual pion in the proton.  \label{pidis}}
\end{figure}

The 12\,GeV upgrade at JLab permits a new opportunity to study structure functions at extremely high $x$.  Although the greater energy that will be available at JLab will not be sufficient to avoid target-mass corrections for the nucleon at the higher values of $x$, it does present the prospect of obtaining precise data on the proton and neutron structure functions.  It might also represent a new opportunity to study the pion structure function, although assumptions concerning virtual pions would have to be invoked \cite{Wijesooriya:2001}.  This is because, in determining meson structure functions from deep inelastic scattering, it is necessary to consider scattering from virtual mesons, as illustrated in Fig.\,\ref{pidis}.

An electron ion collider (EIC) is aimed at polarization studies of the nucleon structure function at lower values of $x$.  Nevertheless, it will permit access to the virtual pion and kaon structure function through measurements of the forward nucleon structure functions \cite{Holt:2000cv}.  For example, the scattering process illustrated in Fig.\,\ref{pidis} can be simulated using the RAPGAP Monte Carlo program \cite{Jung:1993gf}.  These processes include DIS from an exchanged pion or pomeron \cite{Pirner:1998bf,Holtmann:1996be}.  A comparison of results from RAPGAP with HERA data show reasonable agreement for fast outgoing neutrons \cite{Adloff:1998yg}.

\begin{figure}[t]
\includegraphics[width=0.75\textwidth]{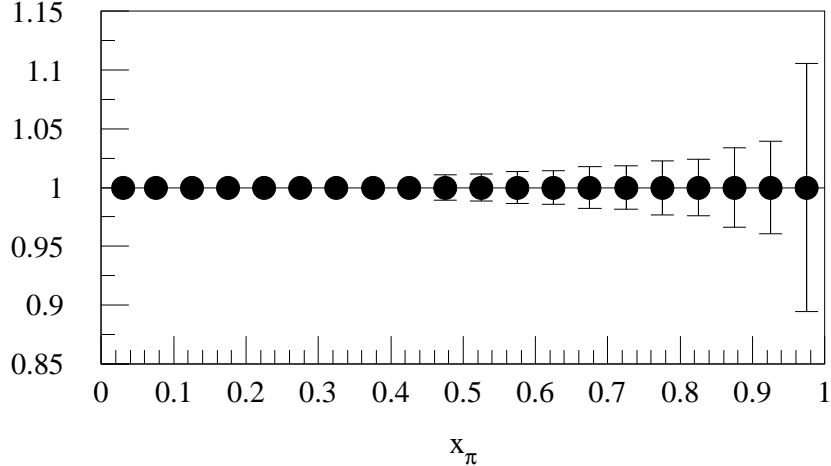}
\caption{The expected statistical error for a future experiment to measure the $\pi^-$ structure function at an electron-ion collider.  [Figure adapted from \protect\cite{Holt:2000cv}.] \label{pionerror}}
\end{figure}

The expected accuracy of a possible DIS experiment from a virtual pion was estimated using events in which a ``spectator'' neutron was identified \cite{Holt:2000cv}.  In general, the neutron is scattered less than 50 mrad from the nominal proton beam axis.  Events were cut on $Q^2 > 1~\textrm{GeV}^2$.  For a $5\,$GeV electron beam on a $25\,$GeV proton beam, the expected errors are shown in Fig.\,\ref{pionerror}.  A modest luminosity of $10^{32}~\textrm{cm}^{-2}\textrm{s}^{-1}$ was assumed for a run lasting $10^6~\textrm{s}$.  Such an experiment should be feasible because a similar experiment was conducted at HERA \cite{Chekanov:2007}.  This experiment gave the very interesting result that the shape of the sea distribution for the virtual pion is the same as that for the proton.

The $K^+$ structure function can be measured by considering deep inelastic scattering from the kaon cloud surrounding a proton.  The basic Feynman diagram would be the same as in Fig.\,\ref{pidis}, with the pion replaced by a kaon and the neutron replaced by a $\Lambda$.  The probability for scattering from the $K^+$ cloud surrounding the proton should be comparable to that for the $\pi^+$ because the $KN\Lambda$ coupling constant is comparable to that of the $\pi NN$ vertex.  In fact, one would only expect about a factor of two reduction in the vertex function for the kaon compared to the pion.

The difficulty with this process is in the detection of the $\Lambda$.  The $\Lambda$ decays predominantly (64\%) to a proton and a $\pi^-$.  Thus, a special forward proton spectrometer as well as a forward pion spectrometer would be necessary.  This should be feasible since the ZEUS and H1 experiments at HERA have already successfully employed forward proton spectrometers.

It is also likely that an EIC will permit measurements of the $F_2^n/F_2^p$ ratio at extremely high $x$.  The technique should be analogous to the method that would be used for the pion and kaon structure function, with a deuteron target beam in this case.  The scattered electron would be detected in coincidence with either a forward-going spectator proton or neutron, depending on whether the DIS occurred on a neutron or proton, respectively.  In this case, a $5\,$GeV electron beam on a $25\,$GeV deuteron beam at a relatively modest luminosity should be sufficient.

\begin{figure}[t]
\includegraphics[width=0.80\textwidth]{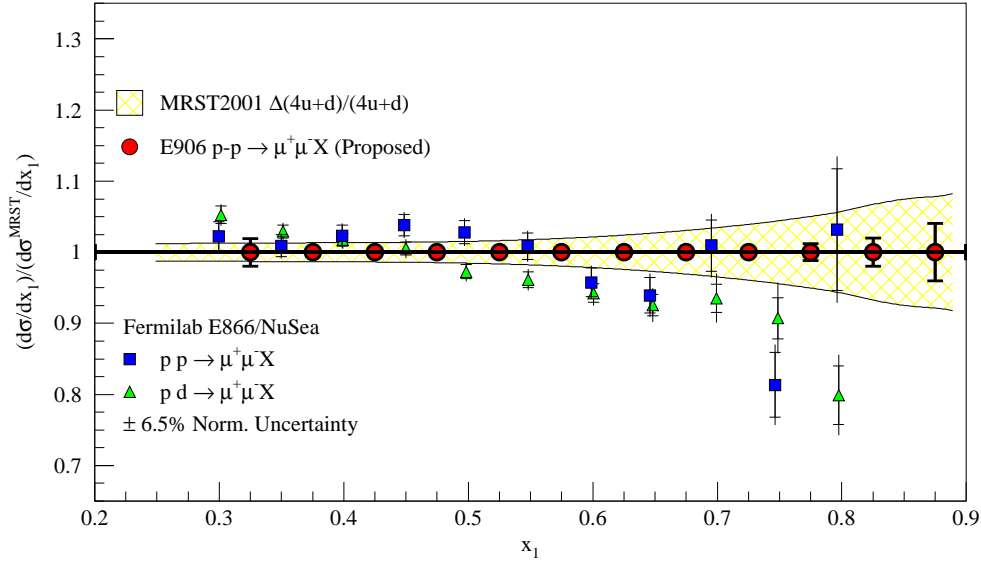}
\caption{(Color online) The absolute Drell-Yan cross section is proportional to $4u(x)+d(x)$ in the beam proton.  The data are from the FNAL E866 experiment. The solid circles are projected data from experiment E906 at FNAL. [Figure adapted from \cite{Webb:2003bj,Webb:2003ps,Reimer:Figure}.]
\label{highxdyE906}}
\end{figure}

Reliable results at extremely high $x$ are also expected from forthcoming Drell-Yan measurements.  For example, the solid circles in Fig.\,\ref{highxdyE906} indicate the error limits expected for the proposed Drell-Yan experiment at FNAL (E906) \cite{Reimer:Figure}.  The assumptions that went into this simulation were a $120\,$GeV proton beam from the FNAL main injector and a total integrated luminosity of $5 \times 10^{33} {\rm cm}^{-2}$.

Finally, the Japanese Proton Accelerator Research Center facility, J-PARC, currently under construction at Tokai, Japan, could make polarized Drell-Yan studies possible.  Serious consideration is being given to experiments aimed at measuring parton distribution functions in mesons.  In addition, the Facility for Antiproton and Ion Research, FAIR, under construction at GSI in Darmstadt would make it possible to directly measure the Drell-Yan process from high-x antiquarks in the antiproton annihilating with quarks in the proton. 
%\newpage

%\setcounter{figure}{0}
%\setcounter{table}{0}
\setcounter{equation}{0}
\section{Present-day parametrizations of the PDFs}
%\section{Present day parametrizations of the distribution functions}
\label{sec:param}
Excellent reviews of the parametrization of distribution functions exist.  In fact, a good starting point for most of the distribution functions can be found at the Durham website (\href{http://durpdg.dur.ac.uk/HEPDATA/PDF}{http://durpdg.dur.ac.uk/HEPDATA/PDF}).  At this website, one can access compilations of data, codes and grids associated with the following distribution functions:
MRST/MSTW -- Sec.\,\ref{sec:mrst};
CTEQ -- Sec.\,\ref{ss:cteq};
GRV/GJR -- Sec.\,\ref{sec:GRV};
and ALEKHIN -- Sec.\,\ref{ss:relief},
as well as an online PDF calculator.  Here we consider primarily the CTEQ, MRST and the GRV parametrizations, emphasizing the differences between them.  Another very useful website is that of CTEQ
(\href{http://www.phys.psu.edu/~cteq/}{http://www.phys.psu.edu/{\~\,}\rule{-0.2em}{0ex}cteq/}).
Here the CTEQ QCD handbook \cite{Brock:1993sz} can readily be accessed.  The recent status of the MRST/MSTW PDFs can be found in \cite{Thorne:2009ky}, wherein the MSTW2008 distribution functions \cite{Martin:2009iq} are recommended.

The primary source of variations between the parametrizations are the different:
\begin{itemize}
\item data sets used in the fits;
\item selections of data within the data sets;
\item pQCD choices -- e.g., evolution order, factorization scheme, renormalization scale, $\alpha_s$;
\item parametric forms for the PDF;
\item theoretical assumptions about the $x \rightarrow 1$ behavior.
\end{itemize}
Although there are also distinct treatments of heavy flavors, assumptions of sea flavor asymmetry and $x \rightarrow 0$ behavior, these do not have a large impact on the valence region.

Most parametrizations begin with valence-like input, which means that at some infrared scale $Q_0 \lesssim 1\ GeV$ all distribution functions are represented as
\begin{equation}
\label{eq:grj}
  xf(x,Q_0^2) \sim  x^{\alpha_f}(1-x)^{\beta_f},
\end{equation}
where $\alpha_f>0$ and $\beta_f>0$ are fit parameters, so that even those of the sea and glue distributions are nonzero but finite at the infrared boundary.  Then, typically, the distribution functions are evolved to higher scales, commensurate with experiment, and the parameters fitted to obtain agreement with data.  Some detailed starting distributions for the various parametrizations are given in the identified sections below.

\subsection{MRST and MSTW}
\label{sec:mrst}
The data sets included in Martin-Roberts-Stirling-Thorne (MRST\,2002) \cite{Martin:2001es, Martin:2002aw} parametrization are: H1; Zeus, both neutral current and charge current data; BCDMS; SLAC; FNAL E665; CCFR; Drell-Yan data (FNAL E605, E772, E866); and FNAL CDF and D0 data.  The MRST parametrization has a starting scale of $Q_0^2 = 1\,$GeV$^2$ and accepts data down to  $Q^2 = 2\,$GeV$^2$.  Thus, the SLAC data have an influence at large values of $x$.

A parametrization of the parton distribution functions is given by
\begin{equation}
\label{eq:mrstnlo}
  xf(x,Q_0^2) =
A_0 x^{\alpha}(1-x)^{\beta}(1 + \delta x^\gamma +\eta x).
\end{equation}
The parameters at the starting scale are given in Table~\ref{mrst} for the MRST\,2001 parametrization, a NLO fit to the data.

\begin{table}[t]
\caption{For illustration, parameters characterizing the MRST\,2001 NLO parton distribution functions at $Q_0^2 = 1\,$GeV$^2$, defined by Eq.\,(\protect\ref{eq:mrstnlo}). \label{mrst}}
%--We couldn't find a table with a more recent set of parameters
%
\begin{tabular}{|c|c|c|c|c|c|c|}
\hline
parton  & $A_0$ & $\alpha$  &   $\beta$  &  $\gamma$  &  $\delta$   &  $\eta$ \\
\hline
$d_v(x,Q_0^2)$         & 0.040   & 0.27 & 3.88 & 52.73 & 0.5 & 30.65     \\
$u_v(x,Q_0^2)$         & 0.158   & 0.25 & 3.33 & 5.61 & 0.5 & 55.49     \\

\hline
\end{tabular}
\end{table}

\begin{figure}[t]
\includegraphics[angle=-90,width=0.80\textwidth]{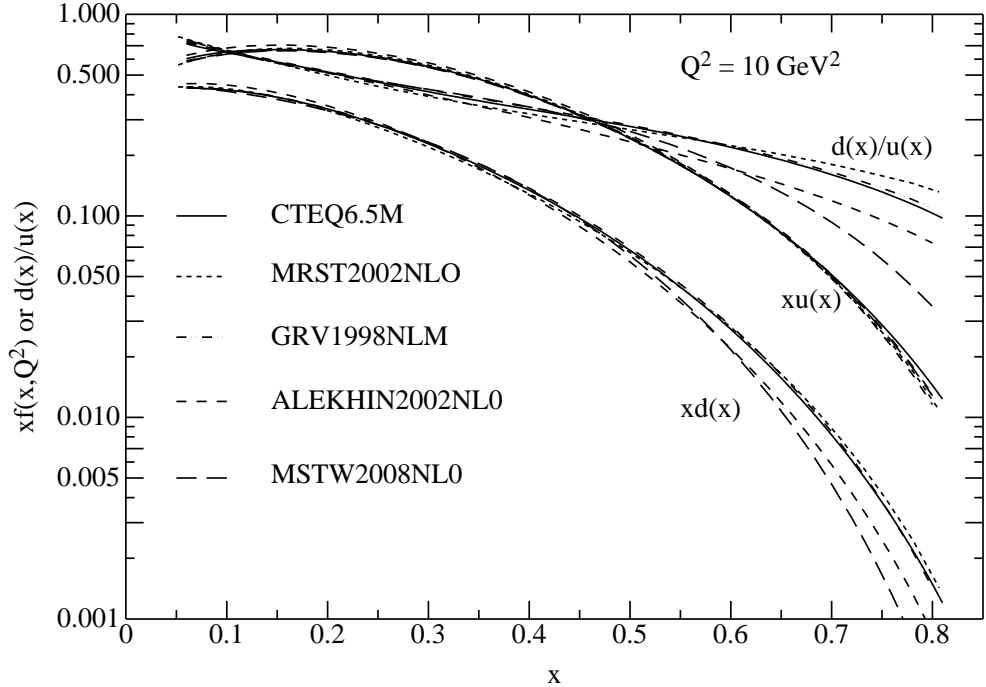}
\caption{Five parton distribution evaluations at NLO in the $\overline{MS}$ factorization scheme at $Q^2 = 10 \,$GeV$^2$ in the valence region.  There are three families of curves: the $d(x)/u(x)$ ratios, the $xu(x)$ distributions, and the $xd(x)$ distributions.\label{nlocomp}}
\end{figure}

The chief difference between the MRST\,2003 set \cite{Martin:2003sk} and MRST2002 is the exclusion of data below $Q^2 = 10\,$GeV$^2$ and below $x = 0.005$; whereas, the MRST\,2004 analysis \cite{Martin:2004ir} included new HERA data at moderate values of $x$ and high $Q^2$.
The MRST\,2004 NNLO set was the first to use the full NNLO splitting functions.
The newer MSTW parametrization \cite{Watt:2008hi,Martin:2009iq} represents an update of MRST.  This update has a number of new theoretical features aimed at the NNLO parametrization; e.g., NNLO corrections to the Drell-Yan data.  This parametrization also includes NuTeV and Chorus data, the CDFII data, HERA inclusive jet data as well as direct high-$x$ data on the $F_L$ structure function.  The most interesting feature of these updates is apparent in the valence region of Fig.\,\ref{nlocomp}.  The NLO MSTW evaluation gives a much smaller $xd(x)$ distribution at very high $x$ than the previous work.  Nevertheless, it is apparent in the differences between the parametrizations at large $x$ that the $d$ quark distribution is poorly constrained above $x=0.6$.  This further emphasizes the need for new data in the very high $x$ region.

\subsection{CTEQ parameterization}
\label{ss:cteq}
The most recent parametrizations from the Coordinated Theoretical-Experimental Project on QCD (CTEQ) is the CTEQ6 series \cite{Pumplin:2001ct, Pumplin:2002vw, Stump:2003yu, Pumplin:2007wg, Lai:2007dq, Nadolsky:2008zw}.  The CTEQ collaboration omits data for $Q^2 \le 4\,$GeV$^2$.  In particular, CTEQ6 omits SLAC data as well as some high-$Q^2$ H1 data.  CTEQ uses 10\% systematic errors in quadrature with the statistical errors for the Drell-Yan data in comparison with 5\% systematic errors assumed by MRST\,2002.  CTEQ uses a starting scale of $Q_0^2 = 1.69\,$GeV$^2$.
%Uses CDF data and some neutrino data ... just like MRST

The functional form of the CTEQ6 parametrization of the distribution functions is given by:
\begin{equation}
\label{eq:cteq6}
  xf(x,Q_0^2) =
A_0 x^{A_1}(1-x)^{A_2}e^{A_3x}(1 + e^{A_4}x)^{A_5}.
\end{equation}
The parameters at the starting scale are given in Table\,\ref{cteq6m} for the CTEQ6M parametrization, a NLO fit to the data.

\begin{table}[t]
\caption{Parameters at $Q_0^2=1.69\,$GeV$^2$ for the CTEQ6M parton distribution functions, defined by  Eq.\,(\ref{eq:cteq6}).\label{cteq6m}}
\begin{tabular}{|c|c|c|c|c|c|c|}
\hline
parton  & $A_0$ & $A_1$  &   $A_2$  &  $A_3$  &  $A_4$   &  $A_5$ \\
\hline
$d_v(x,Q^2_0)$ & 1.4473   & 0.6160 & 4.9670 & -0.8408 & 0.4031 & 3.0000  \\
$u_v(x,Q_0^2)$ & 1.7199   & 0.5526 & 2.9009 & -2.3502 & 1.6123 & 1.5917   \\
$g(x,Q_0^2)$  & 30.4571   & 0.5100 & 2.3823 & 4.3945 & 2.3550 & -3.0000     \\
$(\bar{u}+\bar{d})(x,Q_0^2)$ & 0.0616 & -0.2990 & 7.7170 & -0.5283 & 4.7539 & 0.6137     \\
$s(x,Q_0^2) = \bar{s}(x,Q_0^2)$  & 0.0123 & -0.2990 & 7.7170 & -0.5283 & 4.7539 & 0.6137    \\

\hline
\end{tabular}
\end{table}

\subsection{GRV/GJR distribution functions}
\label{sec:GRV}
The Gluck-Reya-Vogt (GRV) parton distribution functions were developed in a series of publications throughout the 1990s \cite{Gluck:1991ng,Gluck:1993im,Gluck:1994uf,Gluck:1998xa}.  They are dynamical distributions, which are generated radiatively from valence-like inputs at a low resolution scale.  The latest of this series makes use of the 1994-95 HERA data for \mbox{$Q^2 \ge 2 \,$GeV$^2$} as well as the SLAC, BCDMS, NMC and E665 data with  $Q^2 \ge 4 \,$GeV$^2$ and the simply extracted ratios $F_2^n/F_2^p$ from the NMC, BCDMS and E665 experiments.  This analysis takes into account the Drell-Yan data and the $u_v/d_v$ ratios extracted from the CERN CDHSW and WA21 neutrino data.

The GRV parton distribution functions are parametrized as
\begin{equation}
\label{eq:grvlo}
  xf(x,Q_0^2) =
A_0 x^{\alpha}(1-x)^{\beta}(1 + \delta \surd x +\eta x).
\end{equation}
In Table\,\ref{grv98lo} we report the parameters at the starting scale for the GRV98\,LO parametrization, a leading order fit to the data; and the parameters at the starting scale for the GRV98\,NLO parametrization, a next-to-leading-order fit to the data in the $\overline{MS}$ scheme.  It is interesting that the value of $\beta$ for $u_v$ and $d_v$ systematically increases in going from LO to NLO.  This is expected and consistent with the effect of gluon radiation reducing the hardness of the parton distribution.

\begin{table}[t]
\caption{\label{grv98lo} \emph{Left panel} -- Parameters for the GRV98\,LO parton distribution functions at $Q_0^2 = 0.26\,$GeV$^2$, defined by  Eq.\,(\ref{eq:grvlo}), where $\Delta \equiv \bar{d}-\bar{u}$ and $xs(x,Q^2_0)=x\bar{s}(x,Q^2_0)$.
\emph{Right panel} -- Parameters for the GRV98\,NLO parton distribution functions at $Q_0^2 = 0.40\,$GeV$^2$.}
\begin{tabular}{|c|c|c|c|c|c|c|}
\hline
parton  & $A_0$ & $\alpha$  &   $\beta$  &  $\delta$    &  $\eta$ \\
\hline
$d_v(x,Q^2_0)$         & 0.761   & 1.48 & 3.62 & -1.8  & 9.5     \\
$u_v(x,Q^2_0)$         & 1.239   & 0.48 & 2.72 & -1.8 &  9.5     \\
$x\Delta(x,Q^2_0)$     &  0.23   & 0.48  & 11.3 & -12.0 & 50.9   \\
$x(\bar{u}+\bar{d})(x,Q^2_0)$  &  1.52   & 0.15  & 9.1  & -3.6 & 7.8   \\
$xg(x,Q^2_0)$     &  17.47   & 1.6  & 3.8 & - & -   \\
\hline
\end{tabular}
%\end{table}
%
%\begin{table}[t]
%\caption{\label{grv98nlo} Parameters for the GRV98\,NLO parton distribution functions at $Q_0^2 = 0.40\,$GeV$^2$, defined by Eq.\,(\ref{eq:grvlo}), where $\Delta \equiv \bar{d}-\bar{u}$ and $xs(x,Q^2_0)=x\bar{s}(x,Q^2_0)$.}
%
\begin{tabular}{|c|c|c|c|c|c|c|}
\hline
parton  & $A_0$ & $\alpha$  &   $\beta$  &  $\delta$    &  $\eta$ \\
\hline
$d_v(x,Q^2_0)$         & 0.394   & 1.43 & 4.09 & -  & 18.2     \\
$u_v(x,Q^2_0)$         & 0.632   & 0.43 & 3.09 & - &  18.2     \\
$x\Delta(x,Q^2_0)$     &  0.20   & 0.43  & 12.4 & -13.3 & 60.0   \\
$x(\bar{u}+\bar{d})(x,Q^2_0)$  &  1.24   & 0.20  & 8.5  & -2.3 & 5.7 \\
$xg(x,Q^2_0)$     &  20.80   & 1.6  & 4.1 & - & -   \\
\hline
\end{tabular}
\end{table}

These distributions are characterized by a relatively low starting scale for evolution: LO, $Q_0=0.5\,$GeV; and NLO, $Q_0=0.63\,$GeV.  One might question whether therefrom it is valid to employ pQCD evolution equations.  In this connection, we observe that it is the combination
$\alpha(Q_0)/[2\pi]$ which appears in the evolution equations.  At leading-order, \mbox{$\alpha(Q_0=0.5\,{\rm GeV})/[2\pi] \approx 0.17$} and hence $(\alpha(Q_0)/[2\pi])^2 < 0.03$.  Therefore both the procedure and a physical interpretation of the distributions at this scale might be meaningful if higher-twist and essentially nonperturbative effects are not too important.  At present one cannot answer whether the latter is true.  Notwithstanding this, the procedure alone can be useful as a fitting and correlating tool.

A more recent analysis \cite{Gluck:2007ck} extends the GRV98 analysis.  The distribution functions are generated from a radiated valence-like positive input with $Q_0 \le 1\,$ GeV, where valence-like refers to Eq.\,(\ref{eq:grvlo}).  With these constraints, the predictions for $F_L$ remain positive throughout the accessible kinematic region.  Constraining the starting distribution to be valence-like leads to smaller quoted uncertainties in the very low-$x$ region than those analyses without this constraint.

\subsection{Distributions in relief}
\label{ss:relief}
A comparison of five modern parton distributions on the valence domain at $Q^2 = 10 \,$GeV$^2$ is shown in Fig.~\ref{nlocomp}.  All five evaluations are in remarkably good agreement for the $xu(x)$ distribution.  Despite the large uncertainty in the $d/u$ ratio discussed in sec.~\ref{neutronstructurefunction}, three of the distributions (MRST\,2002NLO, CTEQ6.5M, GRV98) are in remarkably good agreement for the $d/u$ ratio.  The newest evaluations, MSTW\,2008 \cite{Martin:2009iq}, differ here largely because of a more flexible parametrization of the $d$ quark distribution, and the inclusion of new data from FNAL on the lepton charge asymmetry from W decays and the Z rapidity distribution.

The ALEKHIN\,2002 parametrization \cite{Alekhin:2002fv} is a somewhat difference case.  This parametrization is based on the SLAC, BCDMS, NMC, and HERA data existing at the time.  The cuts on the data included $2.5\,$GeV$^2< Q^2 < 300\,$GeV$^2$ and $x < 0.75$.  The model function was parametrized for $Q_0^2 = 9\,$GeV$^2$, much higher than most other forms.  The cuts in $Q^2$ and $x$ give this parametrization more emphasis on the high-$x$ SLAC data than CTEQ or MRST.  This could possibly explain the differences in the high-$x$ $d$-quark distribution.  Later ALEKHIN parametrizations include Drell-Yan data \cite{Alekhin:2006zm} as well as even lower momentum transfer data \cite{Alekhin:2008ua}.  However, the differences between these updated PDFs and the ALEKHIN\,2002 distributions do not exceed one standard deviation.

Notwithstanding differences between the Groups' PDFs, it is evident from the figure and the tables that all fits are consistent with
\begin{equation}
\label{donuempirical}
\frac{d(x)}{u(x)} \stackrel{x\sim 1}{ \sim } 0\,.
\end{equation}
However, this empirical statement has large uncertainties, some of which are discussed in \cite{Owens:2007kp}.  It reflects primarily upon the fact that $u$-quarks dominate proton cross-sections and hence it is difficult to tightly constrain the proton's $d$-quark distribution.

MRS \cite{Sutton:1991ay} and GRV \cite{Gluck:1991ey, Gluck:1999xe} also produced analyses of the parton distribution function in the pion.  The pion distribution functions were extracted from Drell-Yan and prompt photon data.  As we discussed in Sec.\,\ref{exptpionstructure}, a more recent NLO re-analysis of the FNAL E615 pionic Drell-Yan data \cite{Wijesooriya:2005ir} indicated that there was more curvature in the pion distribution at very large values of $x$ than given by either of these fits.  As displayed in Fig.\,\ref{xuvxpicall}, this result strengthens the disagreement between the data and contact-interaction models, but does not produce as much curvature as expected from either pQCD \cite{Ezawa:1974wm,Farrar:1975yb} or nonperturbative DSE calculations \cite{Hecht:2000xa}.  This discrepancy persists to this day and is one of the outstanding problems in understanding the application of QCD to real-world data.  We therefore discuss it at length in Sec.\,\ref{sec:theory}. 
%\newpage

%\subsection{MRST and MSTW}
%\subsection{CTEQ}

%\setcounter{figure}{0}
%\setcounter{table}{0}
\setcounter{equation}{0}
\section{Theoretical interpretation of the distribution functions}
% Title: Parton Distributions Authors: M. Dittmar, S. Forte, A. Glazov, S. Moch (convenors), G. Altarelli, J.  Anderson, R. D. Ball, G. Beuf, M. Boonekamp, H. Burkhardt, F. Caola, M.  Ciafaloni, D. Colferai, A. Cooper-Sarkar, A. de Roeck, L. Del Debbio, J.  Feltesse, F. Gelis, J. Grebenyuk, A. Guffanti, V. Halyo, J. I. Latorre, V.  Lendermann, Gang Li, L. Motyka, T. Petersen, A. Piccione, V. Radescu, M.  Rogal, J. Rojo, C. Royon, G. P. Salam, D. Salek, A. M. Stasto, R. S. Thorne,  M. Ubiali, J. A. M. Vermaseren, A. Vogt, G. Watt, C. D. White
% http://arxiv.org/abs/0901.2504

\label{sec:theory}

\subsection{Evolving insight from models}
\label{theoryopen}
We noted at the outset that while parton distribution functions are basic elements of factorized perturbative QCD, they are essentially nonperturbative; viz., they can only be calculated using a nonperturbative framework.  Their calculation is a problem that has been with us for more than thirty years; e.g., \cite{Politzer:1974}.  The interim has seen the proposal and elucidation of numerous models for the phenomena of low-energy QCD, many of which have been employed to estimate the valence-quark distribution functions.

\subsubsection{MIT bag model}
\label{MITbagtheory}
Soon after it had been proposed, the MIT bag model \cite{Chodos:1974je,Chodos:1974pn} was being used to calculate structure functions \cite{Jaffe:1974nj}.  It is still being used to make estimates, now of generalized parton distribution functions; e.g., \cite{Scopetta:2005fg,Pasquini:2005dk}.

The basic difficulties with this approach were noted at the outset.
The bag is treated as a static cavity; namely, a spherical chamber with fixed radius, $R_0$.  Amongst other things, this places a lower bound on the domain of $x$ for which distribution functions can be calculated; i.e., $x\gtrsim 1/[2 M R_0] \sim 0.1$, where $M$ is the nucleon's mass.  Moreover, the bag boundary cannot realistically be static.  It should respond to the action of currents on the quarks it contains and that would affect the form of the distribution functions.
The target is treated as being at rest but the framework is not Poincar\'e covariant and hence momentum is not conserved through the intermediate Compton scattering state.  This entails that the calculated structure functions do not possess the correct support; e.g., they do not vanish for $x>1$.
In addition, in an independent particle picture arising from a spin-flavor symmetric Hamiltonian the ratio in Eq.\,(\ref{eq:npratio}) is particularly simple; viz.,
\begin{equation}
\label{F2nF2pSU6}
\forall x: \; \frac{F_2^n(x)}{F_2^p(x)}=\frac{2}{3}\,, \; \{SU(6)-{\rm symmetric}\}\,,
\end{equation}
because the pointwise behavior of the distributions is the same for both $u$ and $d$ quarks.  This prediction conflicts with data, Fig.\,\ref{f2nratio}.
Moreover, while the momentum sum rule is satisfied, that is achieved with all the target's momentum being carried by the quarks, whereas, experimentally, the quarks' momentum fraction should only be $\sim 50$\%; i.e., $\langle x\rangle \sim \sfrac{1}{2}$.

From a modern perspective, the last of these outcomes is not seen as a critical impediment.  The distribution functions are known to be scale dependent in QCD and, in principle, once determined at a particular scale, QCD predicts their pointwise form at any and all other scales.  It should nonetheless be noted that, owing to the role played by gluons in binding the compound hadron, there is no scale at which the momentum fraction carried by valence-quarks can equal one.

In practice, one views a model computation as providing the distribution functions at some infrared resolving scale, $Q_0$.  The evolution equations,
%\cite{Dokshitzer:1977, Gribov:1972, Lipatov:1974, Altarelli:1977},
discussed in Sec.\,\ref{sec:QCDscalingV}, are then used to obtain their form at some other scale, typically $Q=2\,$GeV.  In applying a model, $Q_0$ is fixed \emph{a posteriori} by requiring that the evolved distribution functions yield $\langle x\rangle \approx 0.4$,
which is an empirical feature of the PDFs described in Sec.\,\ref{sec:param}, as evident, e.g., in Fig.\,2 of \cite{Gluck:1991ng}.
%--Find reference to this number. I had one, talking about 0.47->0.42 but an hour on 19/12/08 couldn't rediscover it.
In contrast to modeling, however, in a truly nonperturbative solution of QCD the scale $Q_0$ would be known \emph{a priori}.

On the other hand, the bag model does yield some insight and positive results.  For example, as treated, Bjorken scaling is recovered for $W_1$ and $(\nu/M^2) W_2$ in Eq.\,(\ref{eq:hadron}).  Moreover, in the Bjorken limit, Eq.\,(\ref{eq:dis}), $W_L$ vanishes and $\nu W_L$ scales and is calculable.  Thus one preserves the Callan-Gross relation.
The structure functions obey quantum number sum rules, such as the Adler relation:
\begin{equation}
\int_0^1 dx \frac{1}{x}\left[ F_2^{\nu p}(x) - F_2^{\bar\nu p}(x)\right] = 2\,,
\end{equation}
where $F_2^{\nu p,\bar\nu p}$ are neutrino scattering analogues of the usual $F_2^p$.  In addition, $F_1$ and $F_2/x$ are peaked at $x_0 = 1/N$, where $N$ is the number of quanta in the target: if three valence quarks are all the model possesses, then these functions peak at $x_0=1/3$.  The width of the peak is related to the model's confinement radius.
%Structure functions peak at xi_0=1/N, N=number of occupied modes  in the target.  Spread about xi0 attributed to confinement.  N increases,structure functions->0-width and one structure functions revert to those of a collection of free quarks

%Modern perspective?
Primary amongst the problems identified above, in part because it also affects other models, is the absence of translational invariance and hence momentum conservation.
In the usual mean-field approach the quarks move independently within a spherical cavity of fixed radius and definite location.  The Hamiltonian associated with this problem is not translationally invariant.  From this arises the so-called center-of-mass problem: particles moving independently within the volume cannot constitute an eigenstate of the total momentum, in particular, they cannot properly describe a hadron at rest.\footnote{A perhaps even greater difficulty is that in a interacting relativistic theory, which the model is aimed to be, Lorentz boosts change the particle number of a given state.}

\begin{figure}[t]
\includegraphics[clip,width=0.50\textwidth]{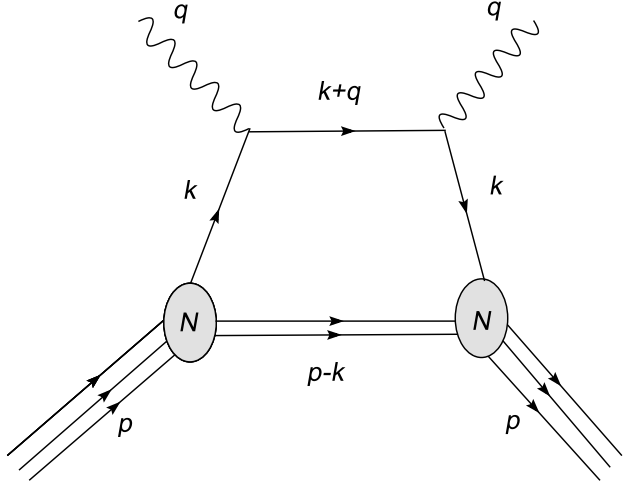}
\caption{``Handbag diagram'' contribution to the virtual photon-nucleon forward Compton scattering amplitude: $\gamma^\ast(q) N(p) \to \gamma^\ast(q) N(p)$.  In this figure the intermediate state (or nucleon remnant in the DIS process) is assumed to be a scalar or axial-vector diquark, with masses $M_s$ and $M_v$, respectively. \label{Comptondq}}
\end{figure}

In general, the removal of spurious center-of-mass motion from the hadron's wave function is not a well-defined problem and has no unique solution \cite{Lipkin:1958zz}.  The illness is typical of shell-like models.  A number of prescriptions exist, which serve to mitigate the undesirable effects that this spurious motion has on physical observables.  Amongst these, the Peierls-Yoccoz projection \cite{Peierls:1957er} has been most widely discussed.  It projects exact eigenstates of the center-of-mass momentum by forming a linear superposition of at-rest bag states at different locations.  Artifacts remain, however; e.g., some expectation values exhibit unexpected momentum dependence.  A remedy may be found in the Peierls-Thouless projection \cite{Peierls:1961} but that is difficult to implement.  NB.\ The Peierls-Yoccoz projection is a nonrelativistic prescription and hence cannot properly account for the boost experienced by the intermediate state in the forward Compton scattering process.  This means that for an initial state at rest, the procedure can only be internally consistent for $|\vec{k}|<M_{qq}$, in the notation of Fig.\,\protect\ref{Comptondq}, which corresponds to $x\lesssim 0.6$.

In a little used alternative \cite{Szymacha:1984fz,Jasiak:1997ac}, one may augment the independent particle (bag model) Hamiltonian for the composite state by a fictitious attractive potential, which depends on the center-of-mass coordinate; viz.,
\begin{equation}
H = \frac{P^2}{2 M} + H_{\rm BM} \to \tilde H = \frac{P^2}{2 M}+ V(R) + H_{\rm BM} = H_{\rm CM} + H_{\rm BM},
\end{equation}
where $H_{\rm BM}$ is a bag-model Hamiltonian.  With the potential tuned such that the spectra of $H_{\rm CM}$ and $H_{\rm BM}$ are approximately the same, then one has an equivalence between the nonlocal-composite and a translationally-invariant pointlike state.  Unlike the Peierls-Yoccoz prescription, this method ensures that the valence-quark distribution functions are correctly normalized \cite{Jasiak:1997ac}; namely, Eqs.\,(\ref{uvdvnorm}) are satisfied independent of model details.

With a prescription for removing the spurious center-of-mass motion one can be confident that computed valence-quark distribution functions will possess the correct support; i.e., are nonzero only for $0\leq x < 1$.  Typical of such calculations are \cite{Steffens:1993pc,Steffens:1994hf}, which employ the Peierls-Yoccoz projection following \cite{Signal:1988vf} and achieve a fair degree of phenomenological success.

In this connection it is notable that having overcome the problem of incorrect support, one can ask for more; e.g., models of the nucleon can be classified, and perhaps retained or discarded, according to whether or not they comply with the perturbative QCD constraint on the valence-quark distribution function \cite{Brodsky:1995,Avakian:2007xa}
\begin{equation}
\label{valenceconstraint}
q_v(x) \stackrel{x\sim 1}{\sim} (1-x)^{2 n -1+ 2 \,\delta \lambda}\, ,
\end{equation}
where $n$ is the minimal number of fermion spectator lines in the quark-level scattering process, and $\delta \lambda= |\lambda^q-\lambda^H|$ is the difference in helicities between the struck parton and the hadron; e.g., $\delta \lambda$ is \emph{zero} for a struck-quark with helicity parallel to that of the $J=\sfrac{1}{2}$ hadron and \emph{one} for a struck-quark with helicity antiparallel.  These counting rules are first-principles predictions of QCD, in which the power increases logarithmically under evolution to an higher scale, and they satisfy ``Gribov-Lipatov reciprocity'' \cite{Gribov:1971zn,Gribov:1972}; namely, that at leading-order, spacelike and timelike parton cascades are identical so that structure and fragmentation functions are algebraically related.
%http://www.lpthe.jussieu.fr/~yuri/reviews/yudok-ITEP-06-corr.pdf

For the proton, Eq.\,(\ref{valenceconstraint}) means
\begin{equation}
\label{protonvalenceconstraint}
q_v^p(x)\stackrel{x\sim 1}{\sim} (1-x)^\beta \,, \; \beta=3\,,
\end{equation}
for both $u$- and $d$-quarks, unless nonperturbative effects somehow preclude either the valence $u$-quarks or $d$-quark from sharing in the proton's helicity on a domain that includes the neighborhood of $x=1$.  NB.\ Subsequently, in connection with Eq.\,(\ref{xdomainlarge}), we will describe a quantitative bound on the $x$-domain within which Eq.\,(\ref{protonvalenceconstraint}) should be observable.

A naive consideration of the isospin and helicity structure of the proton's light-front quark wave function at $x\simeq 1$ leads one to expect that $d$-quarks are five-times less likely than $u$-quarks to possess the same helicity as the proton they comprise.  This leads obviously to the prediction \cite{Farrar:1975yb}
\begin{equation}
\label{donuFarrar}
\frac{d(x)}{u(x)} \stackrel{x\to 1}{=} \frac{1}{5}\,.
\end{equation}
However, as we saw in Sec.\,\ref{sec:param}, whilst fitted PDFs produce $\beta_u\approx 3$, they yield $4\lesssim \beta_d\lesssim 5$, the feature which leads to Eq.\,(\ref{donuempirical}).  This may be understood as an indication that the probability for a $d$-quark at $x=1$ to possess the same helicity as the proton is actually much less than the naive expectation.  Given the large uncertainties in the fits, however, it does not necessarily require that $\beta_d>\beta_u$.  A reliable nonperturbative approach to nucleon structure is required in order to determine the relative strength of the $\beta_d=3$ and $\beta_d=5$ components in the proton.

As explained in \cite{Brodsky:1995} and we subsequently elucidate,\footnote{A quantitative definition of $Q_0$ is presented in Sec.\,\protect\ref{sec:DSE}; and specifically in connection with Eq.\,(\protect\ref{qvpikappa}).} Eq.\,(\ref{valenceconstraint}) is applicable at the infrared resolving scale, $Q_0$.  The exponent is a lower bound and increases under QCD evolution to a larger momentum scale.  This is described in Sec.\,\protect\ref{sec:QCDscalingV}.  Equation~(\ref{protonvalenceconstraint}) can therefore serve as a stringent discriminator.  It is satisfied with bag model valence-quark wave functions that fall as $1/|\vec{p}|^2$ for large momentum $|\vec{p}|$ \cite{Margolis:1992gf}, whereas the oft used independent particle wave functions with Gaussian decay fail this test [see Eq.\,(\ref{CQMxfvF}) and associated discussion].

The computation described in \cite{Steffens:1994hf} is formulated with four parameters: a bag radius $R$; the masses of the two valence-diquark spectator states in an impulse approximation to the forward Compton scattering amplitude -- scalar, $M_s$, and axial-vector, $M_v$ (see Fig.\,\ref{Comptondq}); and the infrared resolving scale at which the computation is supposed to be valid, denoted by $Q_0$ above and $\mu$ therein.\footnote{A comparison between bag model results and those of other, kindred non-topological solition models is presented in \protect\cite{Bate:1992zx}.  Marked quantitative differences can be found in spin-dependent structure functions.}  Naturally, the calculated valence-quark distribution functions depend on these parameters, which may therefore be fixed by applying leading- or next-to-leading-order evolution to the calculated distributions and requiring a good least-squares fit to the MRS parametrization \cite{Martin:1992zi} of experimental data at $Q^2=10\,$GeV$^2$.

\begin{figure}[t]

\centerline{\includegraphics[clip,width=0.6\textwidth,angle=-90]{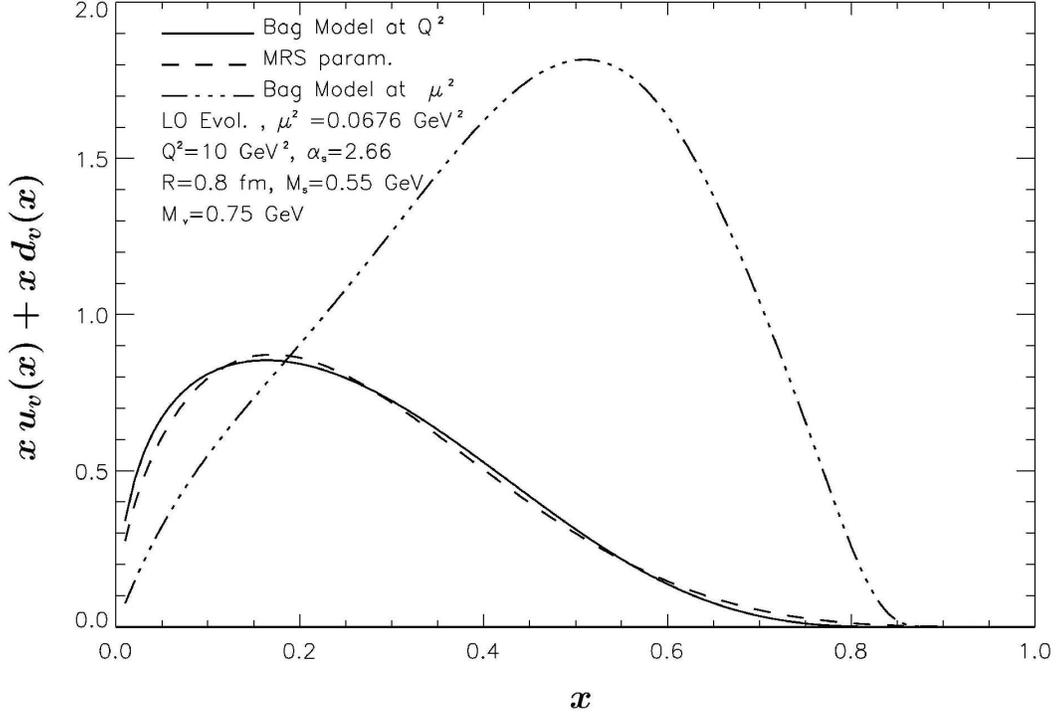}}
\caption{A bag model result for the $x$-weighted sum of valence-quark distributions in the proton.  The model parameters are: bag radius $R=0.8\,$fm; scalar and axial-vector diquark masses, $M_s=0.55\,$GeV, $M_v=0.75\,$GeV; and infrared resolving scale $Q_0=\mu=0.26\,$GeV.  [Figure adapted from \protect\cite{Steffens:1994hf}.] \label{ST941}}
\end{figure}

At leading-order, Fig.\,\ref{dglap}, the values $R=0.8\,$fm, $M_s=0.55\,$GeV, $M_v=0.75\,$GeV, $Q_0=\mu=0.26\,$GeV yield the results depicted in Fig.\,\ref{ST941}.  The diquark masses are roughly 20\% lower than estimates derived from QCD's Dyson-Schwinger equations, Eq.\,(\ref{diquarkmass}) \cite{Burden:1996nh,Maris:2002yu}.  Of more material concern is the small value of $Q_0$, in which case the evolution equations involve the coupling $\alpha(Q_0)/[2\pi] = 0.4$; cf.\ the values considered reasonable in Sec.\,\ref{sec:GRV}.

Since this value of $\alpha(Q_0)$ might be considered too large for the fitting-by-evolution procedure to be valid, \cite{Steffens:1994hf} repeated the analysis using the next-to-leading-order evolution equations, Eqs.\,(\ref{eq:evol1}) -- (\ref{eq:evol5}), and in this way obtained $R=0.8\,$fm, $M_s=0.7\,$GeV, $M_v=0.9\,$GeV, $Q_0=\mu=0.34\,$GeV.  While the value of $Q_0$ is still small, these diquark masses agree with contemporary estimates and now $\alpha(Q_0)/[2\pi] = 0.1$.  In comparison with Fig.\,\ref{ST941}, the NLO prescription cannot significantly change the evolved distribution because that is the quantity fitted.  On the other hand, there is a significant difference in the valence-quark distribution at the new infrared scale $Q_0=0.34\,$GeV: $x [u_v(x;Q_0)+d_v(x;Q_0)]$ peaks at $x=0.4$ (reduced by 30\%), has a maximum value of 1.4 (reduced by 20\%) and material support only for $x<0.8$ (domain contracted by 10\%).
These differences owe primarily to the accelerated production of sea-quarks through valence-quark depletion in NLO evolution as compared with the rate of this effect in LO evolution.  The change in diquark masses serves to fine-tune the initial condition: more massive intermediate states produce a $Q_0$ distribution that peaks at smaller $x$, a feature required to fit the data parametrization when evolving from a larger value of $Q_0$.

As one might have expected, the computation reviewed here does not ensure correct normalization of the valence-quark distribution functions: roughly 20\% of the strength is missing.  The authors argue plausibly that this defect owes to modeling and parametrization of the intermediate state as opposed to its internally consistent calculation.  They assert that the missing strength is probably localized in the sea-quark region; viz., $x < 0.3$, and one might therefore be justified in supposing that this weakness affects only modestly the pointwise behavior of the valence-quark distribution on the domain of interest.

Upon the domain within which perturbative QCD is unarguably valid the parametrizations of valence-quark distribution functions are infinite at $x=0$ while $ x q_v(x;Q^2_{\rm pert})=0$.  The divergence is associated with parton splitting, see Fig.\,\ref{dglap}, and may be understood intuitively by observing that infinitely many $x\sim 0$ partons are required to carry any fixed fraction of the proton's momentum.  On the other hand, by assumption, model calculations, such as that just described, generate $q_v(x;Q^2_0)<\infty$ because only a fixed number of valence quarks (three for the proton) interact with the photon probe, and the valence-quark's structure is not resolved; i.e., there is no mechanism for them to fragment.  While the property $q_v(x;Q^2_0)<\infty$ is not incompatible with the evolution equations, it does disagree with data parametrizations; e.g., see Sec.\,\ref{sec:param}.

% http://www.slac.stanford.edu/spires/find/hep/www?eprint=hep-ph/9610205

\subsubsection{Constituent quark models}
\label{CQMtheory}
% Constituent quarks and parton distributions. http://www.slac.stanford.edu/spires/find/hep/www?j=NUPHA,A614,472
% Quark model predictions for the SU(6) breaking ratio of the proton momentum distributions. http://www.slac.stanford.edu/spires/find/hep/www?eprint=nucl-th/0209029
% Relativity and constituent quark structure in model calculations of parton distributions. http://www.slac.stanford.edu/spires/find/hep/www?eprint=hep-ph/0409059
%
In the minds of many, potential models framed in terms of constituent-quark degrees-of-freedom are exemplified by \cite{Isgur:1978xj}.  The application of such models to the spectra, and strong and electromagnetic couplings of baryons is reviewed in \cite{Capstick:2000qj}.  It is noteworthy that one can exactly eliminate the spurious center-of-mass motion in nonrelativistic constituent-quark models (CQMs) and hence they can be used effectively as a tool to study the spectrum.  However, one should bear in mind that the absence of Poincar\'e covariance is a problem when studying scattering, in general, and deep inelastic scattering, in particular.

These models are defined by an Hamiltonian, which is typically diagonalized in an harmonic oscillator basis.  The Hamiltonian contains a hyperfine interaction, whose structure is modeled according to a practitioner's taste.  In many cases it is based on notions of one-gluon exchange [e.g., \cite{Isgur:1978xj}], in others, on pseudoscalar meson exchange [e.g., \cite{Glozman:1997ag}].  Herein, it is only important that the Hamiltonian's eigenvectors do not exhibit $SU(6)$ spin-flavor symmetry and hence results other than Eq.\,(\ref{F2nF2pSU6}) are possible.

Relevant here is a relationship that appears empirically to hold between the anomalous magnetic moments of the neutron and proton, $\kappa_{n,p}\,$, and the valence-quark distribution functions [see Sec.\,4.2.3 of \cite{Goeke:2001tz}]; viz.,
\begin{equation}
\label{valencekpkn}
\frac{\kappa_p}{\kappa_n} = -\,\frac{1}{2}\left[\frac{4 \langle x\rangle_{d_v}^{Q_0}+\langle x \rangle_{u_v}^{Q_0}}{\langle x\rangle_{d_v}^{Q_0}+\langle x\rangle_{u_v}^{Q_0}}\right]\,,
\end{equation}
where, as usual,
\begin{equation}
\langle x\rangle_{q_v}^{Q_0} = \int_0^1 dx\, x q_v(x;Q_0)
\end{equation}
is the fraction of the proton's momentum carried by a valence-quark of type $q\,(=u,d,\,{\rm etc.})$, at the scale $Q_0$.   Direct computation shows that even though the numerator and denominator on the right-hand-side (rhs) in Eq.\,(\ref{valencekpkn}) are separately scale-dependent, the ratio is not.  NB.\ We reiterate and emphasize that Eq.\,(\protect\ref{valencekpkn}) holds in a phenomenological sense: it is satisfied to within 1\% by modern parametrizations of parton distribution functions but has not been proven theoretically.

The model described in Sec.\,\ref{MITbagtheory} yields $\langle x\rangle_{u_v} = 2 \langle x\rangle_{d_v}$ and hence, unsurprisingly, $\kappa_p = -\kappa_n$ from Eq.\,(\ref{valencekpkn}).  It is therefore evident that the magnitude of any deviation from this result can serve as one gauge of the amount by which $SU(6)$ symmetry is broken within the nucleon.  Experimentally, $\kappa_p/\kappa_n \approx -0.94\,$, so the breaking is small by this measure.

It has been argued \cite{Traini:1997jz} that in CQMs, if one assumes
\begin{equation}
\frac{1}{(2\pi)^3} |\phi_q(\vec{k})|^2 = m_q n_q(|\vec{k}|)\,,
\end{equation}
where $\phi_q(\vec{k})$ is the momentum-space wave function for a constituent-quark of type $q$ within a nucleon of mass $M$, and $m_q(Q_0)$ and $n_q(|\vec{k}|)$ are, respectively, the constituent-quark's mass and momentum density distribution, then the virtual photon-nucleon Compton scattering amplitude, illustrated in Fig.\,\ref{Comptondq}, can yield ($k^+ = k_0+k_3$, see App.\,\ref{appLF})
\begin{equation}
\label{CQMxfvE}
x q_v(x;Q_0) = \frac{m_q(Q_0)}{M} \int \! d^3k \, n_q(|\vec{k}|) \, \delta(x - \frac{k^+}{M})\,.
\end{equation}
This expression can satisfy the momentum sum rule if sea-quarks and gluons are absent at $Q_0$ and $m_q(Q_0) = M/3$.  However, the distribution does not have correct support: it is nonzero for $x>1$, and nor does it satisfy the normalization conditions, Eqs.\,(\ref{uvdvnorm}).

Both flaws are overcome if one adopts the prescription \cite{Traini:1997jz}
\begin{equation}
\label{CQMxfvF}
x q_v(x;Q_0) = \frac{1}{(1-x^2)} \int \! d^3k \, \frac{k^+}{M} \, n_q(|\vec{k}|) \, \delta(\frac{x}{1-x} - \frac{k^+}{M})\,,
\end{equation}
which exhibits two modifications.
The simpler is the replacement $m_f\to k^+$, which provides the struck parton with an amount $k^+$ of the target's momentum, whereas it was at rest in Eq.\,(\ref{CQMxfvE}).  This ensures that Eqs.\,(\ref{uvdvnorm}) are satisfied.\footnote{In the Bjorken limit, $k^+ = x M$ in the target rest frame.}
More complicated is the modification of the Dirac $\delta$-function's argument.  It was inspired by an observation that only inelastic processes can contribute to the distribution functions.  In this case the intermediate fermion line in Fig.\,\ref{Comptondq} should carry momentum $k+q^\prime$, where $q^\prime=(\nu^\prime,\vec{q}{\,}^\prime)$ with $\nu{\,}^\prime = \nu - \nu_{\rm el}$ and $\nu_{\rm el}=Q^2/[2 M]$.  With this prescription the valence-quark distribution function is defined such that the energy transfer does not include the part absorbed in elastic scattering.  An additional factor of $1/(1-x)^2$ is necessary to preserve the normalization.

Equation~(\ref{CQMxfvF}) is a reasonable definition, which can be applied to a large class of CQMs.  That of \cite{Isgur:1978xj} is included.  In this case the wave functions are Gaussian, from which it follows immediately that the model yields valence-quark distribution functions  that violate Eq.\,(\ref{protonvalenceconstraint}); namely, $q_v(x) \sim \exp (-x^2/(1 - x)^2)/(1 - x)^2$ for $x\simeq 1$.

It is plainly the high-momentum components of constituent-quark relative-momentum wave functions that determine this behavior and Eq.\,(\ref{protonvalenceconstraint}) can be obtained from Eq.\,(\ref{CQMxfvF}) if, and only if,
\begin{equation}
n_q(|\vec{k}|) \stackrel{|\vec{k}|\to \infty}{\sim} \frac{1}{|\vec{k}|^{7}}.
\end{equation}
(NB.\ The behavior $1/|\vec{k}|^6$ gives $(1-x)^2$, while $1/|\vec{k}|^8$ gives $(1-x)^4$.)  Hence, while the constituent-quark wave functions in the CQMs of \cite{Bijker:1994yr,Ferraris:1995ui,Giannini:2001kb} provide greater support at high-momentum than those of \cite{Isgur:1978xj}, in all cases it is still insufficient to give agreement with Eq.\,(\ref{protonvalenceconstraint}) via Eq.\,(\ref{CQMxfvF}).
% Would be nice to know if any CQM can provide wave functions with correct power-law behaviour; i.e., psi(k) ~ 1/|k|^3
We know of no CQM that can.

\begin{figure}[t]
\centerline{\includegraphics[clip,width=0.6\textwidth,angle=-90]{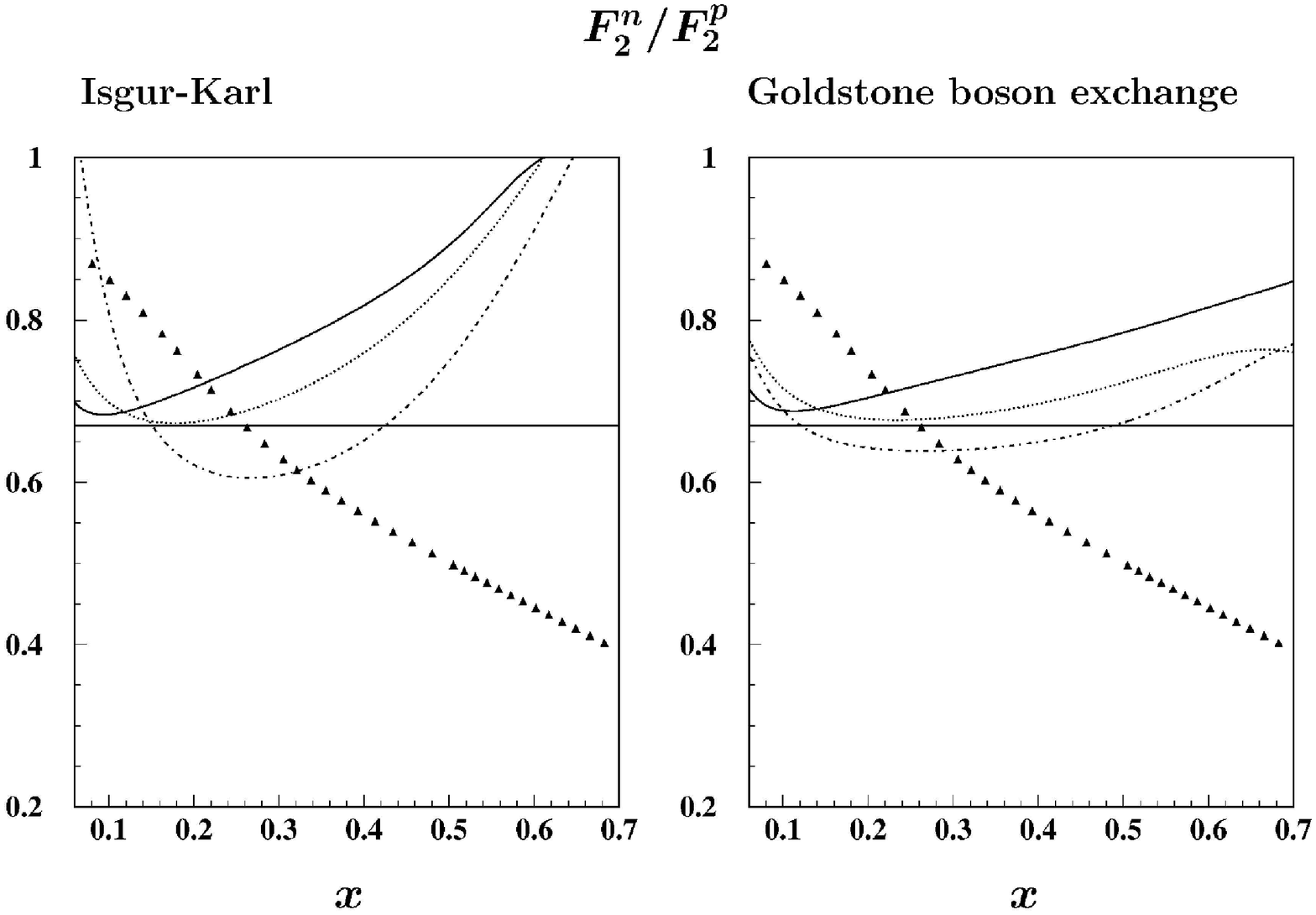}}
\caption{\label{PTB02}
$F_2^n(x)/F_2^p(x)$.  \emph{Left panel} -- computed in the model of \protect\cite{Isgur:1978xj}: \emph{dot-dashed curve}, at hadronic scale $Q_0 \sim 0.4\,$GeV; \emph{dotted curve} - leading-order evolution to $Q_0^2=10\,$GeV$^2$, \emph{solid curve} - next-to-leading-order evolution.  \emph{Right panel} -- computed in the model developed from \protect\cite{Glozman:1997ag}.  In both panels -- \emph{Triangles}: CTEQ5 fit to experimental data \protect\cite{Lai:1999wy}; and \emph{straight line}: Eq.\,(\protect\ref{F2nF2pSU6}).
NB.\ Using Eq.\,(\ref{eq:npratio}) and assuming $s_v(x)=0$ on the valence-quark domain, the large-$x$ behavior in these panels corresponds to $d_v(x)/u_v(x) \gtrsim 1.0$.
[Figure adapted from \cite{Pasquini:2002be}.]}
\end{figure}

Notwithstanding this, an illustration of the impact of $SU(6)$ symmetry breaking via CQMs might be valuable.  Such illustrations are provided in \cite{Traini:1992ue,Pasquini:2002be}.  The more recent study compares distribution functions calculated in the models of \cite{Isgur:1978xj,Glozman:1997ag}.  Results are depicted in Fig.\,\ref{PTB02}.
It is apparent that evolution does not dramatically affect the computed ratio; e.g., compare with the large effect evident in Fig.\,\ref{ST941}.  Hence a comparison between the models' predictions is straightforward.
The Hamiltonian defining each model breaks $SU(6)$ symmetry, albeit in slightly different ways, and therefore both produce curves which deviate from Eq.\,(\ref{F2nF2pSU6}).
Notably, the models exhibit quite different behavior at large-$x$.  This is primarily because, in contrast to \cite{Isgur:1978xj}, a relativistic kinetic energy operator is used in \cite{Glozman:1997ag} and that increases the wave function's support at high-momentum.

Along with the two CQM calculations, Fig.\,\ref{PTB02} depicts the behavior of $F_2^n/F_2^p$ that is inferred from the CTEQ5 parametrizations of experimental data \cite{Lai:1999wy}.  In this connection it is instructive to recall Fig.\,\ref{f2nratio} and the discussion in Sec.\,\ref{neutronstructurefunction}, which emphasizes that $F_2^n/F_2^p$ is well known for $x< 0.7$.  The parametrization ratio can therefore be viewed as a good representation of data on this domain.  The absence of a free neutron target only introduces material uncertainty for $x\gtrsim 0.7$, a domain not reproduced in the figure.

It is therefore significant that neither CQM yields behavior for $F_2^n/F_2^p$ that is consistent with the parametrization ratio, not even for $x\gtrsim 0.4$; i.e., the valence-quark domain, whereupon they are supposed by proponents to provide a veracious description of the nucleon via its salient degrees-of-freedom.
For $0.4< x <0.7$ the ratio in both panels lies well above the data parametrization.  This is in contrast to models wherein the struck quark is partnered by scalar- and axial-vector-diquark intermediate states \cite{Close:1973xw,Carlitz:1975bg,Close:1988br,Meyer:1990fr}, to which we shall return.
In \cite{Pasquini:2002be} it is argued that the large-$x$ behaviour of the the distribution functions is greatly influenced by proper implementation of the Pauli principle.  That is neglected in \cite{Close:1973xw,Carlitz:1975bg,Close:1988br}, which treat the diquarks as an elementary degree of freedom whose quark constituents are not active in the scattering process.  Naturally, this is a weakness of those models, whose impact should be explored.  (NB.\ That can be done within a Faddeev equation treatment of the nucleon; see, e.g., Sec.\,\ref{NJLnucleon} and App.\,\ref{Appendix:FE}.)

With the information described hitherto one can examine Eq.\,(\ref{valencekpkn}) within CQMs.  To this end three different $SU(6)$-breaking Hamiltonians were considered in \cite{Giannini:2002tn}; namely, those associated with the models of \cite{Isgur:1978xj,Ferraris:1995ui,Giannini:2001kb}: Eq.\,(\ref{valencekpkn}) is satisfied to within 1\% in each case.  However, despite their differences, this collection of models belongs to a class\footnote{It appears that this class includes the model of \protect\cite{Glozman:1997ag} if a point-form spectator model (PFSM) is employed for the electromagnetic current operator \protect\cite{Melde:2007zz}.  Of the currents considered in this connection, only the PFSM can produce results in fair agreement with experiment.} whose characteristic is $\kappa_p=-\kappa_n$ precisely.  It seems therefore that Nature employs a different mechanism to break $SU(6)$ symmetry than that expressed in this class of models.

\subsubsection{Pion cloud}
\label{pioncloud}
Chiral symmetry and the pattern by which it is broken are fundamental to the nature of hadron structure.  Neither the MIT bag model nor constituent-quark models express QCD's chiral symmetry correctly.  In working toward a theoretical understanding of distribution functions it is therefore important to explore the influence of correcting this defect.

% Kulagin. nucl-th/9510053
% http://www.slac.stanford.edu/spires/find/hep/www?eprint=hep-ph/9711368
Pseudoscalar mesons come naturally to mind when considering chiral symmetry in QCD and mesons mean antiquarks.  The addition of $q \bar q$ pairs or correlations to the intermediate state that is depicted in Fig.\,\ref{Comptondq} has long been considered a plausible mechanism by which to solve the normalization problem encountered in computing the valence-quark distribution functions \cite{Schreiber:1991tc}.

This idea is canvassed in \cite{Kulagin:1995ia} within the framework of a dispersion representation of the valence-quark distribution function \cite{Kulagin:1994fz}
\begin{equation}
\label{qvdispersion}
q_v(x;Q_0)= \int dw \int_{-\infty}^{k^2_{\rm max}(w,x)} dk^2\, \rho(w,k^2,x;Q_0)\,.
\end{equation}
Here $\rho$ is the probability density associated with an intermediate spectator state of invariant mass $w=(P-k)^2$ and
\begin{equation}
\label{qvvirtuality}
k^2_{\rm max}(w,x) = x \left(M^2-\frac{w}{1-x}\right),
\end{equation}
where $M$ is the nucleon mass, is the maximum kinematically-allowed value of the struck-quark's squared-four-momentum, which is equivalent to its virtuality if the quark is massless.  In connection with Eq.\,(\ref{qvdispersion}), convergence and internal consistency suggest that the probability density can only have material support for $k^2\lesssim -M^2$.  The condition
\begin{equation}
k^2_{\rm max}(w,x) = -M^2 \;\Rightarrow \; w^2 = M^2 (1-x^2)/x\,.
\end{equation}
Hence, in this case it follows that only intermediate states with $w \lesssim M^2$ are important for $x\gtrsim 0.6\,$.
% k_m^2 = -M^2 => w^2 = M^2 (1-x^2)/x
This explains the spectator diquark assumption of Fig.\,\ref{Comptondq}.  However, as $x$ decreases below $x=0.6$, intermediate states with $s>M^2$ will play an increasingly important role and must be included in order to obtain a pointwise-accurate result for $q_v(x;Q_0)$.

\begin{figure}[t]
\centerline{\includegraphics[clip,width=0.35\textwidth,angle=-90]{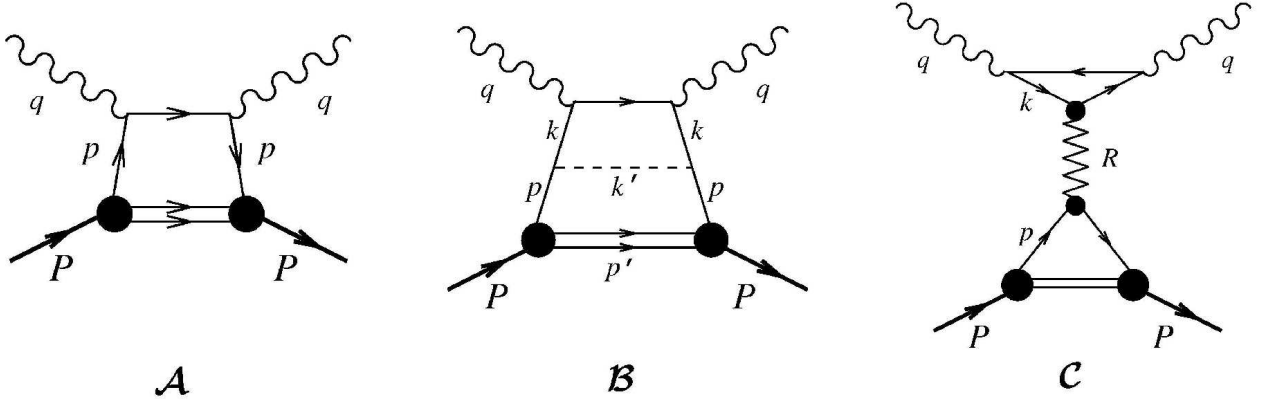}}
\caption{\label{Ku1995ia} Depiction of three invariant-mass sectors conjectured in \protect\cite{Kulagin:1995ia} to contribute to deep inelastic scattering. \emph{Left image} -- ${\cal A}$: low-$w$ -- diquark spectators.  \emph{Central image} -- ${\cal B}$: intermediate-$w$ -- compound spectator system involving a pion (dashed line), constituent-quarks and diquarks.  \emph{Right image} -- ${\cal C}$: large-$w$ -- Regge trajectory model for $q \bar q$ scattering from a constituent-quark.
[Figure adapted from \protect\cite{Kulagin:1995ia}.]}
\end{figure}

A concrete model is used in \cite{Kulagin:1995ia} to illustrate these points.  It is notable that, by combining the components indicated below, this study aims to produce a valence-quark distribution function that is valid at a resolving scale $Q_0=1\,$GeV; i.e., at a scale which may lie within the perturbative domain.  We judge that perturbative QCD evolution from such a scale is sounder than from the manifestly infrared scales employed in Secs.\,\ref{MITbagtheory} and {\ref{CQMtheory}.

The study divides the space of intermediate states into three sectors, which are illustrated in Fig.\,\ref{Ku1995ia} and described here:
\begin{description}
\item[${\cal A}$]: $w \lesssim M^2$.  This is the diagram of Fig.\,\ref{Comptondq}, involving diquark spectators, which the studies described in Secs.\,\ref{MITbagtheory} and \ref{CQMtheory} computed as the sole process contributing to the valence-quark distribution function.  It is represented by a probability density
\begin{equation}
\rho_{\cal A}(w,p^2,x;Q_0) = \sum_{D=0^+,1^+} \delta(w-m_{D}^2) \, f_{Q/N}^D(p^2,x;Q_0)\,,
\end{equation}
with $f_{Q/N}^D(p^2,x;Q_0)$ identified as the probability distribution for finding a constituent-quark in the nucleon with invariant mass $p^2$, proton light-cone momentum fraction $x$ and partnered by a diquark with $J^P=D$.  Faddeev equation models of nucleon structure provide an appropriate framework for its evaluation.  (See, e.g., App.\,\ref{Appendix:FE}.)\\
%%%%
\item[${\cal B}$]: $M^2 \lesssim w \lesssim w_0$, $w_0 \sim 2 \,M^2$.
One model for the addition of $q \bar q$ pairs or correlations to the intermediate state is to dress constituent-quarks with pseudoscalar mesons.\footnote{This can be sensible even when one unfolds and understands the structure of constituent-quarks \protect\cite{Blaschke:1995gr,Oertel:2000jp,Cloet:2008fw}.  Indeed, it is an integral step in a systematic truncation of QCD's Dyson-Schwinger equations.}  In \cite{Kulagin:1995ia} the dressing sum is truncated at just one pion rung, depicted in the central image of Fig.\,\ref{Ku1995ia}.  The remainder is shifted to sector ${\cal C}$.\\
%%%%
\item[${\cal C}$]: $w \geq w_0$.
The large invariant mass spectator component is modeled via a Regge trajectory, with intercept $\alpha_R \approx 1/2$, to describe $q \bar q$ scattering from a constituent-quark.  This piece provides a contribution to the valence-quark distribution function that behaves as $x^{-\alpha_R}$ for small-$x$.  As we reported in Sec.\,\ref{MITbagtheory}, this is a desirable feature.
\end{description}

The model is elaborate and has numerous parameters.  Their values are fixed through comparison with data and an eye to avoiding large discrepancies with legitimate theoretical constraints.  Important amongst the parameters are:
the constituent-quark mass, $m_Q = 0.45\,$GeV;
the diquark masses, with the large values\footnote{These values are roughly 20\% larger than extant computations of the mass-scales associated with diquark correlations, Eq.\,(\protect\ref{diquarkmass}) \protect\cite{Burden:1996nh,Maris:2002yu}.} $m_{0^+}=1.0\,$GeV and $m_{1^+}=1.2\,$GeV appearing to be favored by data in this framework;
momentum space widths used in parametrizations of constituent-quark--diquark Faddeev wave functions, $\Lambda_{0^+}=1.0\,$GeV and $\Lambda_{1^+}=1.2\,$GeV, and exponents for these vertex functions, $n_{0^+}=2.0$ and $n_{1^+}=3.5$, chosen to guarantee Eq.\,(\ref{protonvalenceconstraint});
a hard-cutoff-scale for the constituent-quark--pion vertex, $\Lambda_{Q\pi} \sim 1.0\,$GeV;
and a power-law form factor to model the $p^2$- and $k^2$-dependence of the Regge exchange quark-quark scattering amplitude, which is defined by a mass-scale, $\Lambda_R=0.25\,$GeV, and a power, $n_R=4$.

Combining these ingredients, the valence-quark distribution functions are computed at $Q_0=1\,$GeV from the sum
\begin{equation}
q_v(x;Q_0)  =  Z_q(Q_0) \{ q_{\cal A}(x) + q_{\cal B}(x) - \bar q_{\cal B}(x) + q_{\cal C}(x)- \bar q_{\cal C}(x)\}\,, \; q=u,d\,,
\end{equation}
with the normalization constants fixed by Eqs.\,(\ref{uvdvnorm}) subject to the additional constraints $\int dx\, u_{\cal A}(x) = 2$, $\int dx \, d_{\cal A}(x) = 1$.  The relative normalizations of the contributions in Fig.\,\ref{Ku1995ia} are determined once the model's parameters are fixed; i.e., they are not additional independent quantities.  Hence, in this model, defined at $Q_0=1\,$GeV, the total valence-quark normalization is constituted from the diagrams in Fig.\,\ref{Ku1995ia} as follows:
\begin{equation}
\int_0^1 dx\,\{u_v(x;Q_0)+d_v(x;Q_0)\} = {\cal A} (56\%) + {\cal B} (16\%) + {\cal C} (28\%)\,.
\end{equation}
Moreover,
\begin{equation}
\int_0^1 dx\, x \,\{u_v(x;Q_0)+d_v(x;Q_0)\} = {\cal A} (41\%) + {\cal B} (6\%) + {\cal C} (0.06\%) = 0.47\,,
\end{equation}
which leaves 53\% of the proton's momentum to be carried by non-valence-quark degrees-of-freedom.

\begin{figure}[t]
\centerline{\includegraphics[clip,width=0.35\textwidth,angle=-90]{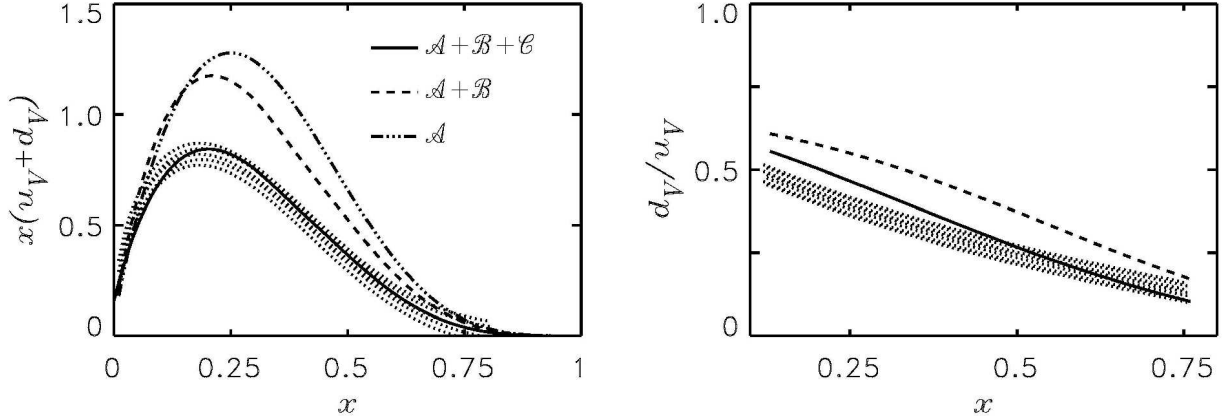}}
\caption{\label{Ku1995iaB}
Results illustrative of the pion-cloud model described in connection with Fig.\,\protect\ref{Ku1995ia}.
\emph{Left panel}: Total valence-quark distribution function.  Complete calculation, combining all sectors [Solid curve]; ${\cal A}+{\cal B}$ [dashed curve]; ${\cal A}$ alone [dash-dotted curve].
\emph{Right panel}: Contribution from sector-${\cal A}$ to the ratio of valence-quark distribution functions.  $m_{0^+}=1.0\,$GeV, $m_{1^+}=1.2\,$GeV [solid curve]; and $m_{0^+}=0.6\,$GeV, $m_{1^+}=0.8\,$GeV [dashed curve].
\emph{Both panels}: Computed results are reported at $Q_0^2=10\,$GeV$^2$, after LO evolution from $Q_0=1\,$GeV, whereat $\alpha/[2\pi] \simeq 0.03$.  Parametrizations of data \protect\cite{Lai:1994bb,Martin:1994kn} are provided for comparison [shaded band].
[Figure adapted from \cite{Kulagin:1995ia}.]}
\end{figure}

Figure~\ref{Ku1995iaB} depicts features of the model's distribution functions.
Although the sector-${\cal A}$ contribution is always dominant on the valence-quark domain, it is evident from the \emph{left panel} that the renormalization effected by adding the sector-${\cal B}$ and -${\cal C}$ contributions plays an important part within the model: it is a surrogate for evolution, shifting support to lower $x$ at the model's mass-scale of $Q_0=1\,$GeV.
The \emph{right panel} exhibits that behavior for the valence-quark flavor-ratio which is typically identified with diquark models: $d_v(x)/u_v(x)$ decreases monotonically with increasing $x$.  The ratio plotted is formed only from the sector-${\cal A}$ contributions.  On the valence-quark domain the full result may be estimated by multiplying the value plotted in Fig.\,\ref{Ku1995iaB} by the ratio of normalization constants; namely, $Z_d/Z_u \approx 2/3$.

It is now clear that by implementing a scheme which resolves some of the structure of a constituent-quark it is possible to shift the model-scale $Q_0$ to a point wherefrom perturbative evolution may plausibly be justified \emph{a priori}.  In this way one can reproduce experimental data and provide an interpretation.  However, the price appears high: with many ingredients needing to be carefully constrained and combined, the predictive power of an elaborate model is diminished.

% NJL model appears in many guises.  The first is as a basis for topological soliton models.  Then treated via BSE/FE.  It is the first framework able to treat the pion's structure function.
% Instanton models are distinct from soliton models.
\subsubsection{Topological soliton models}
\label{TopSolmodel}
Attempts to describe baryons as topological solitons are usually motivated by:
emphasizing the importance of chiral symmetry and the pattern by which it is broken in QCD;
the notion that when a carefully-defined large-$N_c$ limit is considered, QCD is a theory of weakly interacting mesons (and glueballs) and strongly interacting baryons;
and the observation that a simple form of classical nonlinear $\sigma$-model supports strongly-interacting static topological soliton solutions, which can be quantized as fermions.
Given this starting point, which possesses no explicit quark degrees-of-freedom, a connection with the parton model and DIS looks problematic.

% hep-ph/9604295  hep-ph/0302212
% Diakonov http://arxiv.org/PS_cache/hep-ph/pdf/9606/9606314v1.pdf
As in \cite{Jaffe:1974nj}, a straightforward calculation of distribution functions for localized field configurations is possible when the current operator is at most quadratic in the fundamental fields and the associated propagators are free-particle-like.  The first of these conditions is difficult to satisfy in topological soliton models expressed in terms of elementary meson fields because the nucleon is a nonperturbative object involving all orders of the pion field and hence the current operator cannot be quadratic.  This difficulty can be circumvented \cite{Weigel:1996kw,Diakonov:1996sr} by realizing the nonlinear $\sigma$-model through a truncated bosonization of a four-fermion theory \cite{Cahill:1985mh,Ebert:1985kz,Roberts:1987xc,Roberts:1989fu}.  Point coupling models of the Nambu--Jona-Lasinio type are often used.  A regularization prescription and an associated ultraviolet mass-scale, $\Lambda_{\rm UV}$, are an essential part of the definition of this model.  In general one cannot guarantee \emph{a priori} that $\Lambda_{\rm UV}$ plays no role in the Bjorken limit; namely, that $Q^2/\Lambda_{\rm UV}^2$ scaling violations are absent.

It has been argued that within this framework the singlet unpolarized quark distribution function in a nucleon at rest may be written \cite{Diakonov:1996sr}
\begin{equation}
\label{solitonupd}
u(x) + d(x) =  \frac{M N_c}{2\pi}
\int \frac{d^2p_\perp}{(2\pi)^2}
\sum_{n \; {\rm occup.}}
\left. \Phi_n^\dagger(\vec{p})
[ (I_{\rm D} + \gamma^0\gamma^3) ]
\Phi_n(\vec{p})\right|_{p_3= x M - E_n},
\end{equation}
wherein: $\Phi_n$ is an eigenfunction, with energy $E_n$, of a Dirac Hamiltonian constructed using a so-called hedgehog soliton \emph{Ansatz} for the pion field; the sum runs over the Dirac sea and the single discrete (valence) level bound in the soliton background, with $-m_Q < E_v < m_Q$, containing $N_c$ constituent-like quarks; and $x \in [-1,1]$. In defining Eq.\,(\ref{solitonupd}) one must follow the usual soliton model steps of projection and quantization to arrive at states with the correct spin and isospin, and three-momentum.  Moreover, it is necessary to employ a Poincar\'e invariant Pauli-Villars regularization of the underlying Nambu-Jona--Lasinio model, or an equivalent, in order to preserve, amongst other things, completeness of the eigenfunctions and positivity of the rhs in Eq.\,(\ref{solitonupd}).  NB.\ The singlet case is simplest in this approach because effects associated with cranking the topological soliton can be neglected.  The distributions in Eq.\,(\ref{solitonupd}) satisfy
\begin{equation}
q(x) = \left\{\begin{array}{cl}
q(x), & x>0 \\
-\bar q(-x), & x<0 \end{array}\right..
\end{equation}

\begin{figure}[t]
\centerline{\includegraphics[clip,width=0.36\textwidth,angle=-90]{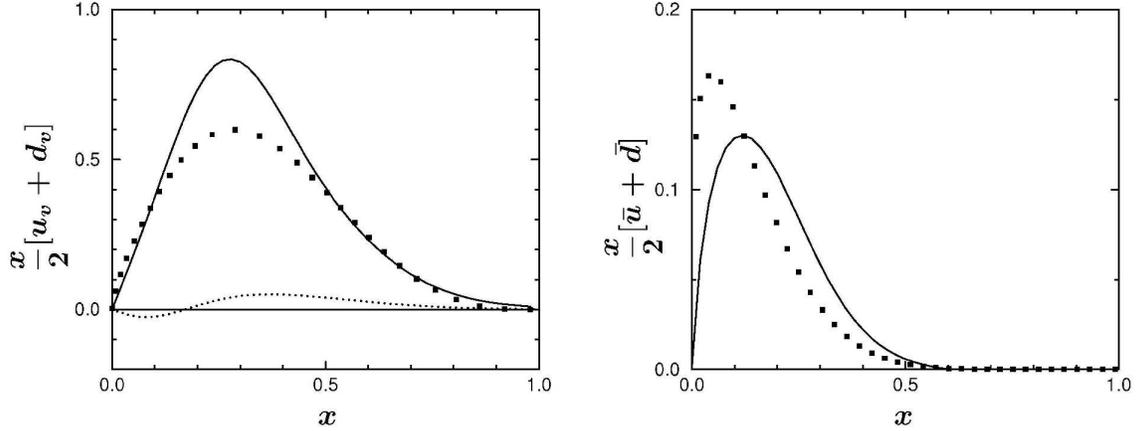}}
\caption{\label{1996srA}
Results obtained in a topological soliton model whose infrared resolving scale is argued to be $Q_0\sim 0.6\,$GeV.
\emph{Left panel}: $(x/2) \{u_v(x;Q_0) + d_v(x;Q_0)\}$, proportional to the valence-quark distribution: contribution from the single discrete (valence) energy level bound in the soliton background [Solid curve] and from the Dirac continuum [dotted curve].
\emph{Right panel}: $(x/2) \{\bar u(x;Q_0) + \bar d(x;Q_0)\}$, proportional to the antiquark distribution.
\emph{Both panels}: squares -- NLO data parametrization drawn from \cite{Gluck:1994}, associated with $Q_0=0.56\,$GeV.
[Figure adapted from \cite{Diakonov:1996sr}: $\Lambda_{\rm UV} = 0.56\,$GeV, the calculated nucleon mass is 1170\,MeV, and $N_c=3$ with $1/N_c$ corrections neglected.]}
\end{figure}

It can be shown algebraically within the model that
\begin{equation}
\label{solitonnorm}
\int_{-1}^{1}dx\, [ u(x;Q_0) + d(x;Q_0)] = \int_0^1 dx\, [ u_v(x;Q_0) + d_v(x;Q_0)] = N_c\,,
\end{equation}
which may be compared with Eqs.\,(\ref{uvdvnorm}).  It is notable that the rhs of Eq.\,(\ref{solitonnorm}) is saturated by the discrete level alone and hence that the Dirac continuum, distorted by the presence of the large pion field, contributes nothing to the soliton's baryon number.  In addition, with the Pauli-Villars regularization,
\begin{eqnarray}
\nonumber \lefteqn{
\int_{-1}^{1}dx\, x [ u(x;Q_0) + d(x;Q_0)]  =:  \int_0^1 dx\,x \Sigma(x;Q_0)}\\
& = & \int_0^1 dx\, x [ u(x;Q_0) + \bar u(x;Q_0) + d(x;Q_0) + \bar d(x;Q_0)] = 1\,,
\end{eqnarray}
so that at the resolving scale for which the model is assumed to be defined; viz., $Q_0 \simeq \Lambda_{\rm UV} = 0.6\,$GeV, the proton's complete momentum is carried by the model's constituent-quarks and -antiquarks.

Distributions typical of a topological soliton model are exemplified in Fig.\,\ref{1996srA}: \emph{left panel}, valence-quark distribution; and \emph{right panel}, the nonzero antiquark distribution.
%The latter receives a significant contribution from the Dirac continuum.
In this model the discrete level contributes to both the DIS valence-quark and antiquark distributions.  Furthermore, the Dirac continuum, distorted by the soliton, produces unequal distributions of quarks and antiquarks, and hence should not be solely identified with the sea.  It will be observed that while the model's antiquark distribution is incorrect in detail, it does have the important property of possessing little support on the valence-quark domain.  Moreover, owing to the nonzero antiquark distribution, which only vanishes as $N_c \to \infty$, the valence-quark distribution peaks at $x\lesssim 1/N_c$.

A comparison between the \emph{left panels} of Figs.\,\ref{Ku1995iaB} and \ref{1996srA} is instructive.  Allowing for the difference in normalization, there is a semi-quantitative similarity between the soliton model result and that produced by the sum of sectors ${\cal A}$ and ${\cal B}$ in the model of \cite{Kulagin:1995ia}; i.e., constituent-quark plus pion (see Fig.\,\ref{Ku1995ia}).  This highlights the omission of sea-quark effects in the soliton model at its resolving scale and also the role that sea-quarks play in general; e.g., with their inclusion, $x q(x)$ has far greater support at low-$x$ and commensurately less support on the valence-quark domain.

It is interesting to compare the approach discussed here with that in  Sec.\,\ref{MITbagtheory}.  In contrast to the static cavity treatment of bag models, wherein the three constituent-quarks contribute only to quark distributions, here they also generate an antiquark distribution owing to the strong pion mean-field, which distorts the negative-energy continuum of the Dirac Hamiltonian.  In the context of Sec.\,\ref{MITbagtheory}, this might be mimicked by incorporating effects of feedback between the constituent-quarks and the bag surface.

In the approach outlined here the baryon number and momentum sum rules are automatically preserved.  However, again owing to a lack of translational invariance in the formulation, the model's distribution functions do not vanish at $x\geq 1$ but behave as $\exp(- {\rm const.}\, N_c\, x)$.

That problem was considered in \cite{Gamberg:1997qk}, which argues that distribution functions in the infinite momentum frame (IMF) may be obtained from those calculated in the nucleon rest frame via the \label{PfnIMF} substitution:\footnote{Recall that in the infinite momentum frame the target's momentum is equivalent to its light-front momentum and hence Bjorken-$x$ coincides with the fraction of the target's light-front momentum carried by a parton.  It is only in the IMF that $q_v(x)$ can be interpreted as a single-parton probability density. \label{fnIMF}}
\begin{equation}
\label{RFtoIMF}
q_{\rm IMF}(x) = \theta(1-x) \frac{1}{1-x} \, q_{\rm RF}(-\ln [1-x]) \,.
\end{equation}
The same formula was derived in formulating a translationally invariant bag model in $1+1$ dimensions \cite{Jaffe:1980qx}, with the operation $x\to -\ln (1-x)$ identified as expressing Lorentz contraction of the bag-like object.  Akin to Eq.\,(\ref{CQMxfvF}), Eq.\,(\ref{RFtoIMF}) maps a function with support on $[0,\infty)$ to one with support restricted to $[0,1]$ and preserves normalization; i.e.,
\begin{equation}
\int_0^1 dx\, q_{\rm IMF}(x) = \int_0^\infty dz\, q_{\rm RF}(z)\,.
\end{equation}
For a typical distribution, it increases the peak height and suppresses the tail.  The effects are large, perhaps too large for favorable comparison with experiment, and soliton rotation was not considered in the derivation \cite{Wakamatsu:1997en}.  While the alternative of merely neglecting the support problem is not meritorious, the prescription of Eq.\,(\ref{RFtoIMF}) is not widely employed.

\subsubsection{Pion structure function in the topological soliton model}
In topological soliton models chiral symmetry is emphasized and a strong pion mean-field is indispensable in order to realize a baryon.  This said, it is natural to ask what the model predicts for the pion's valence-quark distribution functions.  Suppose, as above, one specifies the theory via a bosonization of a point-coupling Nambu--Jona-Lasinio model, and in addition completes its definition by using a two-subtraction Pauli-Villars regularization scheme for the real part of the action.  Under these conditions it has been argued \cite{Gamberg:1999gr} that the pion's valence-quark distribution function is
\begin{equation}
\label{qvSoliton}
q_v^\pi(x;Q_0) = \frac{{\cal P}(m_\pi^2,x;Q_0)}{\int_0^1 dx\,{\cal P}(m_\pi^2,x;Q_0)}
\end{equation}
with
\begin{eqnarray}
{\cal P}(m_\pi^2,x;Q_0) &=& \left.\frac{d}{dq^2} \,q^2 \Pi(q^2;Q_0) \right|_{q^2=m_\pi^2},\\
\Pi(q^2;Q_0) & = & -i \sum_{i=0}^2 \int\frac{d^4 k}{(2\pi)^4}\frac{1}{[-k^2- x(1-x) q^2+m_Q^2+\Lambda_{i\,{\rm UV}}^2]^2}\,,
\end{eqnarray}
where $\{c_i,\Lambda_{i\,{\rm UV}}\}$ are the Pauli-Villars parameters.  It follows that in the chiral limit; viz., $m_\pi^2=0$, an analytic point in this calculation, $q_v^\pi(x;Q_0)$ is a constant whose integral over $x\in [0,1]$ is unity.  Hence
\begin{equation}
\label{qvpiTS}
q_v^{\pi\,{\rm soliton}}(x;Q_0) = 1\,,
\end{equation}
which, strikingly, means that at the model's resolving scale it is equally likely for the $u$-quark to carry any fraction of the pion's momentum: all, none, or some fraction in between, with the same probability.

To place Eq.\,(\ref{qvpiTS}) in context, the parton model prediction for the pion's valence-quark distribution function is \cite{Ezawa:1974wm,Farrar:1975yb,Berger:1979,Brodsky:1979qm}:\footnote{Equation~(\protect\ref{pionpartonmodel}) generalizes to $q_v^{0^{-}}(x;Q_0)\sim (1-x)^{2 n}$ for a pseudoscalar meson with $n+1$ valence-quarks.  Again, the exponent, $2 n$, is a lower bound, saturated at the infrared resolving scale.  The value is increased by QCD evolution.}
\begin{equation}
%q_v^{\pi}(x;Q_0) \stackrel{x\sim 1}{\sim} (1-x)^{2 n -1+ 2 \delta S_z} = (1-x)^2
q_v^{\pi}(x;Q_0) \stackrel{x\sim 1}{\sim} (1-x)^2.
\label{pionpartonmodel}
\end{equation}
Plainly, Eq.\,(\ref{valenceconstraint}) is also applicable to $J=0$ hadrons since $n=1$ and $2\, \delta \lambda = 1$ in this case \cite{Brodsky:2005wx}.  The extra multiplicative factor of $(1-x)$ is present because $x=1$ corresponds to elastic scattering and hence $\sigma_T$ must vanish for a spin-zero target.  That imposes an helicity selection rule at the parton level, which causes $\sigma_T$ to vanish a single power faster as $x\to 1$ than one would anticipate from dimensional counting \cite{LlewellynSmith:1980zw}.
Owing to this, Eq.\,(\ref{pionpartonmodel}) is consistent with the Drell-Yan-West analyses \cite{Drell:1970,West:1970,lepage80}.
The soliton model prediction in Eq.\,(\ref{qvpiTS}) conflicts conspicuously with the parton model result.

\subsubsection{Pion in the Nambu--Jona-Lasinio model}
\label{sec:NJLpion}
Models of the Nambu--Jona-Lasinio (NJL) type have long been used in connection with the strong interaction.  As we described in Sec.\,\ref{TopSolmodel}, the NJL model provides a framework within which quark distribution functions may be defined in topological soliton models.  However, it is more common for NJL studies to emphasize dressed-quark degrees-of-freedom and realize hadrons as poles in $n$-point Green functions.  Following this path the model can be viewed from a Dyson-Schwinger equation (DSE) perspective \cite{Roberts:1994dr,GutierrezGuerrero:2010md}.

The continuing appeal of the NJL model can be explained by two features; namely, it is defined by a simple interaction and exhibits dynamical chiral symmetry breaking (DCSB).  This latter aspect, which the models of Secs.\,\ref{MITbagtheory}--\ref{pioncloud} do not possess, suggests that it might sensibly be used to explore the structure function of the pion.  As we noted above, the pion is special in QCD because it is both a dressed-quark-antiquark bound-state and the Goldstone mode arising from DCSB \cite{Maris:1997hd}.

\begin{figure}[t]
\includegraphics[clip,width=0.50\textwidth]{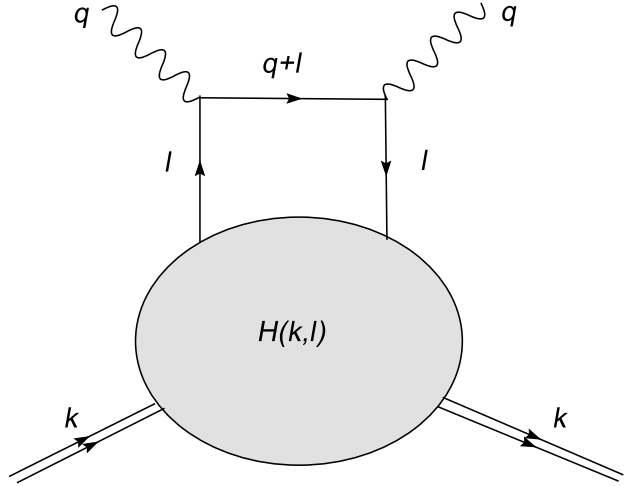}
\caption{Depiction of the general ``handbag diagram'' contribution to deep inelastic scattering from the pion.  As we saw in Fig.\,\protect\ref{Comptondq}, with an explicit expression for the single-flavor removal amplitude, ${\cal H}(k,\ell)$, one defines a model framework within which to complete the calculation.\label{generalHandbag}}
\end{figure}

By analyzing the Bjorken limit of the parton model expression for DIS from a pion target, one can derive a formula for the quark distribution function, which describes the quark-parton number density in the IMF:
% [see footnote on page~\pageref{RFtoIMF}]:
\begin{equation}
\label{operatordefpion}
q_v^\pi(x;Q_0) = \frac{1}{4\pi} \int d\xi^- {\rm e}^{\sfrac{i}{2} \, x \, k^+ \xi^-} \langle \pi(k) |\, \bar \psi_q(0) \gamma^+ \psi_q(\xi^-) | \pi(k) \rangle ,
\end{equation}
where $k^2=m_\pi^2$ or, equivalently, $k^- = (k_\perp^2 +m_\pi^2)/k^+$.  It follows from Eq.\,(\ref{operatordefpion}) that
\begin{equation}
\label{qvpionFF}
\int_0^1\,dx\, q_v^\pi(x;Q_0) = \frac{1}{k_+} \langle \pi(k) |\, \bar \psi_q(0) \gamma^+ \psi_q(0) | \pi(k) \rangle = 1\,,
\end{equation}
independent of $Q_0$, because the last matrix element is simply the contribution of a single quark flavor to the pion's electromagnetic form factor.

Equation~(\ref{operatordefpion}) can be expressed in momentum space as follows:
\begin{eqnarray}
\label{qvHklx}
q_v^\pi(x;Q_0) & = & \int \frac{d^4 \ell}{(2\pi)^4} \,\delta(\ell^+ - x k^+) \, {\rm tr}_{\rm CD} \, \gamma^+ {\cal H}_\pi(k,\ell;Q_0) \\
\label{qvHkl}
& = & \frac{1}{2} \int\frac{d \ell^- d\ell_\perp}{(2\pi)^4} \left. {\rm tr}_{\rm CD} \, \gamma^+ {\cal H}_\pi(k,\ell;Q_0)\right|_{\ell^+ = x k^+}.
\end{eqnarray}
%where $\Gamma^+(\ell,\ell)$ is the plus-component of the fully-dressed quark-photon vertex,
Here the matrix trace applies to color and spinor indices, and ${\cal H}_\pi(k,\ell;Q_0)$ is the amplitude defined by illustration in Fig.\,\ref{generalHandbag}: it is partially-amputated; i.e., the external pion legs are removed.  In this form,
%
%The rhs of Eq.\,(\ref{qvHkl}) is recognizable as the contribution from a single quark flavor to the electromagnetic pion form factor at $Q^2=0$.  It follows therefore that
\begin{equation}
\int_0^1 dx\, q_v^\pi(x;Q_0) = \frac{1}{k^+} \int \frac{d^4 \ell}{(2\pi)^4}\, {\rm tr}_{\rm CD} \, \gamma^+ {\cal H}_\pi(k,\ell;Q_0)
\end{equation}
and one recovers Eq.\,(\ref{qvpionFF}) because the integral on the rhs is merely an expression of the matrix element describing a single flavor's contribution to the form factor.  As we may readily establish, this identity is preserved in the commonly-used rainbow-ladder truncation so long as in constructing ${\cal H}_\pi(k,\ell;Q_0)$ the quark-photon vertex satisfies the Ward-Takahashi identity and the pion Bethe-Salpeter amplitude is canonically normalized \cite{Roberts:1994hh}.
% factor of 2 must be omitted because the flavour trace is missing from rhs. That gives a factor of 2 in the canonical normalisation condition.

Within the NJL model, Eqs.\,(\ref{qvHklx}) and (\ref{qvHkl}) are not yet meaningful.  A regularization scheme must be specified and the result for $q_v^\pi(x;Q_0)$ depends critically on the prescription adopted.  To illustrate this we write explicitly the formal NJL model expression for the rhs of Eq.\,(\ref{qvHkl}):
\begin{eqnarray}
q_v(x;Q_0) & = &
\frac{1}{2} \int\frac{d \ell^- d\ell_\perp}{(2\pi)^4} \left. {\rm tr}_{\rm CD} \, i\gamma_5 g_{\pi \bar qq} S(\ell) \gamma^+ S(\ell) i\gamma_5 g_{\pi \bar qq} S(k-\ell) \right|_{\ell^+ = x k^+},
\end{eqnarray}
where $g_{\pi \bar qq}$ is the Bethe-Salpeter amplitude for the pion\footnote{It is generally true that parton distribution functions are determined by a hadron's Bethe-Salpeter amplitude.  In the light-front quantization of QCD the PDFs can be computed from the absolute squares of light-front wave functions, integrated over the transverse momentum up to the resolution scale $Q_0$ \protect\cite{Brodsky:1997de}.}
%hep-ph/0102051
and $\gamma^+$ is the quark-photon vertex.  Both are independent of the quark-antiquark relative momentum in a symmetry-preserving regularization of the NJL model owing to the momentum-independent interaction.

One feature of a symmetry-preserving regularization is translational invariance, following from which one can use the Ward identity to rewrite
\begin{eqnarray}
q_v(x;Q_0) & = &
\frac{1}{2} \int\frac{d \ell^- d\ell_\perp}{(2\pi)^4} \left. {\rm tr}_{\rm CD} \, i\gamma_5 g_{\pi \bar qq} S(\ell) \gamma^+ S(\ell) i\gamma_5 g_{\pi \bar qq} S(k-\ell) \right|_{\ell^+ = x k^+}\\
\label{qvNJLa}
& = &
- \left. \frac{\partial }{\partial k^2} \Pi(k,x) \right|_{k^2=m_\pi^2},\\
\label{PiNJLa}
\Pi(k,x) & = & k^+ \int\frac{d \ell^- d\ell_\perp}{(2\pi)^4} \left. {\rm tr}_{\rm CD} \, i\gamma_5 g_{\pi \bar qq} S(\ell) i\gamma_5 g_{\pi \bar qq} S(k-\ell) \right|_{\ell^+ = x k^+}.
\end{eqnarray}
Although written in a slightly different form, Eqs.\,(\ref{qvNJLa}) and (\ref{PiNJLa}) are simply a restatement of Eq.\,(\ref{qvSoliton}).  This is natural, given that the NJL model provides the framework within which parton distribution functions are defined in topological soliton models.

At this point it is worth emphasizing that light-front concepts have hitherto appeared only as a result of a change in integration variables and it is generally assumed that the change of variables is nugatory.  Hence one is not actually working within a light-front formulation of QCD.
This makes no difference in perturbative analyses.  However, it does involve significant, hidden assumptions in connection with important nonperturbative phenomena.
For example, in order to evaluate Eq.\,(\ref{PiNJLa}) one requires the light-front form of the nonperturbatively-dressed quark propagator.  That quantity is, however, unknown.  %
Thus, in proceeding from Eq.\,(\ref{PiNJLa}), \cite{Bentz:1999gx}, for example, assumes that a dressed-quark mass, two orders-of-magnitude larger than the $u$-quark current-mass, is generated nonperturbatively and that $g_{\pi \bar qq}$ is related to the dressed-quark mass via a Goldberger-Treiman relation.
Neither of these results has yet been proved in the light-front formulation.  This problem was highlighted and considered in \cite{Bentz:1999gx}, see, e.g., Eq.\,(3.31) therein, but it was not solved, merely defined away.
Thus, an internally consistent treatment of Eq.\,(\ref{PiNJLa}) is wanting, so that a covariant treatment following \cite{Landshoff:1970ff} is preferable.

Nonetheless, if one overlooks these caveats and proceeds as is customary, then it follows from Eq.\,(\ref{PiNJLa}) that in the chiral limit with a translationally invariant regularization:
\begin{equation}
\label{qvpiNJLTI}
q_v^{\pi\,{\rm NJL}_{\rm TI}}(x;Q_0) = q_v^{\pi\,{\rm soliton}}(x;Q_0) = 1\,,
\end{equation}
for all values of $Q_0$, and consequently \cite{Dorokhov:2000gu}:
\begin{equation}
\label{qvpiNJLTIn}
\langle x^n \rangle_{Q_0}^{\pi\,{\rm NJL}_{\rm TI}} := \int_0^1 dx \, x^n\, q_v^{\pi\,{\rm NJL}}(x;Q_0) = \frac{1}{n+1}\,.
\end{equation}
These simple results are themselves problematic and can directly be traced to the absence of a length-scale characterizing the pion's transverse size when a translationally-invariant regularization is employed for a chiral-limit NJL model.  In this case both the interaction and the regularization procedure are momentum-independent, and hence neither can act to bound the relative momentum of the pion's constituents.  This explains why the model disagrees so markedly with the QCD parton model prediction, Eq.\,(\ref{pionpartonmodel}): in QCD there cannot exist a resolving scale at which the relative momentum of the pion's constituents is unbounded.

One need not naively employ a translationally invariant regularization scheme, however. An ultraviolet cutoff, $\Lambda_{\rm UV}$, can serve to mimic asymptotic freedom in QCD, in the sense that it defines an NJL-model quark-antiquark interaction which vanishes for momenta larger than $\Lambda_{\rm UV}$.  This procedure, adopted in \cite{Shigetani:1993dx} for analysis of the virtual photon-pion forward Compton scattering amplitude, yields
\begin{equation}
\label{Eq:NJLxqvpiHC}
q_v^{\pi\,{\rm NJL}_{\rm HC}}(x;Q_0) \propto -g_{\pi \bar qq}^2 \int_0^\infty\! d\kappa\,\theta(\Lambda_{\rm UV}^2 - \kappa_{\rm E})
\frac{\kappa - m_Q^2 - x m_\pi^2}{(\kappa-m_Q)^2} \, \theta( m_\pi^2 x(1-x) - x M^2 - (1-x)\kappa)\,,
\end{equation}
where $\kappa_{\rm E} = - \kappa + x m_\pi^2 - x m_Q^2/(1-x)$ and the constant of proportionality is fixed by requiring $\int_0^1 dx q_v(x;Q_0)=1$.

\begin{figure}[t]
\includegraphics[clip,width=0.75\textwidth]{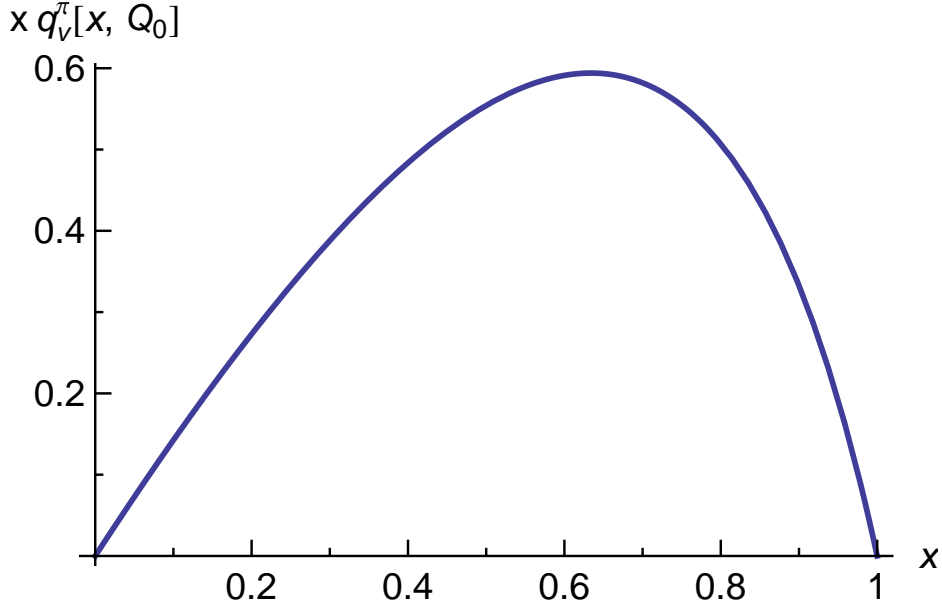}
\caption{Pion's valence-quark distribution function calculated in a Nambu--Jona-Lasinio model that is regularized by using a hard cutoff \cite{Shigetani:1993dx}.  Under one-loop evolution, a resolving scale of $Q_0=0.5\,$GeV optimizes agreement with data parametrizations. \label{NJLxqvpiHC}}
\end{figure}

The valence-quark distribution function obtained from Eq.\,(\ref{Eq:NJLxqvpiHC}), with $\Lambda_{\rm UV}=0.9\,$GeV and $m_Q=0.35\,$GeV, is depicted in Fig.\,\ref{NJLxqvpiHC}.  It differs markedly from Eq.\,(\ref{qvpiNJLTI}).  As will become clear, this could not have been otherwise because the behavior of $q_v^\pi(x)$ for $x\simeq 1$ depends on the precise form of the cutoff function.  If this were not the case, then the valence-quark distribution function would be ineffectual as a discriminator between models for the strong interaction and, furthermore, no behavior could be uniquely identified with a prediction of QCD.  With the hard cutoff
\begin{equation}
q_v^{\pi\,{\rm NJL}_{\rm HC}}(x;Q_0) \stackrel{x\simeq 1}{\propto} (1-x)\,.
\end{equation}
From the distribution function displayed in Fig.\,\ref{NJLxqvpiHC} one obtains
\begin{equation}
\langle x \rangle_{Q_0}^{\pi\,{\rm NJL}_{\rm HC}} =0.39\,,\;
\langle x^2 \rangle_{Q_0}^{\pi\,{\rm NJL}_{\rm HC}} =0.22\,,
\end{equation}
values which are significantly smaller than those obtained in the absence of a bound-state length-scale, Eq.\,(\ref{qvpiNJLTIn}).  It is apparent that in a model whose interaction acts at the resolving scale $Q_0$ to bound the relative momentum of the pion's constituents, the valence-quarks do not carry all the bound-state's momentum.  The remainder may be associated with dressed-gluons, which effect binding within the pion bound state and are invisible to the electromagnetic probe \cite{Hecht:2000xa}. (See also Sec.\,\ref{sec:GRV}.)

The kaon's valence-quark distribution functions were also calculated in \cite{Shigetani:1993dx}.  With a current-quark-mass ratio of $m_s/m_d \sim 25$, they obtained $m_S/m_D \approx 1.5$; and given this value for the ratio of constituent-quark-like masses, it is unsurprising that within the kaon the valence $s$-quark distribution is harder than that of the valence $u$-quark: the heavier valence $s$-quark must typically carry more of the bound-state's momentum.  It follows from momentum conservation that $u_v^K(x)/u_v^\pi(x)<1$ on the valence-quark domain.  This outcome is consistent with experiment \cite{Badier:1980}, as illustrated in Figs.\,\ref{kaon} and \ref{uKpiPCT}.

\subsubsection{Nucleon in the Nambu--Jona-Lasinio model}
\label{NJLnucleon}
%Covariant spectator model ... Kusaka ... http://www.slac.stanford.edu/spires/find/hep/www?eprint=hep-ph/9609277
When emphasizing dressed-quark degrees of freedom, it is natural to describe baryons using a Poincar\'e covariant Faddeev equation, aspects of which are described in App.\,\ref{Appendix:FE}.
Early attempts to compute the valence-quark distribution functions within this framework are described in \cite{Kusaka:1996vm,Mineo:1999eq}.  These studies retained only a scalar diquark correlation and the latter employed a particularly severe truncation of the quark-exchange kernel in Fig.\,\ref{faddeevfigure}; viz., the so-called static-approximation, in which all momentum dependence is ignored in the propagator of the exchanged quark.  This impacts significantly on the results.  The studies also employ a simple current.  In the former, that current is built only from Diagrams~1 and 2 in Fig.\,\ref{vertex}.  These diagrams are also the basis for the current in \cite{Mineo:1999eq} but therein, departing from the purely valence-quark setup, it is augmented by two diagrams argued to provide an estimate of the contribution from a pion cloud.

Unsurprisingly, there are many similarities between the nucleon study in \cite{Mineo:1999eq} and that of the pion in \cite{Bentz:1999gx}.  In particular, a translationally invariant regularization prescription is employed so that, in the absence of a pion cloud, the valence-quarks carry all the nucleon's momentum at the model's resolving scale, in analogy with Eq.\,(\ref{qvpiNJLTIn}).  Some of this momentum is transferred to the pions following their addition to the current.  The amount is related to the pion content of a dressed-quark, which is typically $\sim 10$\% \cite{Cloet:2008fw}.

As found in \cite{Kulagin:1995ia} and illustrated in Fig.\,\ref{Ku1995ia}, \cite{Mineo:1999eq} report that a pion cloud works to soften the valence-quark distribution and increase its support at low $x$.  Unlike \cite{Kulagin:1995ia}, however, \cite{Mineo:1999eq} still need a very low resolving scale; namely, $Q_0 = m_Q = 0.4\,$GeV, if agreement with the parametrizations of \cite{Martin:1994kn} is to be obtained under NLO evolution.  The NJL model, with its momentum-independent interaction,  must naturally produce hard distributions but the effect is compounded by the static-truncation used in \cite{Mineo:1999eq}.  As a result of this truncation, the nucleon's Faddeev amplitude is independent of the quark-diquark relative momentum, just as the pion's Bethe-Salpeter amplitude is independent of the quark-antiquark relative momentum in the NJL model formulated with a translationally invariant cutoff.

\begin{figure}[t]
\centerline{\includegraphics[clip,width=0.4\textwidth,angle=-90]{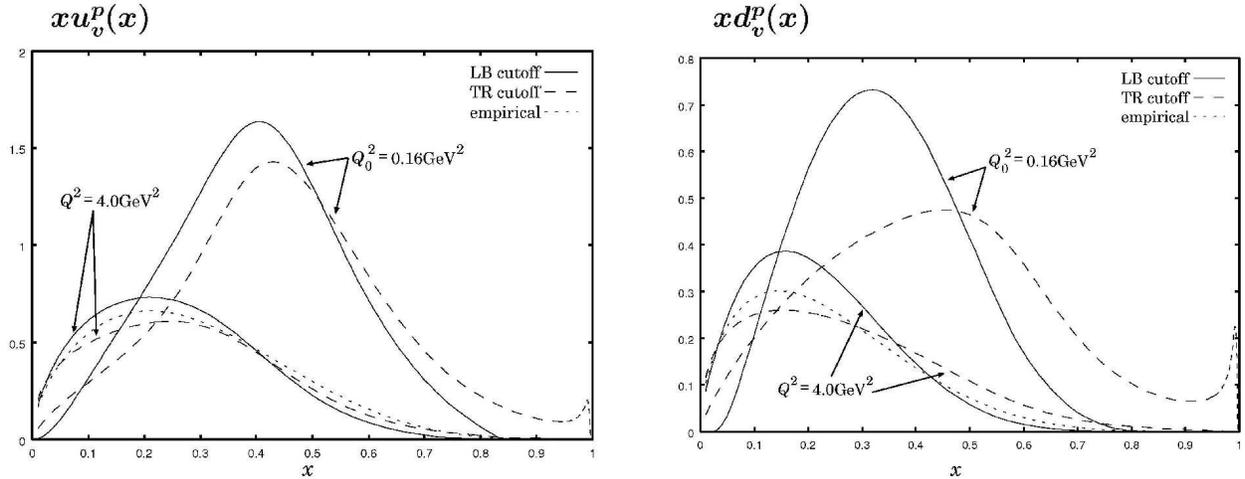}}
\caption{\label{9907043F}
Valence $u$-quark (left-panel) and $d$-quark (right-panel) distributions computed in a scalar-diquark picture of the nucleon based on a NJL model \protect\cite{Mineo:1999eq}.
The results exhibit a marked sensitivity to the manner in which the model is regularized. In this study the ``TR'' scheme is judged best.  Both schemes lead to unphysical behavior on $x\gtrsim 0.8$.
NLO evolution of the model results is performed from $Q_0=0.4\,$GeV.  The parametrization is that of \protect\cite{Martin:1994kn}.
[Figure adapted from \protect\cite{Mineo:1999eq}.]}
\end{figure}

As evident in Fig.\,\ref{9907043F}, the results are also sensitive to details of the regularization scheme.  The favored scheme is that adopted in \cite{Bentz:1999gx}, wherewith a cutoff is placed on $\ell_\perp$ and $q_v(x;Q_0)\neq 0$ on $0\leq x \leq 1$.  However, in this case $q_v(x;Q_0)$ exhibits a disconcerting, sharp and physically unacceptable increase at $x\simeq 0.9$, which must artificially be suppressed before the evolution equations can be used.  In practice, the distribution was forced to behave as $(1-x)^{10}$.  Owing to the behavior produced by both regularization schemes, this model cannot say anything meaningful about the distributions at $Q_0$ for $x\gtrsim 0.8$.  This is a serious limitation because evolution is an area-preserving operation that shifts strength from large- to small-$x$.  Hence, after evolution the behavior in this model of the distributions on the valence-quark domain is primarily determined by artifacts induced by the regularization.  This is neither a sound basis for prediction nor for comparison with parametrizations of data.

Some general features of the distributions are nonetheless physically reasonable.  For example, in this representation of the proton, the $d$-quark appears only as a constituent of a $[u,d]$ scalar diquark with mass $m_Q < m_{0^+}< 2 m_Q$.  Hence, the probability of finding a valence $d$-quark in the proton is obtained by convoluting two probabilities; viz., that of striking a $0^+$ diquark in the nucleon with that of striking a $d$-quark in the scalar diquark.  Naturally, therefore, the valence $d$-quark distribution peaks at smaller $x$ than the valence $u$-quark distribution and is softer for $0.6\lesssim x \lesssim 0.8$.

A realistic picture of the nucleon must include axial-vector diquark correlations, which generate significant attraction \cite{Hecht:2002ej}, and in \cite{Mineo:2002bg} they are added to the model just described.  The Faddeev equation is still solved in the static-truncation but the regularization procedure denoted by ``LB'' in Fig.\,\ref{9907043F} is adopted.  As apparent, it forces the distribution functions to vanish at $x\simeq 0.8$, which, while still unphysical, is less difficult to overlook than the sharp rise produced by the ``TR'' scheme.  The LB regularization procedure defines a model in which the distributions are extremely soft at $Q_0$.  Needless to say, such behavior is inconsistent with Eq.\,(\ref{protonvalenceconstraint}), which specifies the behavior of distribution functions on a domain that includes the neighborhood of $x=1$.
% My meaning is epsilon-neighborhood, which is the one I grew up with.

\begin{figure}[t]
\centerline{
\includegraphics[clip,width=0.55\textwidth,angle=-90]{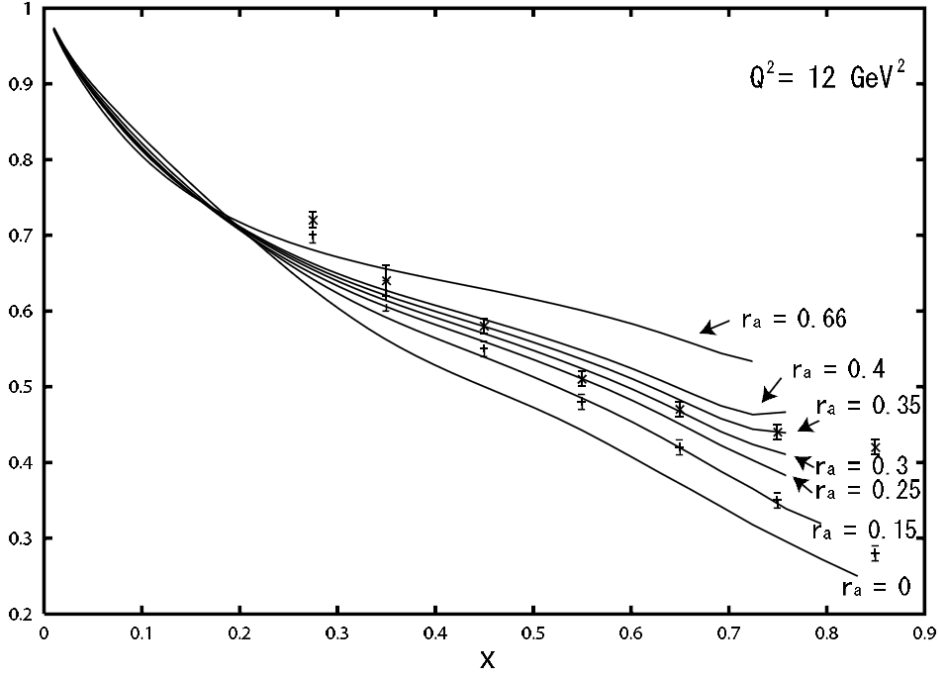}}
\caption{\label{0201082F}
$F_2^n(x)/F_2^p(x)$ computed in \protect\cite{Mineo:2002bg}.
$r_a$ measures the strength of axial-vector diquark correlations in the proton.  It is a free parameter.  Used in conjunction with $m_Q=0.4\,$GeV and $m_{0^+} =0.68\,$GeV, the model-preferred value is $r_a=0.25$, at which the axial-vector correlation contributes just 7\% to the proton's normalization.
$r_a=0$ reproduces the scalar-diquark only model, Fig.\,\protect\ref{9907043F}.  With $r_a=2/3$, the axial-vector diquark contributes 39\% to the proton's normalization.
NLO evolution of the model results is performed from $Q_0=0.4\,$GeV.
An explanation of the experimental points is provided in \protect\cite{Melnitchouk:1995fc}.
[Figure adapted from \protect\cite{Mineo:2002bg}.]}
\end{figure}

The model has other internal difficulties.  Notably, the axial-vector diquark is unbound and, owing to the absence of confinement, the $\Delta$ resonance cannot be formed without requiring the proton to possess what, within the model, is an unrealistically large axial-vector diquark component.  Notwithstanding these problems, one might hope for useful information about the ratio $F_2^n(x)/F_2^p(x)$ on the valence-quark domain.  Figure~\ref{0201082F} displays this ratio's sensitivity to the strength of the proton's axial-vector diquark component.  While the pointwise behavior in $x$ is unlikely to be accurate, we judge that the general trend is a reliable indication of one effect of axial-vector diquark correlations within the nucleon; viz., a $d$-quark in an axial-vector diquark can possess the same helicity as the proton target.

\begin{figure}[t]
\centerline{
\includegraphics[clip,width=0.60\textwidth,angle=-90]{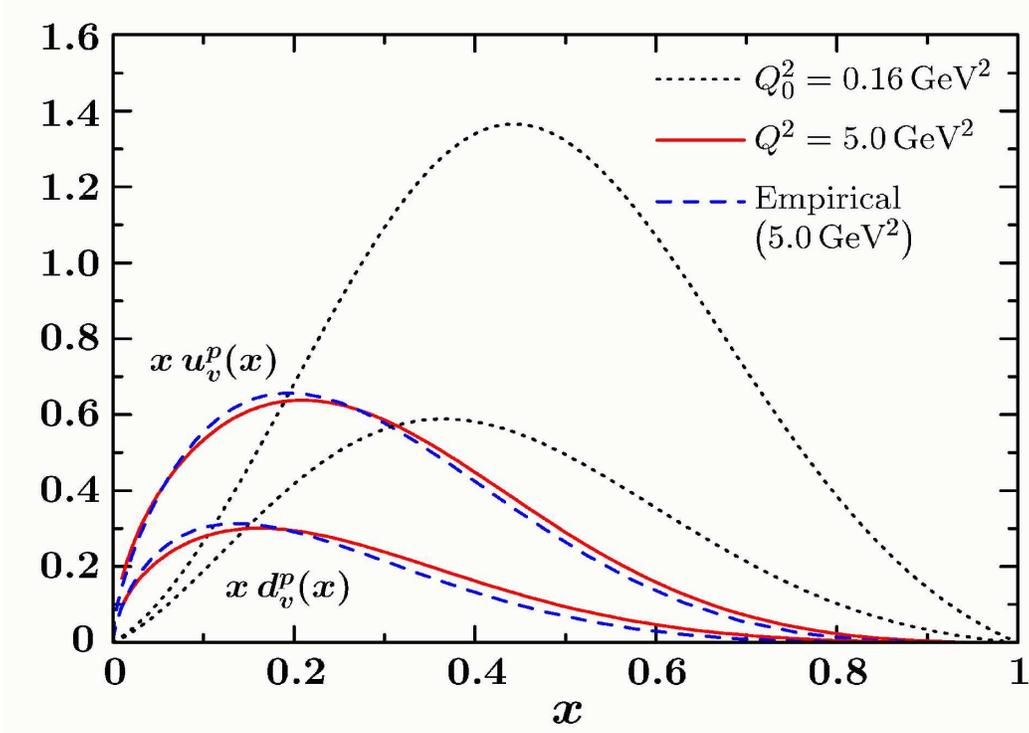}}
\caption{\label{504229F} (Color online)
Valence-quark distribution functions computed in \protect\cite{Cloet:2005pp}, which differs from \protect\cite{Mineo:1999eq} primarily in the choice of regularization prescription.  The marked sensitivity to regularization scheme in NJL-type models is evident through comparison with Fig.\,\ref{9907043F}.
NLO evolution of the model results is performed from $Q_0=0.4\,$GeV.  ``Empirical'' denotes the parametrization of \cite{Martin:2002dr}.
[Figure adapted from \protect\cite{Cloet:2005pp}.]}
\end{figure}

The analysis of \cite{Mineo:1999eq} is updated in \cite{Cloet:2005pp}, wherein the model is redefined through a proper-time regularization procedure.  In addition to the ultraviolet regulator, an infrared cutoff is also used in the reformulation.  That expedient serves to mock-up confinement \cite{Ebert:1996vx} and thereby enables an improved description of the $\Delta$ resonance.

In effecting a soft cutoff at $Q_0$, the proper-time prescription yields much improved behavior of the distribution functions on the valence-quark domain.  This is illustrated in Fig.\,\ref{504229F}, which was computed with: in GeV, $\Lambda_{\rm IR}=0.28$, $\Lambda_{\rm UV}=0.66$, $m_Q=0.4$, $m_{0^+}=0.65$ $m_{1^+}=1.2$; and diquark coupling strengths $r_{s}=0.5$, $r_a=0.08$.  NB.\ The splitting between the scalar and axial-vector diquarks is more than twice the usual size [cf.\, Eq.\,(\protect\ref{diquarkmass})] and the values of $r_s$, $r_a$ suggest that the axial-vector diquark contributes $<2$\% to the proton's normalization.

Both the $u$- and $d$-quark distributions in Fig.\,\ref{504229F} behave as \cite{icloetprivate:2009a}
\begin{equation}
\label{icloetxone}
q_v^{{\rm NJL}_{\rm PT}}(x;Q_0) \stackrel{x\simeq 1}{\sim} (1-x)^{\epsilon}, \; \epsilon\sim 1.5\,,
\end{equation}
a result which, however, still conflicts with the parton model constraint, Eq.\,(\ref{protonvalenceconstraint}): $(1-x)^3$.  On the other hand, for the flavor ratio one finds
\begin{equation}
\label{icloetdvuv}
\frac{d_v^{{\rm NJL}_{\rm PT}}(x;Q_0)}{u_v^{{\rm NJL}_{\rm PT}}(x;Q_0)} \stackrel{x\to 1}{=} 0.2\,,
\end{equation}
which corresponds to $F_2^n/F_2^p \to 0.43$ as $x\to 1$. [See Eq.\,(\ref{eq:npratio}) and cf.\ Fig.\,\ref{0201082F}.]  However, the agreement between Eq.\,(\ref{icloetdvuv}) and the leading-order vector-gluon prediction of \cite{Farrar:1975yb} is incidental.  Instead, the value reflects the role and strength of axial-vector diquarks in the proton.

\subsubsection{Pion in the Instanton liquid}
% -- only useful for pion
The instanton liquid model derives from the notion that topological soliton solutions of the classical Euclidean-space Yang-Mills equations might play a role in describing QCD's ground state \cite{Schafer:1996wv}.  This possibility is being actively discussed \cite{Lian:2006ky,Ilgenfritz:2008ia} and the model meanwhile continues to be used.

In the present context, the instanton liquid model may be viewed as corresponding to a type of NJL model in which the interaction is momentum-dependent.  With this feature it generates, e.g., a dressed-quark mass and pion Bethe-Salpeter amplitude that are momentum-dependent, and thereby overcomes a material weakness of the contact-interaction NJL model described above.  On the other hand, the momentum-dependence differs significantly from that of QCD: it is exponential, not power-law, and hence possesses too little support away from zero relative momentum, which is the domain most relevant to the valence-quark distribution.

In this form the instanton liquid model has been employed to calculate the pion's valence-quark distribution function \cite{Dorokhov:2000gu}.  As should be anticipated, the momentum dependence of the interaction leads to an $x$-dependent distribution at $Q_0=0.55\,$GeV, which vanishes at $x=0,1$, in contrast to that obtained with some regularizations of the contact-interaction NJL model; e.g., Eqs.\,(\ref{qvpiTS}) and (\ref{qvpiNJLTI}).

Following leading-order evolution from $Q_0=0.55\,$GeV to $Q=2\,$GeV, the model's result for the distribution function on $0.2\lesssim x \lesssim 0.9$ is uniformly larger than experiment (see Fig.\,\ref{xuvxpicall}), and the parametrization in \cite{Sutton:1992ay} which is based upon it.\footnote{Recall the critical reanalysis of the conclusions drawn from the E615 experiment, which is reviewed in Sec.\,\protect\ref{exptpionstructure}.}    The authors attribute this to their omission of gluon, sea and higher Fock state contributions at $Q_0$, which is plausible given the content of Fig.\,\ref{Ku1995iaB}.  If the curve is multiplied by 0.65, a factor representing a crude estimate of the renormalization attendant upon inclusion of the omitted effects, then the magnitude is in better agreement with the parametrization.  Following this rescaling, one finds that the large-$x$ behavior is softer than that of the parametrization.  Indeed, in this model
\begin{equation}
\label{dorokhovxone}
q_v^{\pi\,{\rm IL}}(x;Q_0) \stackrel{x\simeq 1}{\sim} (1-x)^{\epsilon}, \; \epsilon\sim 1.53-1.56\,.
\end{equation}
While this behavior also conflicts with the QCD parton model prediction, Eq.\,(\ref{pionpartonmodel}), it is a less striking disagreement than any other result described hitherto.

\subsection{Modern QCD perspective}
Today there are a few approaches to the calculation of parton distribution functions that are firmly founded in QCD.  There is the numerical simulation of lattice-regularized QCD, of course.  There are also frameworks which go beyond modeling by incorporating and expressing effects that are rigorously established features of QCD.  In the best cases these approaches are directly connected with QCD via a symmetry-preserving truncation so that quantitative calculations can be performed with readily quantifiable errors, which can be estimated \emph{a priori}.

\subsubsection{Generalized parton distributions}
\label{GPDtheory}
These quantities (GPDs) were considered in \cite{Dittes:1988xz,Dittes:1994} but have risen to prominence in hadron physics owing to their connection with deeply virtual Compton scattering \cite{Radyushkin:1996nd,Ji:1996nm} and hard meson production \cite{Radyushkin:1996ru}.  Unlike the usual parton distribution functions, GPDs are non-diagonal matrix elements and hence do not have a simple probability interpretation.  Instead, GPDs describe interference between amplitudes that represent different partonic configurations within a hadron and provide a means by which to chart correlations between different momentum states of a given parton.  They are of interest here because the usual parton distribution functions can be expressed in terms of the GPDs.\footnote{NB.\ For our purposes, we need not consider transversity distributions.}
It follows that either: (1) valence-quark distribution functions can be used to constrain the parametrization of GPDs; or (2) complete knowledge of GPDs fixes the valence-quark distributions.  The former link is used in present-day practice and that will not soon change.  Herein we nevertheless elucidate connections between GPDs and PDFs.

\begin{figure}[t]
\includegraphics[clip,width=0.20\textwidth,angle=-90]{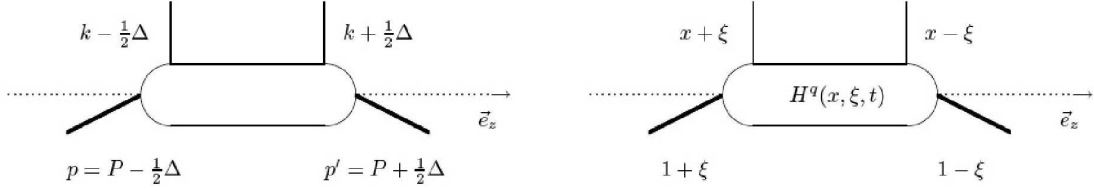}
\caption{``Handbag diagram'' contribution to the virtual Compton scattering process: $\gamma^\ast(q) T(p) \to \gamma(q^\prime) T^\prime (p^\prime)$.  \emph{Left panel} -- Kinematic variables in a symmetric reference frame. \emph{Right panel} -- Light-front representation of $H^q$ in Eq.\,(\protect\ref{defHq}).
[Figure adapted from \protect\cite{Boffi:2007yc}.] \label{DVCSpic}}
\end{figure}

In Fig.\,\ref{DVCSpic} we depict a handbag diagram contribution to a deeply virtual Compton scattering process, in which an high-energy photon with large spacelike virtuality, $Q^2=-q^2$, impacts upon an hadronic target, and produces a real photon and an hadronic final state with invariant mass $p^2\ll Q^2$.  The amplitude depends on the Bjorken variable $x_B = Q^2/(2 p\cdot q)$, defined in Eq.\,(\ref{eq:xbj}), and the momentum transfer $t=(p^\prime - p)^2 = \Delta^2$.  It is assumed that $t\ll Q^2$.

For a spin-$\sfrac{1}{2}$ hadron, $h$, with initial and final helicities $\lambda$ and $\lambda^\prime$, respectively, one can define eight GPDs.  For our purposes it is enough to consider only two of them explicitly; namely, the unpolarized distributions $H_h^q$ and $E_h^q$:
\begin{eqnarray}
\nonumber && \int\frac{dz^-}{4\pi}\, \left. {\rm e}^{\sfrac{i}{2} x z^- P^+}\,
\langle h(p^\prime ,\lambda^\prime) | \bar q(-\sfrac{z}{2}) \gamma^+ q(\sfrac{z}{2}) | h(p ,\lambda)\rangle \right|_{z^+=0,z_\perp=0} \\
&=& \frac{1}{2 P^+}
\bar u_h(p^\prime,\lambda^\prime)\,
[H_h^q(x,\xi,t)\gamma^+ + E_h^q(x,\xi,t)\frac{1}{2 M} \sigma^{+\alpha} \Delta_\alpha]
u_h(p,\lambda), \label{defHq}
\end{eqnarray}
where $x=k^+/P^+$ is the average quark longitudinal-momentum-fraction and \mbox{$\xi= (p^+ - p^{\prime +})/(p^+ + p^{\prime +})$} is the so-called skewness.  The GPDs also depend on the resolving scale $Q_0$.

These two GPDs are relevant herein because in the forward limit; i.e., $p^\prime = p$, one has
\begin{equation}
\label{pdf2gpd}
H_h^q(x=x_B,\xi=0,t=0; Q_0)
= \left\{\begin{array}{cc}
q_h(x_B;Q_0), & x_B>0\\
-\bar q_h(-x_B;Q_0), & x_B<0\\
\end{array}\right. ,
\end{equation}
from which if follows, e.g., that
\begin{equation}
\label{qvh2Hhq}
\int_{-1}^1 dx \, x^{2n} \, H_h^q(x,0,0; Q_0) = \int_0^1 dx\, x^{2n}\, q_v^h(x;Q_0) = \langle x^{2n}\rangle_{Q_0}^h \,.
\end{equation}
Furthermore,
\begin{equation}
\label{HEF12}
\begin{array}{c}
\displaystyle
\int_{-1}^1 dx \, H_h^q(x,\xi,t; Q_0) = F_{1h}^q(t) \;,\\
\displaystyle
\rule{0em}{4ex}\int_{-1}^1 dx \, E_h^q(x,\xi,t; Q_0) = F_{2h}^q(t),
\end{array}
\end{equation}
where, e.g., $F_{1h}^q(t)$ is the contribution to the hadron's Dirac form factor owing to quarks of flavor $q$.  The standard normalization is such that
\begin{eqnarray}
F_{1p}^u(t=0)&=& 2\,,\\
F_{1p}^d(t=0)&=&1\,,\; \\
F_{2p}^u(t=0)&=&\kappa_p^u=2 \kappa_p + \kappa_n= 1.673\,,\\
F_{2p}^d(t=0)&=&\kappa_p^d= \kappa_p + 2 \kappa_n= -2.033\,.
\end{eqnarray}
It is noteworthy that the right-hand-sides in Eqs.\,(\ref{HEF12}) are independent of $\xi$ because integrating over $x$ eliminates all information about the particular light-front direction with respect to which $\xi$ is defined; viz., owing to Lorentz invariance.

Poincar\'e invariance entails that there is only one generalized quark-parton distribution  for the pion \cite{Diehl:2003ny}:
\begin{equation}
H_\pi^q(x,\xi,t; Q_0)
= \int\frac{dz^-}{4\pi}\, \left. {\rm e}^{\sfrac{i}{2} x z^- P^+}\,
\langle \pi^+(p^\prime) | \bar q(-\sfrac{z}{2}) \gamma^+ q(\sfrac{z}{2}) | \pi^+(p)\rangle \right|_{z^+=0,z_\perp=0}.
\end{equation}
Naturally, coupled with the fact that the pion's polarized GPD is identically zero; i.e., $\tilde H_\pi \equiv 0$,\footnote{With the replacement $\gamma^+ \to \gamma^+\gamma_5$ in Eq.\,(\protect\ref{defHq}), one arrives at the definition of $\tilde H_h$.} it follows that $\Delta q_\pi(x)\equiv 0$; i.e., the pion's spin-dependent parton distribution function vanishes. When isospin is a good symmetry, one has
\begin{eqnarray}
&& H_{\pi^+}^{u+d} := H_{\pi^+}^u + H_{\pi^+}^d = H_{\pi^-}^{u+d} = H_{\pi^0}^{u+d},\\
&& H_{\pi^+}^{u-d} := H_{\pi^+}^u - H_{\pi^+}^d = - H_{\pi^-}^{u-d}\,, \\
&& H_{\pi^0}^{u-d} \equiv 0\,.
\end{eqnarray}

The analogues of Eqs.\,(\ref{pdf2gpd}) and (\ref{qvh2Hhq}) are correct, and, moreover,
\begin{equation}
\int_{-1}^1 dx \, H_\pi^q(x,\xi,t; Q_0) = F_{\pi}^q(t)\,,\;
F_{\pi}^u(t)-F_{\pi}^d(t)= 2 F_\pi(t)\,.
\end{equation}

\subsubsection{Lattice QCD}
\label{sec:latticeeQCD}
An operator expression for the pion's quark distribution function is given in Eq.\,(\ref{operatordefpion}).  We have seen in Sec.\,\ref{GPDtheory} that it is generally true; namely, in a spinless hadron ``h'' with total momentum $k$, the distribution function is given by
\begin{equation}
\label{operatordefh}
q^h(x;Q_0) = \frac{1}{4\pi} \int_{-\infty}^\infty d\xi^- {\rm e}^{\sfrac{i}{2} x k^+ \xi^-} \langle h(k) |\, \bar \psi_q(0) \gamma^+ \psi_q(\xi^-) | h(k) \rangle .
\end{equation}
This expression is usually understood as representing the distribution in light-cone gauge; i.e., $A_a^{+}=0$, where $A_a^\mu$ is the gluon field.  Equation\,(\ref{operatordefh}) can readily be generalized to bound-states with spin.

It is straightforward to modify Eq.\,(\ref{operatordefh}) so that the expectation value is gauge invariant and yet unchanged in light-cone gauge; viz., one introduces a path-ordered exponential (Wilson line)
\begin{equation}
{\cal E}(\xi^-) := \mbox{\textbf{P}} \exp\left\{ \sfrac{i}{2} g \int_{\xi^-}^0 dz^- \, \frac{\lambda^a}{2} A_a^+(0,z^-,\vec{0}_\perp)  \right\}
\end{equation}
as follows
\begin{equation}
\label{operatordefhcalE}
q^h(x;{\cal E};Q_0) := \frac{1}{4\pi} \int_{-\infty}^\infty d\xi^- {\rm e}^{\sfrac{i}{2} x k^+ \xi^-} \langle h(k) |\, \bar q(0) \gamma^+ {\cal E}(\xi^-) q(\xi^-) | h(k) \rangle .
\end{equation}
With Eq.\,(\ref{operatordefhcalE}) one has a gauge-invariant expectation value of a bilocal operator evaluated along a light-like line.  It is mathematically precise and provides the pointwise behavior of the distribution function.  However, it cannot be evaluated using the numerical approach of lattice-regularized QCD.

Lattice methods can, however, be used to evaluate expectation values of local operators.  Consider therefore
\begin{equation}
\langle x^n \rangle_{Q_0}^h := \int_0^1 dx \, x^n\, [ q^h(x;{\cal E};Q_0) - (-1)^n \bar q^h(x;{\cal E};Q_0)]\,.
\end{equation}
It is notable that
\begin{equation}
\label{moment0}
\langle x^0 \rangle_{Q_0}^h = \int_0^1 dx \, q_v^h(x;{\cal E};Q_0) = N^h_q\,;
\end{equation}
namely, the number of valence-quarks of type $q$ in hadron $h$.  Furthermore, in general
\begin{eqnarray}
\label{x2n}
\langle x^{2n} \rangle_{Q_0}^h &=&
\int_0^1 dx \, x^{2n} \, [q^h(x;{\cal E};Q_0) - \bar q^h(x;{\cal E};Q_0)] =
\int_0^1 dx \, x^{2n} \, q_v^h(x;{\cal E};Q_0) \\
\label{x2np1}
\langle x^{2n+1} \rangle_{Q_0}^h &=& \int_0^1 dx \, x^{2n+1} \, [q^h(x;{\cal E};Q_0) + \bar q^h(x;{\cal E};Q_0)]\,.
\end{eqnarray}
It is apparent from these two results that the even moments express properties of a nonsinglet distribution whereas the odd moments probe the singlet distribution.  Therefore they are independent and do not mix under evolution.  Importantly, unlike the singlet distribution, a nonsinglet distribution does not receive contributions from disconnected diagrams.

Introducing the Wilson line does not alter the distribution function's domain of support and, moreover, one still has $q^h(x) = - \bar q^h(-x)$.  Hence,
\begin{equation}
\langle x^n \rangle_{Q_0}^h := \int_{-1}^1 dx \, x^n\, q^h(x;{\cal E};Q_0)
= \int_{-\infty}^\infty dx \, x^n\, q^h(x;{\cal E};Q_0) \,.
\end{equation}
Now, upon insertion of Eq.\,(\ref{operatordefhcalE}), one obtains
\begin{equation}
\label{xnLC}
\langle x^n \rangle_{Q_0}^h
= (k^+)^{-n-1}
\langle h(k) |\, \bar \psi_q(0) \gamma^+ [i D^+(0)]^n  \psi_q(0) | h(k) \rangle \,,
\end{equation}
where
\begin{equation}
D_\mu(z) = \frac{\partial }{\partial z_\mu} - i g \frac{\lambda^a}{2} A_a^\mu(z)
\end{equation}
is QCD's covariant derivative.  (NB.\ A connection with Eq.\,(\ref{qvHklx}) is readily established.)  The moment on the left of Eq.\,(\ref{xnLC}) is Poincar\'e invariant but it only has a probabilistic interpretation in the infinite momentum frame.  Owing to Poincar\'e invariance, one may equally write
\begin{equation}
\label{latticemoments}
\{k_{\mu_0} k_{\mu_1} \ldots k_{\mu_n} \}\langle x^n \rangle_{Q_0}^h  =
\langle h(k) |\, \bar \psi_q(0) \{\gamma^{\mu_0} i D^{\mu_1}(0) \ldots i D^{\mu_n}(0) \} \psi_q(0) | h(k) \rangle \,,
\end{equation}
where the parenthesis indicate a symmetrization of indices and the subtraction of traces; e.g., $\{k_{\mu_0} k_{\mu_1}\} = k_{\mu_0} k_{\mu_1} - \delta_{\mu_0 \mu_1} k^2$.  With Eq.\,(\ref{latticemoments}) one has an expression for the moments of quark distribution functions that can be evaluated using modern lattice-QCD methods \cite{Soper:1996sn}.

Computations are readily performed for low-order moments; i.e., $n=0,1,2,3$.  Of course, owing to Eq.\,(\ref{moment0}), there is no information in $\langle x^0 \rangle$: it merely provides a check on the numerical procedure.  One must also bear in mind that discretized spacetime does not possess the full rotational symmetries of the Euclidean continuum.  Hence, for $n\geq 4$ it is impossible to define lattice-operators that are precisely equivalent to those of Eq.\,(\ref{latticemoments}): the lattice analogues cannot transform irreducibly and computations therefore require a numerical determination of the coefficients that describe mixing with operators of lower dimension.

Suppose therefore that one focuses on the $n\leq 3$ moments; namely, those most directly accessible within the lattice framework.  A typical lattice-QCD result for any one of these moments will contain, in addition to statistical errors, systematic errors from the following sources \cite{Guagnelli:2004ga}: the so-called ``quenched approximation'' -- which is not an approximation, but a truncation, that is still common today but will become less so in future; nonperturbative renormalization -- the method used to obtain a renormalization group invariant moment from which continuum results can be inferred; discretization -- extrapolation to zero lattice spacing; finite volume -- extrapolation to infinite lattice size; and chiral extrapolation -- inferring the value of a moment at the real-world current-quark masses from that computed at large, unphysical values, which is discussed, e.g., in \cite{Detmold:2003tm}.  The extent to which these errors have been accounted for and the nature of the procedure employed for each type of error must be considered when comparing the value of a moment obtained from a lattice simulation with its empirical or phenomenologically determined value.

\begin{table}[t]
\begin{center}
\begin{tabular}{|l|c|c|c|c|}\hline
$\langle x^n \rangle_{Q_0}^\pi$ & $Q_0\,$(GeV) & $ n=1$ & $n=2$ & $n=3$ \\\hline
Par.92 & 2.0 & ~$0.24\;\; \pm  0.01 \;\;$~ & ~$0.10 \;\; \pm  0.01\;\;$~ & ~$ 0.058 \pm 0.004$~ \\\hline
Emp.05 & 5.2 & ~$0.217\pm 0.011$~ &  ~$0.087\pm 0.005$~ &  ~$ 0.045\pm 0.003$~\\\hline
Lat.\,07 & 2.0 & ~$0.271 \pm 0.010$~ & ~$0.128 \pm 0.008$~ & ~$0.074 \pm 0.010$~ \\\hline
Lat.\,97$^{\rm Q}$ & 2.4 & ~$0.273\pm 0.012$~ & ~$0.107 \pm 0.035$~ & ~$0.048 \pm 0.020$~ \\\hline
Lat.\,03$^{\rm Q}$ & 2.4 & ~$0.24\;\; \pm 0.02\;\;$~ & ~$0.09\;\; \pm 0.03\;\;$~ & ~$0.043 \pm 0.015$~ \\\hline
DSE\,01 & 2.0 & ~$0.24$\rule{4.1em}{0ex}~ & ~$0.098$\rule{3.5em}{0ex}~ & ~$0.049$\rule{3.5em}{0ex}~ \\\hline\hline
$\langle x^n \rangle_{Q_0}^\rho$ & $Q_0\,$(GeV) & $ n=1$ & $n=2$ & $n=3$ \\\hline
Lat.\,97$^{\rm Q}$ & 2.4 & ~$0.334\pm 0.021$~ & ~$0.174 \pm 0.047$~ & ~$0.066 \pm 0.039$~ \\\hline
\end{tabular}
\caption{\label{pionmoments} Lowest nontrivial moments of the pion's valence $u=d$-quark distribution function.  The rows report:
``Par.92'' -- Moments at $Q_0=2\,$GeV determined from the fits in \protect\cite{Sutton:1991ay};
and ``Emp.05'' -- Empirical result determined in \protect\cite{Wijesooriya:2005ir} at $Q_0=5.2\,$GeV.
%-- right now the following is really SMRS
%``Emp.05$^\ast$'' -- Moments obtained from the fitted distribution function in \protect\cite{Wijesooriya:2005ir} after one-loop evolution to $Q_0=2.0\,$GeV;
%
[NB.\ After evolution, these results are consistent with those in ``Par.92''.]
``Lat.\,07'' -- Lattice-QCD computation reported in \protect\cite{Brommel:2006zz} with nonperturbative evolution to $Q_0=2.0\,$GeV (statistical and renormalisation error added in quadrature);
``Lat.\,97$^{\rm Q}$'' -- Quenched computation reported in \protect\cite{Best:1997qp}, with renormalization scale $Q_0=2.4\,$GeV fixed via one-loop, chiral-limit lattice perturbation theory (statistical error only); and
``Lat.\,03'' -- Reanalysis of the results in \protect\cite{Best:1997qp} using a different extrapolation to physical current-quark masses \protect\cite{Detmold:2003tm}.
The row labeled ``DSE\,01'' is the result obtained using the Dyson-Schwinger equation framework in \protect\cite{Hecht:2000xa}.
}
\end{center}
\end{table}

Lattice results are available for the first three moments of the pion distribution function; e.g., \cite{Brommel:2006zz,Best:1997qp}, with the results reported in Table\,\ref{pionmoments}.
The more recent study \cite{Brommel:2006zz} employed $O(a)$-improved Clover-Wilson fermions with two flavors of dynamical quarks and pion masses $430$, $600$, $800\,$MeV.  A linear extrapolation in $m_\pi^2$ was used to infer a result at the physical pion mass.  The authors estimated rudimentarily that finite-size effects are of the order of 10\% at the smallest pion mass used.  Since disconnected diagrams were neglected in calculating all the moments and the simulation pion masses are large, it is reasonable to identify the results with moments of the pion's valence-quark distribution.  This interpretation turns lattice artifacts to advantage in order to circumvent practically a problem of principle; viz., it is plain from Eqs.\,(\ref{x2n}), (\ref{x2np1}) that odd moments of the valence-quark distribution are not readily accessible via the vector current operator.  The valence-quark distribution is directly accessible though moments of the more complicated $V-A$ current operator.
%
% connection between q_v and phi_LC: Brodsky http://www.scielo.br/scielo.php?script=sci_arttext&pid=S0103-97332004000200003
%
% http://www.slac.stanford.edu/spires/find/hep/www?eprint=nucl-ex/0509012
% Reimer et al at 5.2 GeV 1 0.217±0.011  2 0.087±0.005  3 0.045±0.003
%
%http://www.slac.stanford.edu/spires/find/hep/www?j=POSCI,LAT2007,140
% pion n-1,2,3 2-flavours m_pi .ge. 340
% Clover Wilson Nf=2
% The preliminary numbers for the moments of the quark PDF in MS at m = 2GeV are 1 = 0.271(2)(10), 2 0.128(6)(5) and 3 0.074(9)(4), with quoted uncertainties only from statistical and renormalisation errors.  Nonperturbative renormalisation

%http://www.slac.stanford.edu/spires/find/hep/www?eprint=hep-lat/9703014
% Best et al. Quenched, Wilson, heavy-pions with linear extrapolation in hopping parameter to chiral limit, errors only statistical
The earlier study \cite{Best:1997qp} used Wilson fermions, in a quenched simulation, with current-quark masses that correspond to $m_\pi \approx 480$, $650$, $780\,$MeV.  Values for the moments at a physical light-quark current-mass were obtained through linear extrapolation in $m_\pi^2$.  This could plausibly therein lead to overestimation of the first three moments by $10$-$20$\% \cite{Detmold:2003tm} (see Table\,\ref{pionmoments}), and therefore also in \cite{Brommel:2006zz}.

It is curious and disturbing that there is little material difference between the results in Rows~3-5: all agree within quoted errors, even though they were obtained over a ten-year time-span in which lattice methods improved greatly.  In our view, for this reason and because the pion is so important to strong interaction physics, the evaluation of the pion's moments is a problem worth renewed attention.

% a_n polarisation averaged, d_n polarised, r_n operators involving gamma_5
Low moments of the $\rho$-meson's valence-quark distribution functions are also inferred in \cite{Best:1997qp}.  There are three different structure functions for the spin-1 $\rho$:
unpolarized, with moments denoted by $a_n$;
polarized, with moments denoted by $r_n$;
and spin-projection, with moments denoted by $d_n$.  The spin-projection structure function is absent for a spin-$\sfrac{1}{2}$ hadron.  In the infinite momentum frame it measures the difference between quark distributions in $J_z=1$ and $J_z=0$ targets.
Herein, however, we are concerned with unpolarized distributions and so only $a_{1,2,3}$ are listed in Table\,\ref{pionmoments}.  With the parameters of this lattice simulation, the $\pi$ and $\rho$ moments differ noticeably and the first moments do not agree within statistical errors.  It appears to us, therefore, that it is obtuse to claim that the results in \cite{Best:1997qp} support an assumption that $F_1^\rho(x) \approx F_1^\pi(x)$, as was done therein.

\begin{table}[t]
\begin{center}
\begin{tabular}{|l|c|c|c|}\hline
$\langle x^n \rangle_{Q_0}^{u-d}$ & $ n=1$ & $n=2$ & $n=3$ \\\hline
Emp.\,90 & ~$0.16\;\;\;\;\;\;\;\;\;\;\;\;\;\;$~ &  ~$0.054\;\;\;\;\;\;\;\;\;\;\;\;$~ &  ~$ 0.023\;\;\;\;\;\;\;\;\;\;\;\;\;$~\\\hline
Emp.\,02 & $0.157 \pm 0.009$ & & \\\hline
Lat.\,02$^{\rm Q}$ & ~$0.251 \pm 0.018$~ & ~$0.098 \pm 0.068$~ & ~$0.028 \pm 0.049$~ \\\hline
Lat.\,02 & ~$0.269 \pm 0.023$~ & ~$0.145 \pm 0.069$~ & ~$0.078 \pm 0.041$~ \\\hline
Lat.\,04$^{\rm Q}$ & ~$0.245 \pm 0.009$~ & ~$0.083 \pm 0.017$~ & ~$0.059 \pm 0.018$~ \\\hline
Lat.\,08 & ~$0.157\pm 0.010$~ &  &  \\\hline
\end{tabular}
\caption{\label{nucleonmoments} Lowest nontrivial moments of the isovector nucleon distribution function, all evolved to $Q_0=2\,$GeV, albeit using different methods.  Whilst they are all moments of nonsinglet distributions, in principle only $n=2$ is connected with a difference of valence-quark distributions.
The rows report:
``Emp.90'' -- Empirical result determined from the parametrizations of $x (u_v(x)+d_v(x))$ and $x d_v(x)$ in \protect\cite{Kwiecinski:1990ru}, which were fitted to then-extant data via next-to-leading-order evolution from a starting scale of $Q_0=2.0\,$GeV
[these values are little changed in more modern parametrizations, e.g., using MRST\,2002, available through the U.\,Durham data base \cite{Durham:pdf}, $\langle x\rangle^{u_v-d_v}_{2\,{\rm GeV}}=0.157 \pm 0.009$, as indicated in the row labeled ``Emp.02''];
``Lat.\,02$^{\rm Q}$'' and ``Lat.\,02''-- Lattice-QCD computations reported in \protect\cite{Dolgov:2002zm}, the former in quenched-QCD, both with perturbative evolution to $Q_0$ and linear chiral extrapolation (statistical error only);
``Lat.\,04$^{\rm Q}$'' -- Quenched-QCD study described in \protect\cite{Gockeler:2004wp}, with renormalization scale, $Q_0$, fixed via nonperturbative renormalization and linear chiral extrapolation (error is statistical combined with estimates of systematic errors arising from renormalization, operator choice, and continuum and chiral extrapolations);
``Lat.\,08'' -- Analysis reported in \protect\cite{Hagler:2007xi}, with renormalization scale, $Q_0$, fixed via nonperturbatively-improved one-loop renormalization and extrapolation via covariant baryon chiral perturbation theory \protect\cite{Dorati:2007bk} [before extrapolation: $\langle x \rangle^{u-d} \simeq 0.205\pm 0.015$, statistical error only].
}
\end{center}
\end{table}

%http://www.slac.stanford.edu/spires/find/hep/www?eprint=hep-lat/0103006
More attention has focused on the nucleon, with the first computations in \cite{Martinelli:1988rr} and a brief contemporary overview in \cite{Zanotti:2008zm}, but it remains true that disconnected diagrams are neglected in almost all extant studies.  It follows that only results for isovector quantities are directly comparable with those determined empirically; i.e.,
\begin{equation}
\langle x^n \rangle^{u-d}_{Q_0} := \langle x^n \rangle^{p-n}_{Q_0}.
\end{equation}
For the reason noted above, calculations are restricted to at-most the first three nontrivial moments.  We list some results in Table\,\ref{nucleonmoments}.

We present two sets of determinations from \cite{Dolgov:2002zm}.  Both were obtained with Wilson fermions.  However, one set was computed in quenched-QCD; whilst the other was calculated using two degenerate flavors of dynamical sea quarks, with four current-quark mass values that provide for $m_\pi/m_\rho \in [0.69,0.83]$.  Perturbative renormalization was employed to quote results at $Q_0=2\,$GeV.  The studies possess significant but unquantified finite-size and -volume errors, and a linear extrapolation in $m_\pi^2$ was used to infer results at physical current-quark masses.

The fifth row of results in Table\,\ref{nucleonmoments} is compiled from \cite{Gockeler:2004wp}, a quenched-QCD study that employed an $O(a)$-improved Wilson action and fully nonperturbative renormalization to reach $Q_0=2.0\,$GeV.  The authors argued that finite-volume errors were unimportant, and were attentive in estimating systematic errors arising from from nonperturbative renormalization, operator choice, and continuum and linear chiral extrapolations.

\begin{figure}[t]
\includegraphics[clip,width=0.75\textwidth]{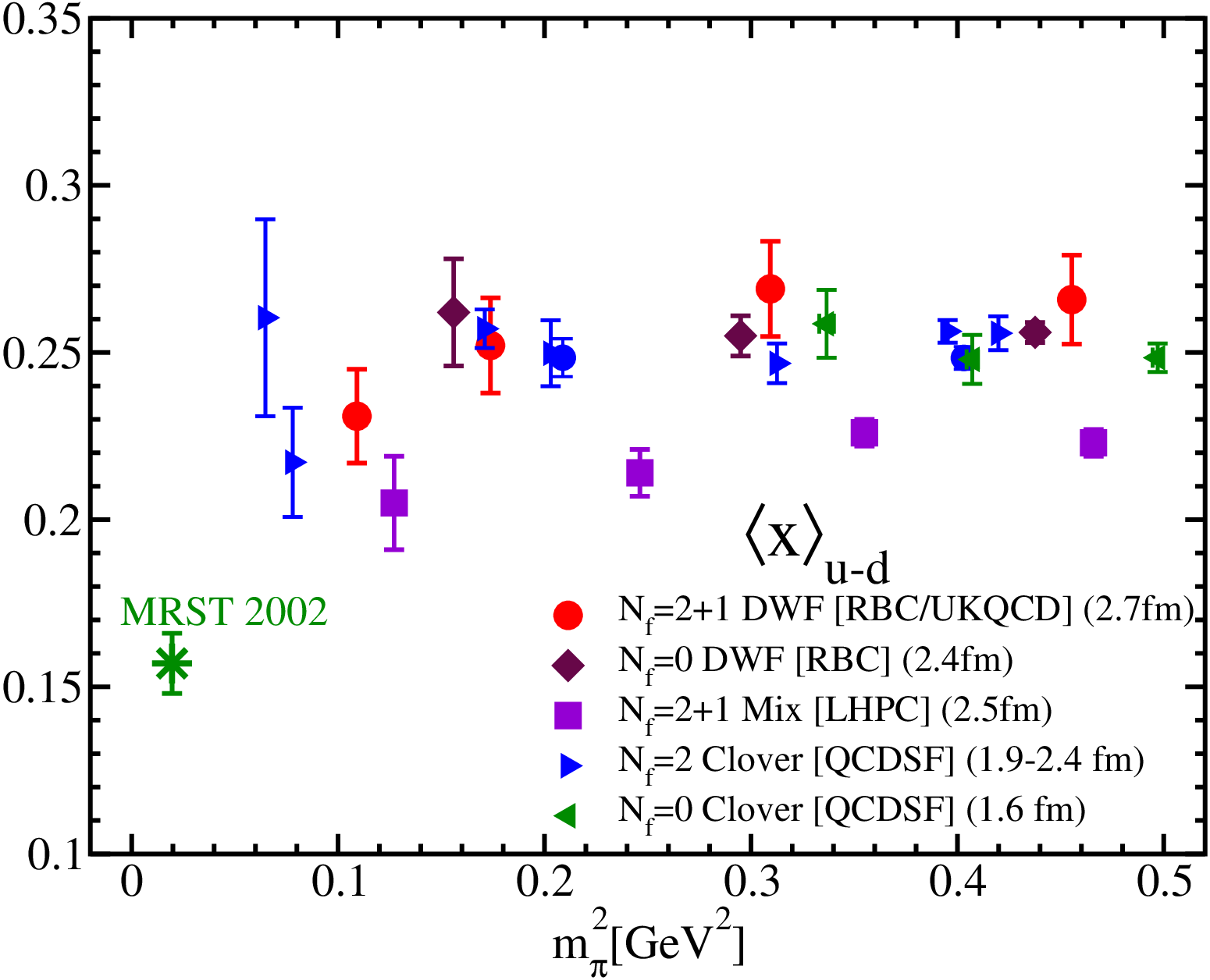}
%\vspace*{40ex}
\caption{(Color online) Current-quark-mass-dependence of $\langle x \rangle^{u-d}_{Q_0}$, $Q_0=2\,$GeV, computed directly through numerical simulations of lattice-regularized QCD:
Circles -- \protect\cite{Ohta:2008kd};
Diamonds -- \protect\cite{Orginos:2005uy};
Boxes -- \protect\cite{Hagler:2007xi};
Right-triangles -- \protect\cite{Brommel:2008tc};
and Left-triangles -- \protect\cite{Gockeler:2004wp}.
The empirical value is marked by an asterisk.
The Clover and Domain Wall Fermion results are in excellent agreement, except at the lightest pion mass, a discrepancy which can likely be attributed to finite size effects.
The mixed-action results appear to differ from the others by a normalisation factor, which is perhaps understandable given that \protect\cite{Hagler:2007xi} does not use the fully nonperturbative renormalization procedure employed in the other studies.
[Figure adapted from \protect\cite{Zanotti:2008zm}.]
\label{JAMES}}
\end{figure}

We also list the estimate made in \cite{Hagler:2007xi} for the $n=1$ moment, which was the only moment considered therein.  This analysis employed a mixed-action approach; namely, domain wall valence-quarks but $2+1$ flavors of staggered sea-quarks, with light-quark masses corresponding to $m_\pi = 350\,$MeV.  A nonperturbatively-improved perturbative renormalization scheme was used to quote a result at $Q_0=2\,$GeV.  It was argued that finite-volume errors are negligible with respect to statistical errors, whilst errors associated with nonzero lattice spacing ($a=0.124\,$fm) were not discussed.  In order to quote a result at the physical pion mass, covariant baryon chiral perturbation theory was employed \cite{Dorati:2007bk}.  As anticipated in \cite{Detmold:2001jb}, an extrapolation nonlinear in $m_\pi^2$ can have an enormous impact: in this instance, a 25\% reduction in the value of the moment.

A consideration of Fig.\,\ref{JAMES} and the results listed in Table\,\ref{nucleonmoments} demonstrates that agreement between the empirical value of $\langle x \rangle^{u-d}$ and that obtained using lattice-QCD is only possible if there is significant nonlinearity in the $m_\pi^2$-dependence of $\langle x \rangle^{u-d}$ on $m_\pi^2<0.1\,$GeV$^2$.  On the other hand, given that a chiral expansion of physical quantities is invalid for $m_\pi^2 \gtrsim 0.2\,$GeV$^2$ \cite{Chang:2006bm} and reliable lattice results are restricted to this domain, it is unsurprising that extant results from lattice-QCD do not exhibit nonlinearity.

It is evident from the discourse in this subsection that contemporary lattice-QCD provides little aid in understanding the vast amount of DIS data, in general, and the valence-quark distributions, in particular.  Practitioners express hope that this will change.  However, it should be noted that low moments of parton distribution functions are only very weakly sensitive to the pointwise behavior of $q^h(x;Q_0)$ on $x\gtrsim 0.4$.  For example, consider two valence-quark distributions with large-$x$ behavior $(1-x)^{1.15}$ \cite{Sutton:1991ay} and $(1-x)^{1.80}$ \cite{Hecht:2000xa}.  The exponents differ by 60\% but the $n=3$ moments, which are the highest accessible in modern lattice-QCD calculations, are indistinguishable within experimental error.  One must compute the $n=5$ moments before the difference exceeds 20\%.  Plainly, pointwise calculations of the distribution functions are vital in order to probe the valence region.

\subsubsection{Dyson-Schwinger equations}
\label{sec:DSE}
As elucidated in \cite{Roberts:1994dr}, the Dyson-Schwinger equations (DSEs) are a system of coupled integral equations (DSEs) that have long been used in nuclear and particle physics.  The DSEs provide a nonperturbative approach to QCD in the continuum and are particularly well suited to the study of QCD because of asymptotic freedom, which entails that model-dependence can always, in principle, be restricted to infrared momenta; viz., $p\lesssim 1\,$GeV.  Today, some elements of the approach are well-constrained even on that domain owing to positive feedback between DSE and lattice-QCD studies.  Applications of DSEs to hadron physics are reviewed in \cite{Roberts:2000aa,Alkofer:2000wg,Maris:2003vk,Fischer:2006ub,Roberts:2007jh}.

The physics of hadrons is ruled by two \emph{emergent phenomena}; namely, confinement and dynamical chiral symmetry breaking.  Confinement is the empirical fact that no quarks have been detected in isolation.  Dynamical chiral symmetry breaking (DCSB), which is responsible, amongst many other things, for the large mass splitting between parity partners in the spectrum of light-quark hadrons, which is present even though the relevant Lagrangian current-quark masses are small, explains the origin of constituent-quark masses and underlies the success of chiral effective field theory.  Neither confinement nor DCSB is apparent in QCD's Lagrangian but yet they play a dominant role in determining the observable characteristics of real-world QCD.

With respect to confinement, it is important to appreciate that the static potential measured in quenched lattice-QCD is not related in any simple way to the question of light-quark confinement.  It is a basic feature of QCD that light-quark creation and annihilation effects are nonperturbative and thus it is impossible in principle to compute a potential between two light quarks \cite{Bali:2005fu}.

On the other hand, confinement can be related to the analytic properties of QCD's Schwinger functions \cite{Roberts:2007ji,Krein:1990sf,Roberts:1994dr}, which are the basic elements of the DSE approach.  Hence the question of light-quark confinement can be translated into the challenge of charting the infrared behavior of QCD's \emph{universal} $\beta$-function.\footnote{This function may depend on the scheme chosen to renormalize the theory but it is unique within a given scheme \protect\cite{Celmaster:1979km}.}  Solving this well-posed problem is an elemental goal of modern hadron physics.  It can be addressed in any framework enabling the nonperturbative evaluation of renormalization constants.

Through the DSEs, the pointwise behaviour of the $\beta$-function determines the pattern of chiral symmetry breaking.  Moreover, the DSEs connect the $\beta$-function to experimental observables, so that a comparison between computations and observations of, e.g., the valence-quark distribution functions, can be used to learn about the evolution of the $\beta$-function into the nonperturbative domain.

%\begin{figure}[t]
%
%\centerline{%\hspace*{2em}%
%\includegraphics[clip,width=0.8\textwidth]{FigsMS/Mp2Jlab.eps}}

%\caption{\label{gluoncloud} Dressed-quark mass function, $M(p)$: solid curves -- DSE results, obtained as explained in Refs.\,\protect\cite{Bhagwat:2003vw,Bhagwat:2006tu}, ``data'' -- numerical simulations of unquenched lattice-QCD \protect\cite{Bowman:2005vx}.  One observes the current-quark of perturbative QCD evolving into a constituent-quark as its momentum becomes smaller.  The constituent-quark mass arises from a cloud of low-momentum gluons attaching themselves to the current-quark.  This is dynamical chiral symmetry breaking: an essentially nonperturbative effect that generates a quark mass \emph{from nothing}; namely, it occurs even in the chiral limit. [Adapted from \protect\cite{Bhagwat:2007vx}.]}
%\end{figure}

A notable and relevant recent success of the DSE approach to hadron physics is its provision of an understanding of DCSB via QCD's gap equation.  The dressed-quark propagator can be written in a number of equivalent forms; e.g.,\footnote{In order to maintain easy contact with existing literature, in this subsection we employ a Euclidean metric.  In concrete terms that means: for Dirac matrices, $\{\gamma_\mu,\gamma_\nu\} = 2\delta_{\mu\nu}$, $\gamma_\mu^\dagger = \gamma_\mu$, $\gamma_5 = \gamma_4 \gamma_1\gamma_2\gamma_3$, tr\,$\gamma_5\gamma_\mu\gamma_\nu\gamma_\rho\gamma_\sigma= - 4 \varepsilon_{\mu\nu\rho\sigma}$; and $a \cdot b = \sum_{i=1}^4 a_i b_i$.  A timelike vector, $p_\mu$, has $p^2<0$.  More information can be found in Sec.\,2.1 of \protect\cite{Roberts:2000aa}.}
\begin{eqnarray}
\nonumber
 S(p) & =&  -i \gamma\cdot p \,\sigma_V(p^2,\zeta^2) + \sigma_S(p^2,\zeta^2)
 = \frac{1}{i \gamma\cdot p \, A(p^2,\zeta^2) + B(p^2,\zeta^2)} \\
&= & \frac{Z(p^2,\zeta^2)}{i\gamma\cdot p + M(p^2)} \,.
%
%& = &  - i \gamma\cdot p \,\sigma_V(p^2,\zeta^2) + \sigma_S(p^2,\zeta^2) \,.
\label{Sgeneral}
\end{eqnarray}
It is important that the mass function, $M(p^2)=B(p^2,\zeta^2)/A(p^2,\zeta^2)$, illustrated in Fig.\,\ref{gluoncloud}, is independent of the renormalisation point, $\zeta$. Furthermore, this propagator and, indeed, all colored Schwinger functions are gauge covariant.  Therefore, when contracted in the expression for a color-singlet and gauge-invariant observable, all information about the evolution of the quark mass with momentum, displayed in Fig.\,\ref{gluoncloud}, is retained.  Hence, this manifestation of DCSB has an impact on cross-sections that is truly observable.

That impact is greatly amplified in observables involving the pion, which, as QCD's Goldstone mode, is innately connected with DCSB.  Indeed, the DSEs explain the pion as, simultaneously, both a Goldstone mode and an intricate bound-state of a dressed-quark and -antiquark, described by a Bethe-Salpeter amplitude:
\begin{equation}
\Gamma_\pi(k;P) = \gamma_5 \left[ i E_\pi(k;P) +
\gamma\cdot P F_\pi(k;P) \rule{0mm}{5mm}+ \gamma\cdot k \,k \cdot P\, G_\pi(k;P) +
\sigma_{\mu\nu}\,k_\mu P_\nu \,H_\pi(k;P) \right]. \label{genpibsa}
\end{equation}
Through the axial-vector Ward-Takahashi identity, which is the statement of chiral symmetry and the pattern by which it is broken, the DSEs yield \cite{Maris:1997hd} a set of Goldberger-Treiman relations for the pion, which are exact in the chiral limit; viz.,
\begin{eqnarray}
\label{bwti}
f_\pi E_\pi(k;0)  &= &  B(k^2)\,, \\
 F_R(k;0) +  2 \, f_\pi F_\pi(k;0)                 & = & A(k^2)\,,
 \label{fwti}\\
G_R(k;0) +  2 \,f_\pi G_\pi(k;0)    & = & 2 A^\prime(k^2)\,,\\
\label{gwti}
H_R(k;0) +  2 \,f_\pi H_\pi(k;0)    & = & 0\,,
\end{eqnarray}
where: the functions on the right-hand-sides are defined in Eq.\,(\ref{Sgeneral}); $f_\pi$ here represents the pion's leptonic decay constant; and $F_R$, $G_R$, $H_R$ are associated with terms in the axial-vector vertex that are regular in the neighborhood of the pion pole.

The first of these relations, Eq.\,(\ref{bwti}), states that in the chiral limit the dominant, pseudoscalar part of the pion's Bethe-Salpeter amplitude is completely determined by the scalar piece of the dressed-quark self-energy.  This exact result in QCD expresses an extraordinary fact; viz., owing to DCSB, the solution of the two-body problem in the pseudoscalar channel is known, almost completely, once the quark one-body problem is solved.\footnote{The following three relations establish that the pion necessarily has pseudovector components because the dressed-quark wave function renormalization is not identically one.  This fact has an extremely important impact on the pion's electromagnetic form factor at large momentum transfer \protect\cite{Maris:1998hc,GutierrezGuerrero:2010md}.}  In combination with the feature that the dressed-quark propagator has a spectral representation when considered as a function of current-quark mass \cite{Langfeld:2003ye}, Eq.\,(\ref{bwti}) entails that the chiral-limit dressed-quark scalar self-energy provides an accurate pointwise approximation to $E_\pi(k;P)$ at  physical light-quark current-masses.

It will now be evident that, through the mesonic analogue of Fig.\,\ref{Comptondq}, illustrated in Fig.\,\ref{fighandbagmeson}, the leading $x$-dependence of the physical pion's valence-quark distribution function is determined by the chiral-limit dressed-quark scalar self-energy.  As with the electromagnetic pion form factor, the leading-order contribution is accurate apart from small corrections to the anomalous dimension \cite{Maris:1998hc,Maris:2000sk}.  This has far reaching consequences, which can be revealed by considering the gap equation.

Let us imagine a theory in which the kernel of the gap equation behaves as
\begin{equation}
\label{modelalpha}
\frac{\alpha(q^2)}{q^2} \stackrel{q^2 \gg M_D^2}{\propto} \left(\frac{1}{q^2}\right)^{1+\kappa} ,
\end{equation}
where $\kappa>0$ and $M_D^2$ is some intrinsic, characteristic scale.  Then, using the method of \cite{Higashijima:1983gx,Roberts:1989mj}, one finds
\begin{equation}
\label{Bk2kappa}
B(k^2) \stackrel{k^2 \gg M_D^2}{\propto} \left(\frac{1}{k^2}\right)^{1+\kappa} .
\end{equation}
The case of QCD is similar: the kernel is
\begin{equation}
\frac{\alpha(q^2)}{q^2} \stackrel{q^2 \gg \Lambda_{\rm QCD}^2}{=}
\frac{1}{q^2} \frac{\lambda \pi}{\ln [q^2/\Lambda_{\rm QCD}^2]},
\end{equation}
where $\lambda = 12/[33-2 n_f]$, with $n_f$  the number of active fermion flavours, and
\begin{equation}
B(k^2) \stackrel{k^2 \gg \Lambda_{\rm QCD}^2}{\propto} \frac{\lambda}{k^2} \left(\frac{1}{\ln [k^2/\Lambda_{\rm QCD}^2]}\right)^{1-\lambda}.
\label{uvBk2}
\end{equation}
The general pattern should now be clear; namely, the momentum-dependence of the chiral-limit dressed-quark self-energy replicates, up to an anomalous dimension, that of the quark-quark interaction and this behavior is manifest for $k^2 \gg M_D^2$.

With this perspective, the analysis of \cite{Ezawa:1974wm} can now be understood to predict that, in a theory with an interaction of the type in Eq.\,(\ref{modelalpha}), the pion's valence-quark distribution function must evolve according to
\begin{equation}
\label{qvpikappa}
q_v^\pi(x;Q_0) \stackrel{x\sim 1}{\propto} (1-x)^{2 (1+\kappa)} \;\; \forall Q_0 \gg M_D\,.
\end{equation}
This indicates and interprets the scale at which ``counting-rule'' behavior should be evident; viz., it is the mass-scale at which the asymptotic form of the evolution of the chiral-limit dressed-quark scalar self-energy is manifest.

One can readily translate the preceding argument into a QCD prediction; namely,
\begin{equation}
\label{qvpiQCD}
q_v^\pi(x;Q_0) \stackrel{x\sim 1}{\propto} (1-x)^{2+\gamma}, \;\; Q_0 \simeq 1\,\mbox{\rm GeV},
\end{equation}
where $0<\gamma\ll 1$ is an anomalous dimension, because, as evident in Fig.\,\ref{gluoncloud}, 1\,GeV$^2 \approx 20\,\Lambda_{\rm QCD}^2$  is the mass-scale at which the chiral-limit dressed-quark mass-function has plainly assumed its asymptotic form, Eq.\,(\ref{uvBk2}).  Given Eq.\,(\ref{qvpikappa}), it is unsurprising that the exponent ``2'' in Eq.\,(\ref{qvpiQCD}) is a lower-bound.  As we have indicated, the additional logarithmic correction to the momentum-dependence of the mass-function, a manifestation of asymptotic freedom, leads truly to an exponent greater-than two at the counting-rule resolving scale.  Moreover, this exponent increases under QCD evolution to $Q>Q_0$.

It is now instructive to reconsider Eq.\,(\ref{qvpikappa}) in connection with models of the NJL-type, Sec.\,\ref{sec:NJLpion}.  Such a model, in the chiral limit and regularized in a translationally invariant manner, may be realized through an analytic continuation of Eq.\,(\ref{modelalpha}) to $\kappa=-1$.  One then reads the mass-function from Eq.\,(\ref{Bk2kappa}): $\forall k^2$, $B(k^2) = m_Q$; i.e., a constant.  The pion's valence-quark distribution function in this model follows from Eq.\,(\ref{qvpikappa}):
$q_v^{\pi\,{\rm NJL}_{\rm TI}}(x) = 1$.  In the absence of any regularization scale, this is the chiral-limit prediction at all resolving scales.  Of course, in concrete applications of such models, an ultraviolet regularization scale is finally introduced: typically $\Lambda_{\rm UV} \sim 1\,$GeV.  The model's practitioners describe this scale as the boundary between the nonperturbative and perturbative domains.  As the only scale in the model's description of a chiral limit pion, it should properly define the model's intrinsic resolving scale.  Hence, one arrives at the prediction
\begin{equation}
\label{qvpikappaNJL}
q_v^{\pi\,{\rm NJL}_{\rm TI}}(x;Q_0=\Lambda_{\rm UV}) = 1\,.
\end{equation}

Our modernization of the analysis in \cite{Ezawa:1974wm} serves to eliminate much of the ambiguity in understanding what is meant by the ``QCD prediction'' for $q_v^{\pi}(x)$.  We have provided a clean definition of the resolving-scale at which this prediction should be valid.  Moreover, with Eqs.\,(\ref{qvpikappa}) -- (\ref{qvpikappaNJL}) and the associated discussion, we have demonstrated concretely that the \emph{pointwise} behavior of the pion's quark distribution function on the valence-quark domain is a sensitive probe of the nature of the quark-quark interaction.  To be concrete, if the conclusion of \cite{Conway:1989fs} is confirmed, then the theory underlying hadron physics is not QCD.  The arguments can be generalized to other hadrons.

\begin{figure}[t]
\centerline{\includegraphics[height=0.5\textwidth]{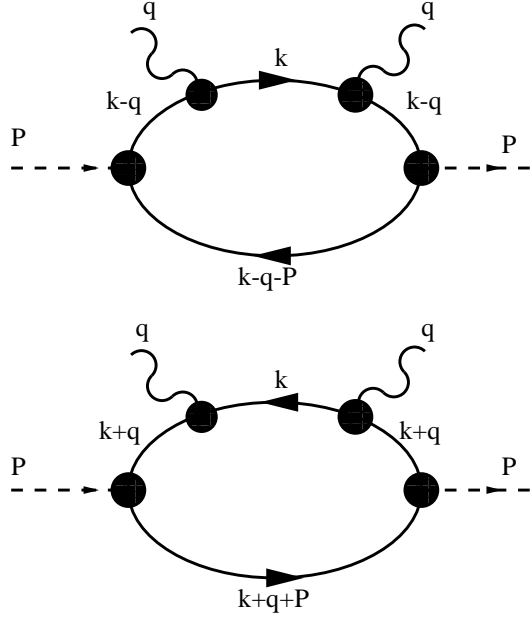}}

\caption{\label{fighandbagmeson} ``Handbag'' contributions to the virtual
photon-pion forward Compton scattering amplitude.  $\pi$, dashed-line;
$\gamma$, wavy-line; ${\cal S}$, internal solid-line, dressed-quark
propagator, Eq.\ (\protect\ref{Sgeneral}).  The filled circles represent the
pion's Bethe-Salpeter amplitude, $\Gamma_\pi$ in Eq.\,(\protect\ref{genpibsa}), and
the dressed-quark-photon vertex, $\Gamma_\mu$ discussed in App.\,\protect\ref{Appendix:FE},
depending on which external line they begin/end.  [Figure adapted from \protect\cite{Hecht:2000xa}.]}
\end{figure}

Additional information can be explicated through a recapitulation of a calculation of the pion's valence-quark distribution function within the DSE framework.  The diagrams in Fig.\,\ref{fighandbagmeson}, an explicit realization of Fig.\,\ref{generalHandbag}, provide the starting point for the computation in \cite{Hecht:2000xa}.  The upper diagram represents the renormalized matrix element
\begin{eqnarray}
\nonumber
T^+_{\mu\nu}(q,P)
& = & {\rm tr}\int  \! \frac{d^4 k}{(2\pi)^4} \left[ \tau_-\Gamma_\pi(k_\Gamma;-P) \right. \\
&& \left. \times \, S(k_t)\, ieQ\Gamma_\nu(k_t,k)
\,\underline{S(k)}\,ieQ\Gamma_\mu(k,k_t)\,S(k_t)\, \tau_+\Gamma_\pi(k_\Gamma;P)\,\underline{S(k_s)}\right] ,
\label{Tmunu}
\end{eqnarray}
where: $\tau_{\pm} = \sfrac{1}{2}(\tau_1 \pm i \tau_2)$ is a combination of Pauli matrices; ${\cal S}(\ell)= {\rm diag}[S_u(\ell),S_d(\ell)]$, with $S_u = S_d = S$, assuming isospin symmetry; and $k_\Gamma= k-q-P/2$, $k_t=k-q$, $k_s=k-q-P$.  The fully-dressed quark-photon vertex, $\Gamma_\mu(\ell_1,\ell_2)$, appeared in Eq.\,(\ref{qvHkl}) and here $Q= {\rm diag}(2/3,-1/3)$ is the quark-charge matrix.  The matrix element represented by the lower diagram is the crossing partner of Eq.\,(\ref{Tmunu}) and is obvious by analogy.  The hadronic tensor relevant to inclusive deep inelastic lepton-pion scattering can be obtained from the forward Compton process via the optical theorem:
\begin{equation}
\label{Wmunu}
W_{\mu\nu}(q;P)= W^+_{\mu\nu}(q;P)+ W^-_{\mu\nu}(q;P)= \frac{1}{2\pi} {\bf\sf
Im}\left[ T^+_{\mu\nu}(q;P) + T^-_{\mu\nu}(q;P)
\right].
\end{equation}

In the Bjorken limit one finds explicitly \cite{Hecht:2000xa}
\begin{equation}
\label{WF1F2}
W^+_{\mu\nu}(q;P) = F_1^+(x) \, t_{\mu\nu} + F_2^+(x) \, \frac{q_\mu^t
q_\nu^t}{2 x} \,,
\end{equation}
$t_{\mu\nu} =\delta_{\mu\nu} - q_\mu q_\nu/q^2$, $q_\mu^t = q_\mu + 2 x
P_\mu$, and
\begin{equation}
\label{CGrel}
F_2^+(x)= 2 x F_1^+(x)\,,\; F_{1,2}^+(x) \to 0\;{\rm as}\;x\to 1\,.
\end{equation}
Combining these results with their analogues for $W^-_{\mu\nu}$, one recovers Bjorken scaling of the deep inelastic cross section; namely, the cross section depends only on $x$, and not separately on $P \cdot q$ and $q^2$.  One may therefore write
\begin{equation}
F_2^{e\pi}(x) = F_2^+(x) + F_2^-(x)
= \frac{4}{9} [ x u(x) + x \bar u(x) ]
+ \frac{1}{9} [ x d(x) + x \bar d(x) ] + \ldots,
\end{equation}
where the ellipsis denotes contributions from heavier quarks, which are small.  We emphasize that this result is only valid at lowest order in the strict Bjorken limit.  It is not preserved under QCD evolution, see Sec.\,\ref{sec:QCDscalingV}.

In an explicit calculation of the handbag diagrams, Fig.\,\ref{fighandbagmeson}, the resolving scale is exposed through the act of taking the imaginary part when employing the optical theorem in Eq.\,(\ref{Wmunu}).  At this point a snapshot is taken of the dressed-quark propagators underlined in Eq.\,(\ref{Tmunu}).  Their dressing functions are sampled at the particular renormalization point and this defines the resolving scale $Q_0$.  One can recover the analysis of \cite{Ezawa:1974wm}, augmented by perturbative QCD evolution, by pursuing this route at a resolving scale deep into the perturbative domain.

On the other hand, \cite{Hecht:2000xa} employ algebraic \emph{Ans\"atze} for the elements in Eq.\,(\ref{Tmunu}) whose form is known even on the nonperturbative domain.  They are determined by studies of meson properties \cite{Roberts:1994hh,Burden:1995ve,Maris:1998hc} and exhibit behavior that is broadly consistent with that of QCD's $n$-point functions; viz., the ultraviolet power-laws are explicitly expressed but the logarithms are suppressed in order to achieve a level of simplicity.  Hence, following the path described above, \cite{Hecht:2000xa} delivers a representation of the pion's distribution functions at an infrared resolving scale through the definitions
\begin{equation}
\label{F2uvx}
u_v^\pi(x;Q_0) := \frac{9}{2} \, F_1^+(x;Q_0)\,,\;
d_v^\pi(x;Q_0) := 9 \, F_1^-(x;Q_0)\,.
%F_2^+(x;Q_0)  =  \frac{4}{9}\,x\, u_v^\pi(x;Q_0) \,,\;
%
%F_2^-(x;Q_0)  =  \frac{1}{9}\,x\, \bar d_v^\pi(x;Q_0)\,;
\end{equation}
%...This is the DIS-scheme (almost) where F2 <=> Sum[q(x)] at Q0.

How should Eqs.\,(\ref{F2uvx}) be understood?  In \cite{Hecht:2000xa}, the computation is interpreted within the setting of a rainbow-ladder DSE truncation.  Hence, sea-quark contributions are absent because they cannot appear without nonperturbative dressing of the quark-gluon vertex \cite{Cloet:2008fw,Chang:2009ae,Chang:2009zb}.  Thus Eqs.\,(\ref{F2uvx}) describe valence-quark distribution functions and one should have
\begin{equation}
\label{uvnorm}
\int_0^1\,dx\,u_v^\pi(x;Q_0) = 1 = \int_0^1\,dx\,\bar d_v^\pi(x;Q_0)\,;
\end{equation}
viz., the $\pi^+$ contains one, and only one, $u$-valence-quark and one $\bar d$-valence-quark.
Note that one has $\bar d_v^{\pi^+}(x;Q_0)=u_v^{\pi^+}(x;Q_0)=d_v^{\pi^-}(x;Q_0)$ in this calculation.
%--What and why did I write previously about G-parity?

It is a deficiency of this and kindred calculations that the model's resolving scale is not determined \emph{a priori}.  That can be overcome by calculating the moments of the pion's distribution via Eq.\,(\ref{qvHklx}), in which case the resolving scale is identical to the renormalization point used in computing the dressed Schwinger functions.  The cost of that approach, however, is a loss of direct knowledge about the distribution's pointwise evolution.  In \cite{Hecht:2000xa}, $Q_0$ was chosen so that the computed distribution, when evolved to $Q_0^\prime=2\,$GeV using leading-order formulae, produced first and second moments in agreement with those reported in \cite{Sutton:1991ay}.  This yields\footnote{The quantitative similarity between this and the mass-scale for LO-evolution in Sec.\,\protect\ref{sec:GRV} is noteworthy.} $Q_0=0.54\,$GeV so that Eqs.\,(\ref{uvnorm}) are satisfied with a valence-quark mass $\check M=0.30\,$GeV$\approx M(Q_0)$.

It is apparent that this procedure views the distributions defined in Eqs.\,(\ref{F2uvx}) as infrared boundary value input for the valence-quark evolution equations.  This is strictly valid only if all nonperturbative corrections are negligible for $Q>Q_0$, a constraint whose faithful implementation would require $Q_0>1\,$GeV.  On the other hand, it might be an efficacious approximation so long as $\alpha(Q_0)^2/(4\pi^2) \ll 1$.  In this connection, it is notable that
\begin{equation}
\label{Q0crit}
\frac{\alpha(Q_0)^2}{4\pi^2} = 0.017\,,\; Q_0=0.54\,{\rm GeV};
\end{equation}
whereas $\alpha(0.5\,Q_0)^2/(4\pi^2)=0.60$.  Plainly, there is a narrow domain upon which this perspective may be reasonable.  NB.\ This view mirrors that described in Sec.\,\protect\ref{sec:GRV}.  Furthermore, and significantly, these considerations do not affect the pointwise behavior of the distribution function at $Q_0$.

\begin{figure}[t]
\centerline{\includegraphics[height=0.5\textwidth]{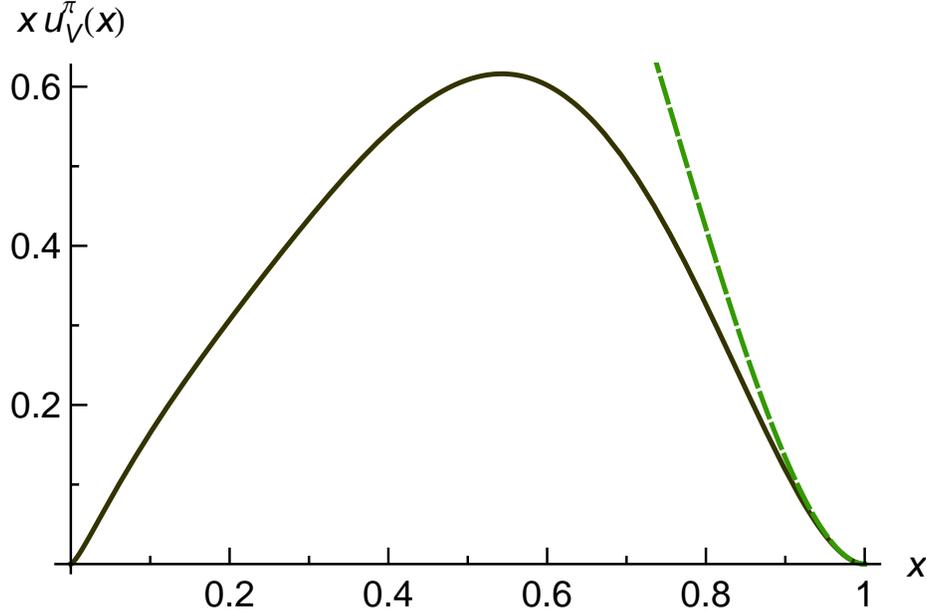}}

\caption{\label{hechtxuvxQ0} (Color online)
\emph{Solid curve} -- Pion's valence-quark momentum distribution computed in \protect\cite{Hecht:2000xa}.  \emph{Dashed curve} -- $A (1-x)^{\eta_2}$, with the parameters drawn from Eq.\,(\protect\ref{paramxuvx}).}
\end{figure}

The valence-quark distribution function computed by \cite{Hecht:2000xa} is depicted in Fig.\,\ref{hechtxuvxQ0}.  From this form one finds the momentum-fraction carried by the valence-quarks; viz.,
\begin{equation}
\langle x_q\rangle^\pi_{Q_0=0.54\,{\rm GeV}}=\int_0^1\!dx\,x\,[u_v^\pi(x;Q_0)+\bar d_v^\pi(x;Q_0)] = 0.72\,.
\end{equation}
The remainder is carried by the gluons; namely, $\langle x_g\rangle^\pi_{Q_0}=0.28$, which are invisible to the electromagnetic probe, since the sea distribution is zero at $Q_0$.  The pion is a bound-state of finite extent formed through the exchange of dressed-gluons between a dressed-quark and -antiquark, and hence gluons are necessarily always present.  It is therefore plain that at no resolving scale could all the pion's momentum be carried by the valence-quarks.  %
For comparison, the parametrized valence-like pion parton distributions in \protect\cite{Gluck:1999xe} yield a gluon momentum-fraction of\,\footnote{%
A novel perspective on the magnitude of an hadron's gluon momentum fraction is discussed in \protect\cite{Chen:2009mr,Ji:2009fu,Chen:2009dg}.}
$\langle
x_g\rangle^\pi_{Q_0}=0.29$ at $Q_0= 0.51\,$GeV.
%
%and a sea-quark fraction $\langle x_{\bar q}\rangle^\pi_{Q_0}=0.15$.
%
In \cite{Hecht:2000xa}, the second and third moments are
\begin{equation}
\langle x^2_{u+\bar d}\rangle^\pi_{Q_0=0.54\,{\rm GeV}}= 0.36\,,\;
\langle x^3_{u+\bar d}\rangle^\pi_{Q_0=0.54\,{\rm GeV}}= 0.21\,.
\end{equation}

The low moments are primarily determined by the distribution function's behavior at small-$x$ (see the close of Sec.\,\ref{sec:latticeeQCD}).  Sensitivity to the nature of QCD's interaction is found at large $x$, which corresponds to large relative momentum between the dressed-quark and -antiquark.  Two questions come immediately to mind, namely: what should one expect for the $x$-dependence of the distribution in Fig.\,\ref{hechtxuvxQ0} at large-$x$; and what truly constitutes the large-$x$ domain?

The first question is readily answered.  In this case $Q_0^2 \gtrsim 5\,\Lambda_{\rm QCD}^2$, which corresponds to a scale whereat the chiral limit mass function is dropping rapidly but does not yet exhibit the behavior associated with its truly asymptotic momentum-dependence: $Q_0=0.54\,$GeV does not lie beyond the inflexion point of the chiral-limit mass-function, see Fig.\,\ref{gluoncloud}.  One would therefore anticipate that, for $x\sim 1$, $u_v(x;Q_0=0.54\,{\rm GeV} )\approx (1-x)^\eta$, with $\eta \lesssim 2$.  It was found in \cite{Hecht:2000xa} that the distribution in Fig.\,\ref{hechtxuvxQ0} is pointwise accurately interpolated by the function
\begin{equation}
\label{formxuvx}
x u_v^\pi(x;Q_0) = A \, x^{\eta_1} \, ( 1 - \epsilon \sqrt x + \gamma x) \,(1-x)^{\eta_2}
\end{equation}
with the fit-parameters taking the values
\begin{equation}
\label{paramxuvx}
\begin{array}{ccccc}
A &\eta_1 &\epsilon & \gamma & \eta_2 \\
11.24 & 1.43 & 2.44 & 2.54 & 1.90
\end{array}\,.
\end{equation}
These parameters depend on $Q_0$ and the value of $\eta_2$ is fully consistent with expectation.

The second question posed above can now also be answered quantitatively using Eqs.\,(\ref{formxuvx}) and (\ref{paramxuvx}).  The dashed-curve in Fig.\,\ref{hechtxuvxQ0} is that component of Eq.\,(\ref{formxuvx}) which dominates the $x$-dependence of $u_v^\pi(x;Q_0)$ at ``large $x$.''  On the domain
\begin{equation}
\label{xdomainlarge}
{\cal L}_x = \{ x | x >0.86\}\,,
\end{equation}
the dominant component agrees at the level of 20\% or better with the full curve.  NB.\ The extent of this domain depends weakly on the mass-scale $M_D$: it is a little larger in a model with a smaller value of $M_D$; and the disagreement increases to 37\% at $x= 0.76$

In \cite{Hecht:2000xa}, using leading-order evolution, the distribution in Fig.\,\ref{hechtxuvxQ0} is evolved to $Q_0=2\,$GeV and $Q_0=4\,$GeV.  The $u$-quark moments at the former scale are presented in Table\,\ref{pionmoments} and the curve at the latter scale is presented in Fig.\,\ref{pionP}.
%\begin{equation}
%\langle x_{u}\rangle^\pi_{Q_0=2\,{\rm GeV}}= 0.24\,,\;
%\langle x^2_{u}\rangle^\pi_{Q_0=2\,{\rm GeV}}= 0.098\,,\;
%\langle x^3_{u}\rangle^\pi_{Q_0=2\,{\rm GeV}}= 0.049\,.
%\end{equation}

\begin{figure}[t]
\centerline{\includegraphics[width=0.7\textwidth]{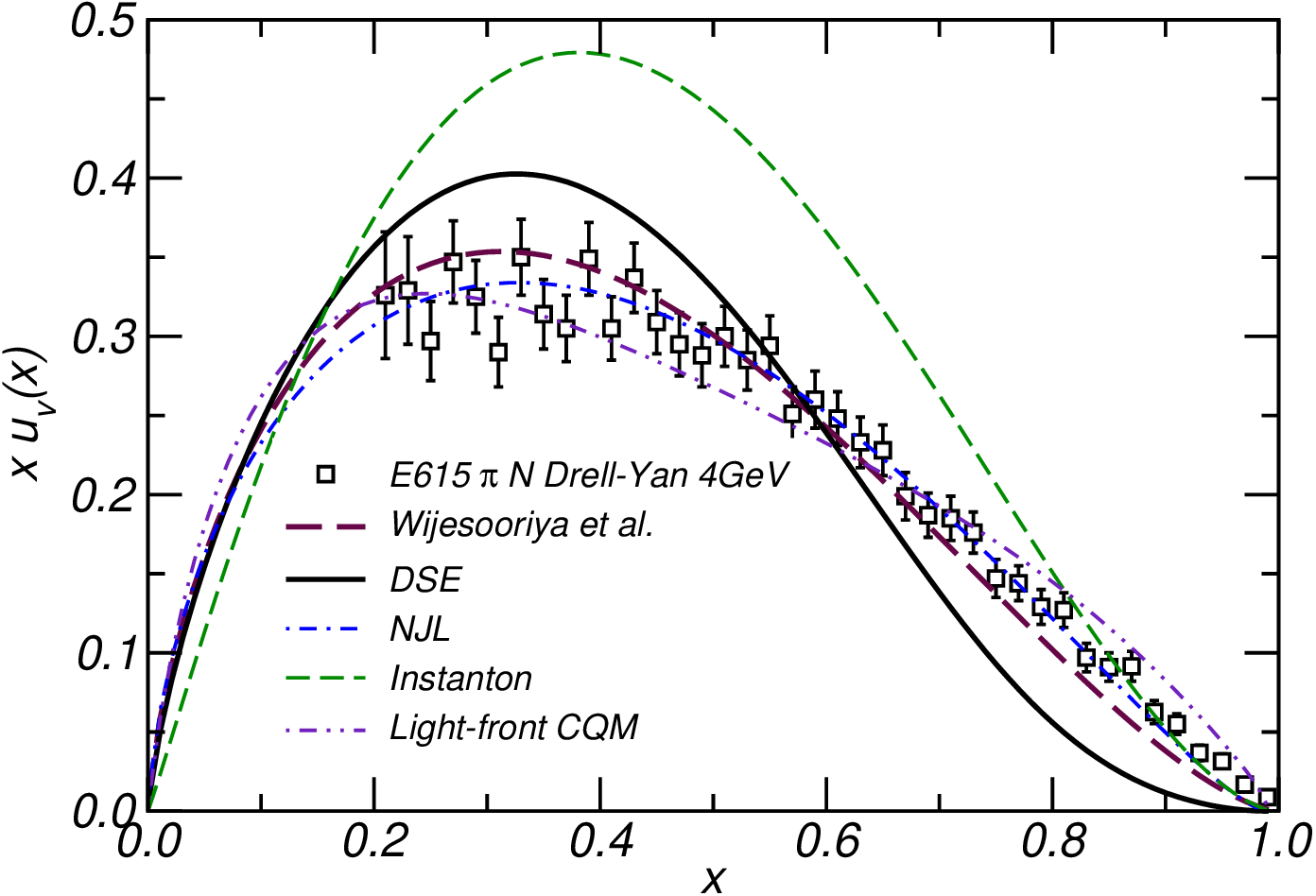}}
\caption{\label{xuvxpicall}
(Color online)
Compilation of results for the valence-$u$-quark distribution in the pion: \emph{Solid curve} -- DSE result \protect\cite{Hecht:2000xa};
\emph{dot-dashed curve}: NJL model (see Sec.\,\protect\ref{sec:NJLpion});
\emph{short-dashed curve}: instanton model \protect\cite{Dorokhov:2000gu};
\emph{dash-dot-dotted curve}: light-front constituent-quark model \protect\cite{Frederico:1994dx};
\emph{circles} -- Drell-Yan data presented in \protect\cite{Conway:1989fs};
and {long-dashed curve} -- reanalysis of that Drell-Yan data described in \protect\cite{Wijesooriya:2005ir}, which is also depicted in Fig.\,\protect\ref{highxdypion}.
All calculations evolved at leading-order to $Q_0=4.0\,$GeV using a four-flavour value of $\Lambda_{\rm QCD}= 0.204\,$GeV, except that from \protect\cite{Wijesooriya:2005ir}, which is reported at $Q_0=5.2\,$GeV.}
\end{figure}

In Fig.\,\ref{xuvxpicall} we display a compilation of results for the valence-$u$-quark distribution in the pion.  The elucidation herein explains why only the DSE prediction exhibits behavior at large-$x$ that is consistent with the QCD parton model, Eq.\,(\ref{qvpiQCD}), first derived in \cite{Ezawa:1974wm,Farrar:1975yb}.  The discussion in \cite{Wijesooriya:2005ir} and Sec.\,\ref{exptpionstructure} shows that one cannot draw firm conclusions about the large-$x$ behavior of the pion's valence-quark distribution function from the single extant $\pi N$ Drell-Yan experiment \cite{Conway:1989fs}.  The status of QCD as the strong interaction piece of the Standard Model will seriously be challenged if an improved experiment, such as that canvassed in \cite{Wijesooriya:2001}, is also incompatible with Eq.\,(\ref{qvpiQCD}).

\begin{figure}[t]
\centerline{\includegraphics[clip,width=0.8\textwidth]{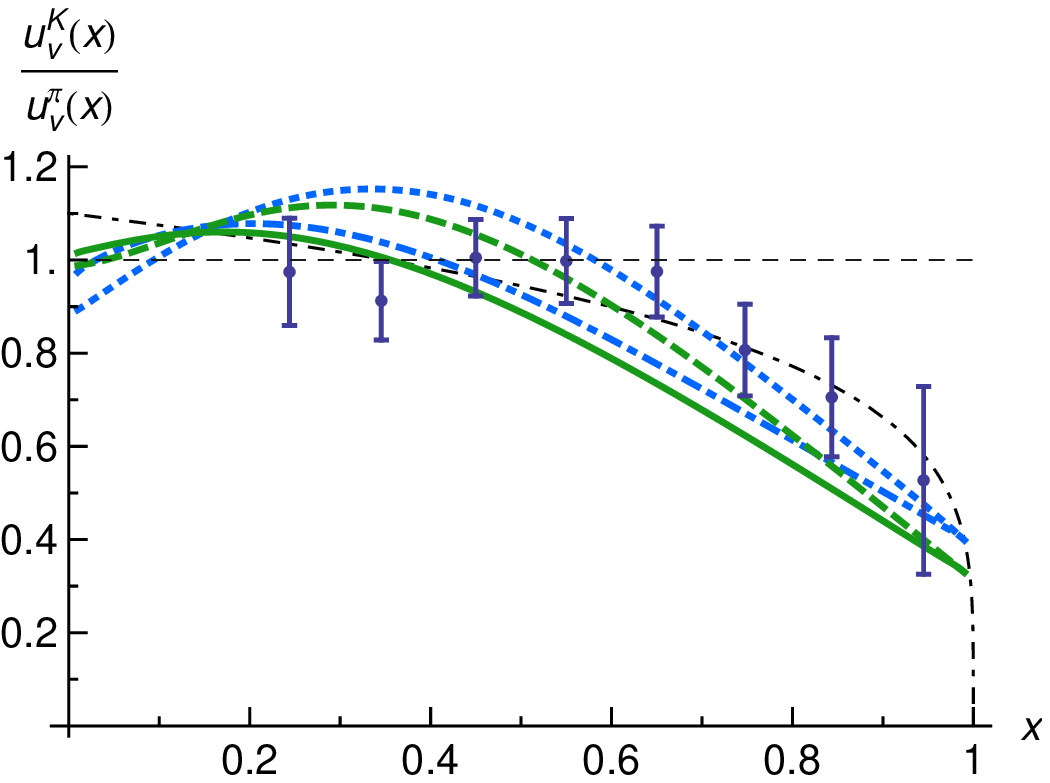}}

\caption{\label{uKpiPCT} (Color online)
\emph{Solid-curve} -- Preliminary Dyson-Schwinger equation prediction for the ratio of kaon-to-pion valence-quark distribution functions at $Q_0=5\,$GeV, evolved at leading-order from the \emph{dashed curve}, which is the DSE prediction at $Q_0=0.57\,$GeV \protect\cite{nguyenphd,pctandyprivate:2009a}.
\emph{Dotted} and \emph{dash-dotted} curves -- our re-evaluation of $u_v^K(x)/u_v^\pi(x)$ using the model in \protect\cite{Shigetani:1993dx};
%Q0=0.5, which is about the same as we use
%
\emph{filled circles} -- the ratio of kaon-to-pion Drell-Yan cross-sections obtained from a sample of dimuon events with invariant mass $4.1<M<8.5\,$GeV \protect\cite{Badier:1980};
\emph{thin dashed line} -- the curve $u_v^K(x)=u_v^\pi(x)$;
and \emph{thin dash-dot curve} -- fit to the data given in Eq.\,(\protect\ref{RDYKpi}).}
\end{figure}

Computations of the valence-quark distribution functions in other mesons are underway \cite{nguyenphd,pctandyprivate:2009a} using the DSE approach that successfully predicted the pion's electromagnetic form factor \cite{Maris:1999bh,Maris:2000sk}.  The impact of the dressed-quark mass function on the ratio $u_v^K(x)/u_v^\pi(x)$ is depicted in Fig.\,\ref{uKpiPCT}.  In comparison with the nonpointlike-pion-regularized NJL result in \cite{Shigetani:1993dx}, one finds that the momentum-dependent mass function markedly affects the separate behaviors of $u_v^\pi(x)$ and $u_v^K(x)$, especially on the valence-quark domain.  However, the preliminary indication is that it does not materially affect the ratio; e.g.,
\begin{equation}
\left.\frac{u_v^K(x)}{u_v^\pi(x)}\right|_{\rm DSE} \stackrel{x\to 1}{=} 0.31 \rule{2em}{0ex} {\rm cf.} \rule{2em}{0ex}
\left.\frac{u_v^K(x)}{u_v^\pi(x)}\right|_{\rm NJL^{\rm HC}} \stackrel{x\to 1}{=} 0.37 \,.
\end{equation}
We anticipated a nonzero value for the ratio because, in the neighborhood of $x=1$, the $u$-quark distribution should have the same pointwise behavior in all pseudoscalar mesons.
%
% Might be different for the TI regularization.
%
These features and predictions add additional emphasis to the need for a much improved measurement of the kaon structure functions.
%our understanding of the existing data \protect\cite{Badier:1980}.
%
%Notwithstanding this,
Before claiming understanding of QCD, it is crucial to verify the predicted large-$x$ behavior, especially those properties which are environment-dependent and those which are not.

The calculation of baryon valence-quark distribution functions is also possible; e.g., using the Poincar\'e covariant Faddeev equation described in App.\,\ref{Appendix:FE}.  This approach capitalizes on the importance of scalar and axial-vector diquark correlations within the nucleon.  In QCD these correlations are essentially nonpointlike \cite{Maris:2004bp,Alexandrou:2006cq}.  While no DSE computations of the pointwise behavior of the nucleon structure functions are yet available, based on the diquark-correlation probabilities presented in Table~2 of the DSE study of nucleon electromagnetic form factors described in \cite{Cloet:2008re} and the fact that the ratio is a fixed-point under evolution, one can estimate
\begin{equation}
\label{donuDSEguess}
\frac{d(x)}{u(x)} \stackrel{x\to 1}{\approx} 0.12 \; \Rightarrow
\frac{F_2^n(x)}{F_2^p(x)} \stackrel{x\to 1}{\approx} 0.36\,,
\end{equation}
%
% scalar dq = 0.602 ... all u-quark
% axial-vector = 0.254 = 2:1 for u:d => 0.169 cf. 0.0847
% mixed = scalar <-> axial vector = 0.144: 1/2 is always u-quark; other piece is 2:1 u:d => u= 0.144 (1/2 + 1/3) & d= 0.144 * 1/2 * 1/3= 0.024
%Ratio = 0.1087/0.891
cf.\ Eqs.\,(\ref{donuempirical}), (\ref{donuFarrar}) and (\ref{icloetdvuv}), and Figs.\,\ref{f2nratio}, \ref{PTB02} and \ref{0201082F}.  In addition, the nonzero value highlights the important role of axial-vector diquark correlations: they enable a truly valence $d$-quark to carry the proton's helicity, which is impossible if scalar diquarks are the only correlations present in the proton's Faddeev amplitude.

%\subsection{Prospects for the future}

%\newpage

%\section{Comments on spin structure}
%\setcounter{figure}{0}
%\setcounter{table}{0}
\setcounter{equation}{0}
\section{Perspective and Prospects}
Understanding the physics of hadrons on the valence-quark domain is a definitive task for hadron physics.  Indeed, a given hadron is defined by its flavor content and that is a valence-quark property.  Whilst significant continuing and new investments in experiment and theory are required in order to acquire this understanding, the potential rewards are great.  For example, this information is essential to both: an elucidation of the Standard Model; and the provision of reliable parton distribution functions for use in the analysis of large collider experiments -- our best tool for uncovering phenomena outside this paradigm.

We have highlighted numerous outstanding and unresolved challenges in the valence region, which experiments must confront.  Important amongst them are the $x>0.8$ region of the nucleon and the pion, and the kaon distribution function.  The upgraded Jefferson Lab facility is well-suited to provide new data for the valence region in the nucleon, from both inclusive and semi-inclusive deep inelastic scattering.  Moreover, a future electron ion collider could provide crucial new data for the mesons.  There is also great potential in using Drell-Yan interactions, at FNAL, J-PARC and GSI, to push into the large-$x$ domain for both mesons and nucleons.

The challenges for theory are exciting and equally great.  The valence-quark domain is the purview of nonperturbative methods in QCD.  Hence, true understanding, in contrast to modeling and parametrization as a simple means of identifying and highlighting key features, requires a widespread acceptance of the need to adapt theory: to the lessons learnt already from the methods of nonperturbative quantum field theory; and a fuller exploitation of those methods.  Thus, before an elucidation of hadron structure can be achieved, theory must accept the most conspicuous of QCD's challenges.  The Standard Model will not otherwise be solved.

\section*{Acknowledgments}
In preparing this article we benefited greatly from constructive comments and input provided by
J.~Arrington,
S.\,J.~Brodsky,
F.~Coester,
I.\,C.~Clo\"et,
D.\,F.~Geesaman,
S.\,E.~Kuhlmann,
A.\,D.~Martin,
W.~Melnitchouk,
\mbox{J.-C.~Peng},
P.\,E.~Reimer,
P.\,C.\,~Tandy,
R.\,D.~Young
and J.\,M.~Zanotti.
This work was supported by
the Department of Energy, Office of Nuclear Physics, contract no.\ \mbox{DE-AC02-06CH11357}.

\newpage

\appendix
\setcounter{equation}{0}
\section{Light-front conventions}
\label{appLF}
Unless otherwise specified, we use the following light-front notation:
\begin{equation}
\ell^+ = \ell^0 + \ell^3\,,\; \ell^- = \ell^0-\ell^3\,,\; \ell_\perp = (\ell^1,\ell^2)\,,
\end{equation}
where $\ell^\mu$ is any quantity that transforms as a contravariant four-vector.  Naturally, $\ell_\pm=\ell^\mp$ and $\ell^2 = \ell^+ \ell^- - \ell_\perp\cdot \ell_\perp$.  It follows that
\begin{equation}
\int {d^4 \ell} \to \frac{1}{2} \int d\ell^-d\ell^+ d^2\ell_\perp\,.
\end{equation}

The spinor describing a spin-$\sfrac{1}{2}$ hadron is orthonormalized according to
\begin{equation}
\bar u(p^\prime,\lambda^\prime) u(p,\lambda) = 2 p^+ (2\pi)^3 \delta_{\lambda \lambda^\prime} \delta(p^\prime - p) \delta^2(p^\prime_\perp - p_\perp),
\end{equation}
where $\lambda$, $\lambda^\prime$ are helicities.

NB.\ If one considers a particle's momentum, then $k^+ \geq 0$ in the light-front frame.  Hence, the only way to make a zero momentum Fock state is for every particle it contains to have $k^+=0$.  Although this defines a set of measure zero in the light-front phase space, such states cannot casually be discarded because operators exist which are singular at $k^+=0$.  For example, to throw away these states is to preclude dynamical chiral symmetry breaking, a keystone of nonperturbative QCD and hadron physics.

\setcounter{equation}{0}
\section{Dressed-quarks, the Faddeev equation and the current}
\label{Appendix:FE}
When emphasizing dressed-quark degrees of freedom it is now natural to describe baryons using a Poincar\'e covariant Faddeev equation.  This approach sits squarely within the ambit of the application of Dyson-Schwinger equations (DSEs) in QCD \cite{Roberts:2000aa,Maris:2003vk,Roberts:2007jh}.

\begin{figure}[t]
\centerline{
\includegraphics[clip,width=0.75\textwidth]{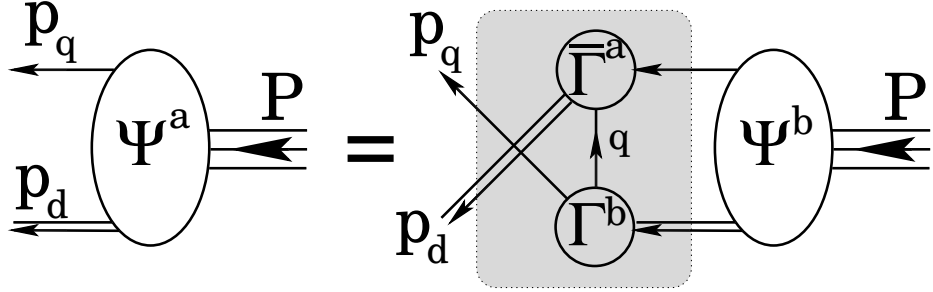}}
\caption{\label{faddeevfigure} Poincar\'e covariant Faddeev equation.  $\Psi$ is the Faddeev amplitude for a nucleon of total momentum $P= p_q + p_d$.  It expresses the relative momentum correlation between the dressed-quark and -diquarks within the nucleon.  The shaded region demarcates the kernel of the Faddeev equation, in which: the \emph{single line} denotes the dressed-quark propagator; $\Gamma$ is the diquark Bethe-Salpeter-like amplitude; and the \emph{double line} is the diquark propagator. [Full explanation provided in App.\,A of \protect\cite{Cloet:2008re}, from which this figure is adapted.]}
\end{figure}

One arrives at the Faddeev equation by noticing that in quantum field theory a nucleon appears as a pole in a six-point quark Green function.  The pole's residue is proportional to the nucleon's Faddeev amplitude, which is obtained from a Poincar\'e covariant Faddeev equation that sums all possible exchanges and interactions that can take place between three dressed-quarks.  A tractable Faddeev equation for baryons was formulated in Ref.\,\cite{Cahill:1988dx}.  Depicted in Fig.\,\ref{faddeevfigure}, it is founded on the observation that an interaction which describes colour-singlet mesons also generates quark-quark (diquark) correlations in the colour-$\bar 3$ (antitriplet) channel \cite{Cahill:1987qr}.

The dominant correlations for ground state octet and decuplet baryons are scalar ($0^+$) and axial-vector ($1^+$) diquarks because, for example, the associated mass-scales are smaller than the baryons' masses \cite{Burden:1996nh,Maris:2002yu}, namely (in GeV)
\begin{equation}
\label{diquarkmass}
m_{[ud]_{0^+}} = 0.7 - 0.8
 \,,\;
m_{(uu)_{1^+}}=m_{(ud)_{1^+}}=m_{(dd)_{1^+}}=0.9 - 1.0\,.
\end{equation}
While diquarks do not appear in the strong interaction spectrum; e.g., Refs.\,\cite{Bender:1996bb,Bender:2002as,Bhagwat:2004hn}, the attraction between quarks in this channel justifies a picture of baryons in which two quarks are always correlated as a colour-$\bar 3$ diquark pseudoparticle, and binding is effected by the iterated exchange of roles between the bystander and diquark-participant quarks.

The kernel of the Faddeev equation is completed by specifying that the quarks are dressed, with two of the three dressed-quarks correlated always as a colour-$\bar 3$ diquark.  As illustrated in Fig.\,\ref{faddeevfigure}, binding is then effected by the iterated exchange of roles between the bystander and diquark-participant quarks.  The Faddeev equation yields the nucleon's mass and amplitude.

\begin{figure}[t]
\begin{minipage}[t]{\textwidth}
\begin{minipage}[t]{0.45\textwidth}
\leftline{\includegraphics[width=0.90\textwidth]{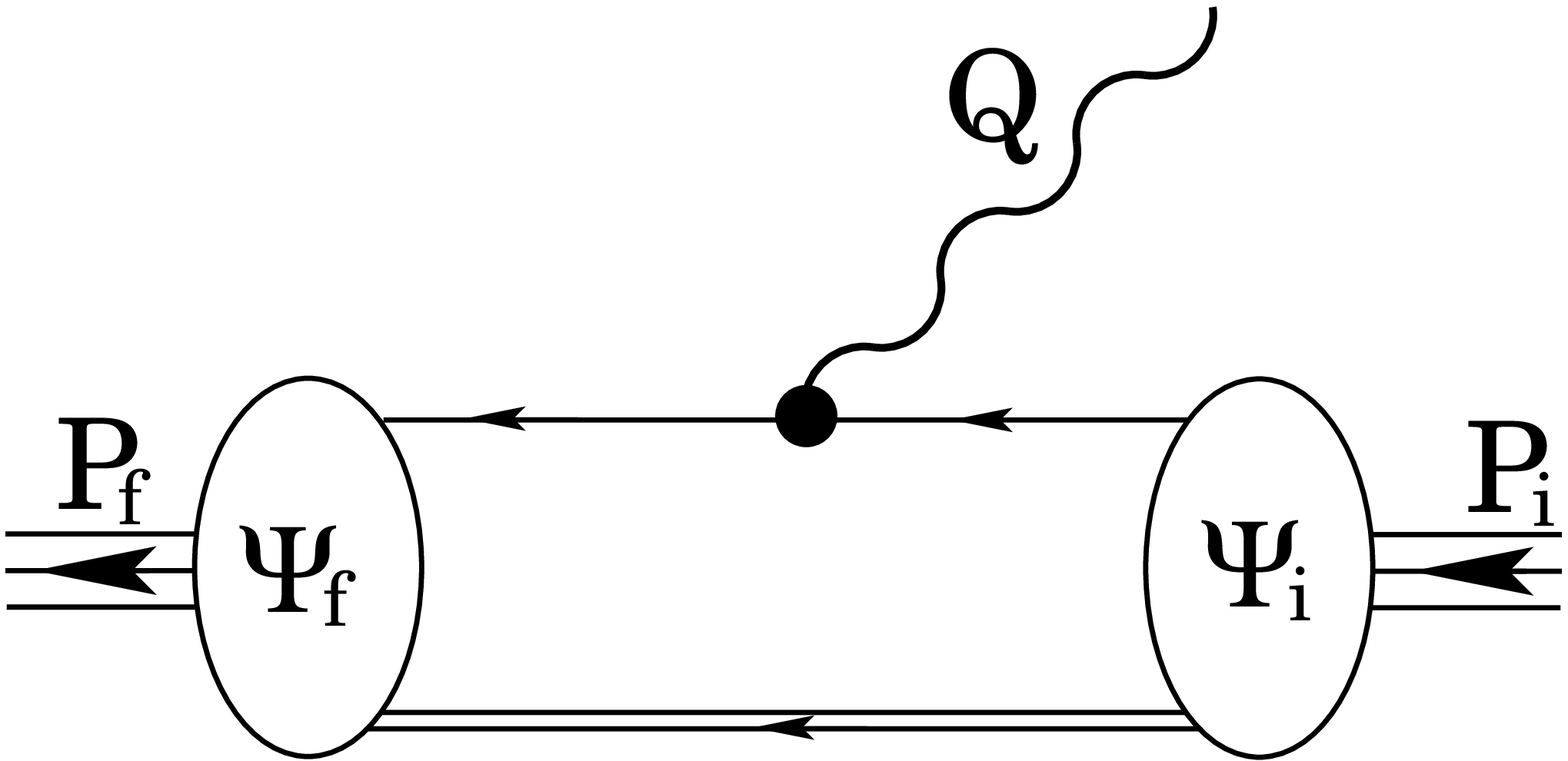}}
\end{minipage}
\begin{minipage}[t]{0.45\textwidth}
\rightline{\includegraphics[width=0.90\textwidth]{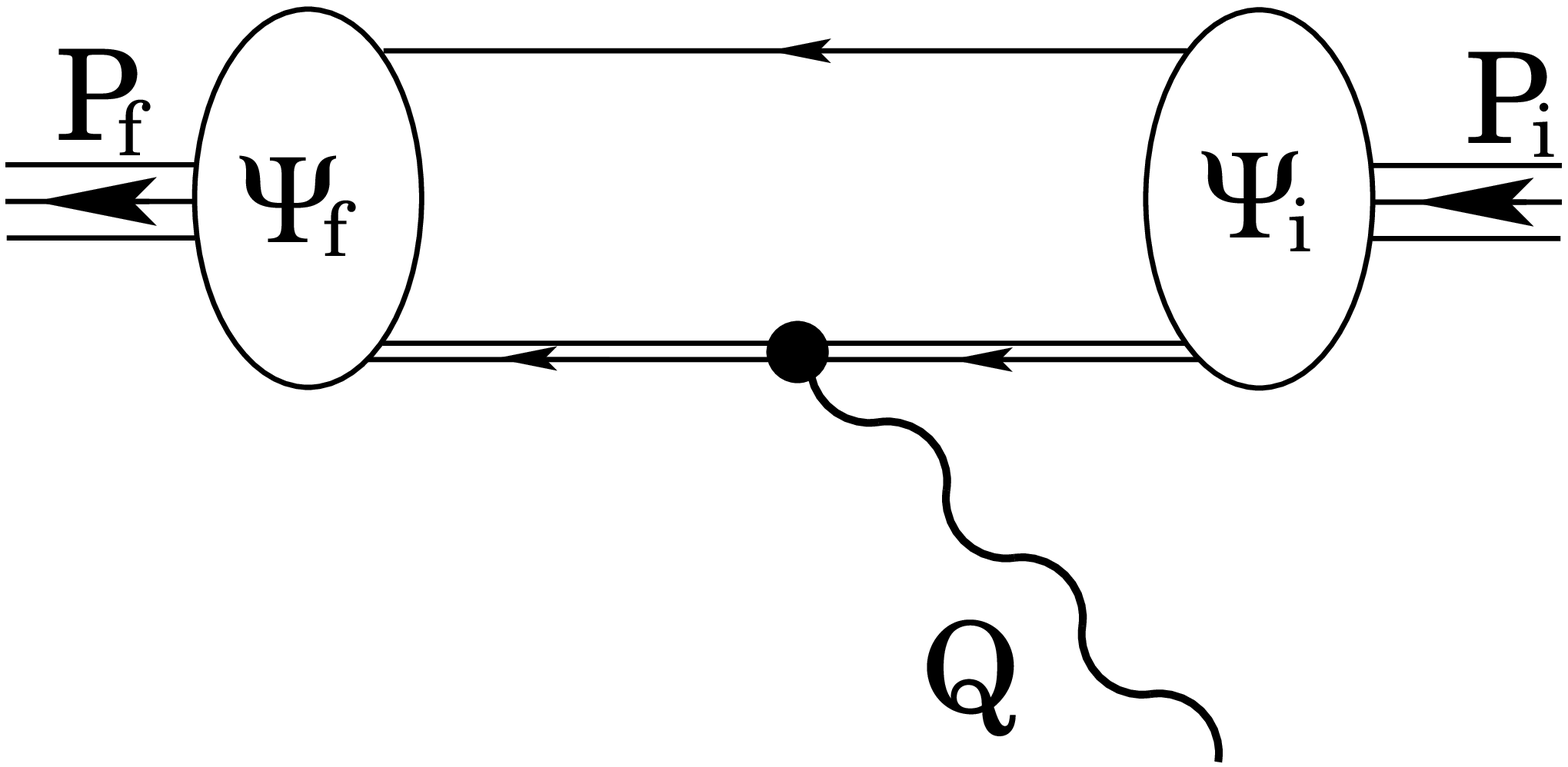}}
\end{minipage}\vspace*{3ex}

\begin{minipage}[t]{0.45\textwidth}
\leftline{\includegraphics[width=0.90\textwidth]{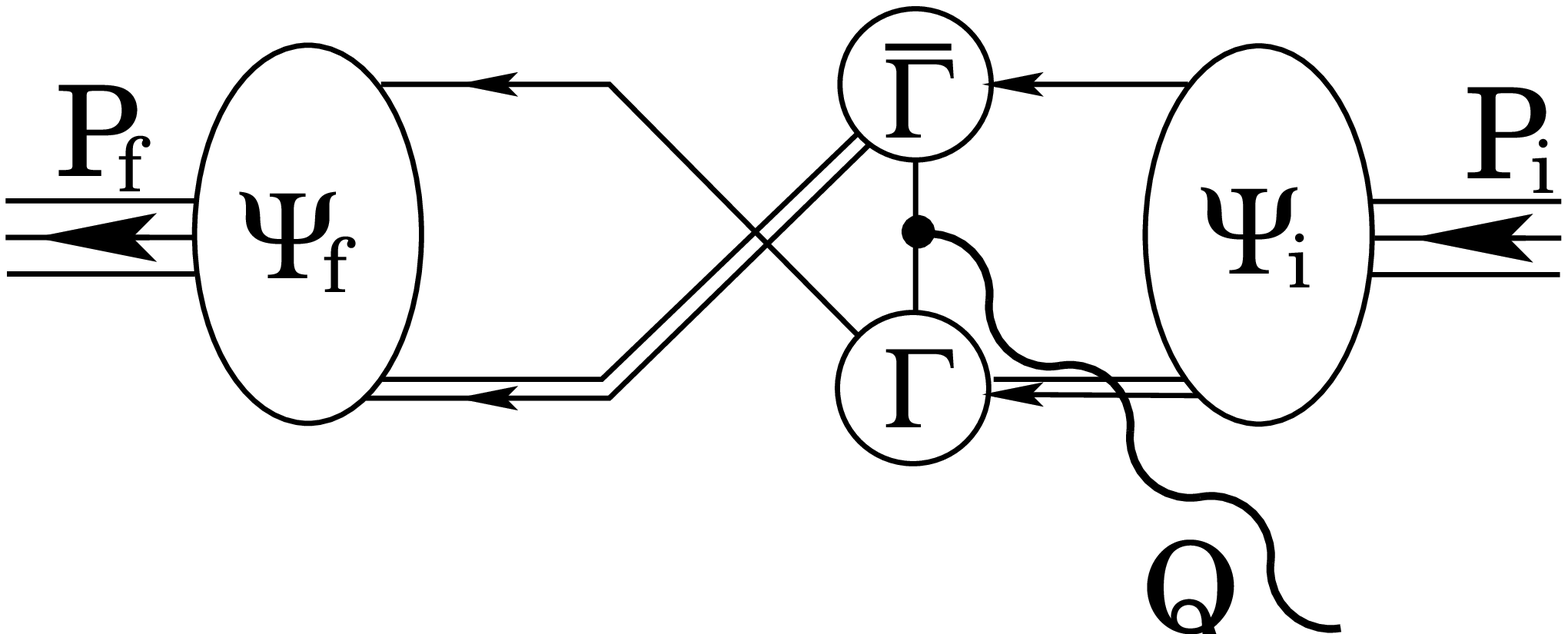}}
\end{minipage}
\begin{minipage}[t]{0.45\textwidth}
\rightline{\includegraphics[width=0.90\textwidth]{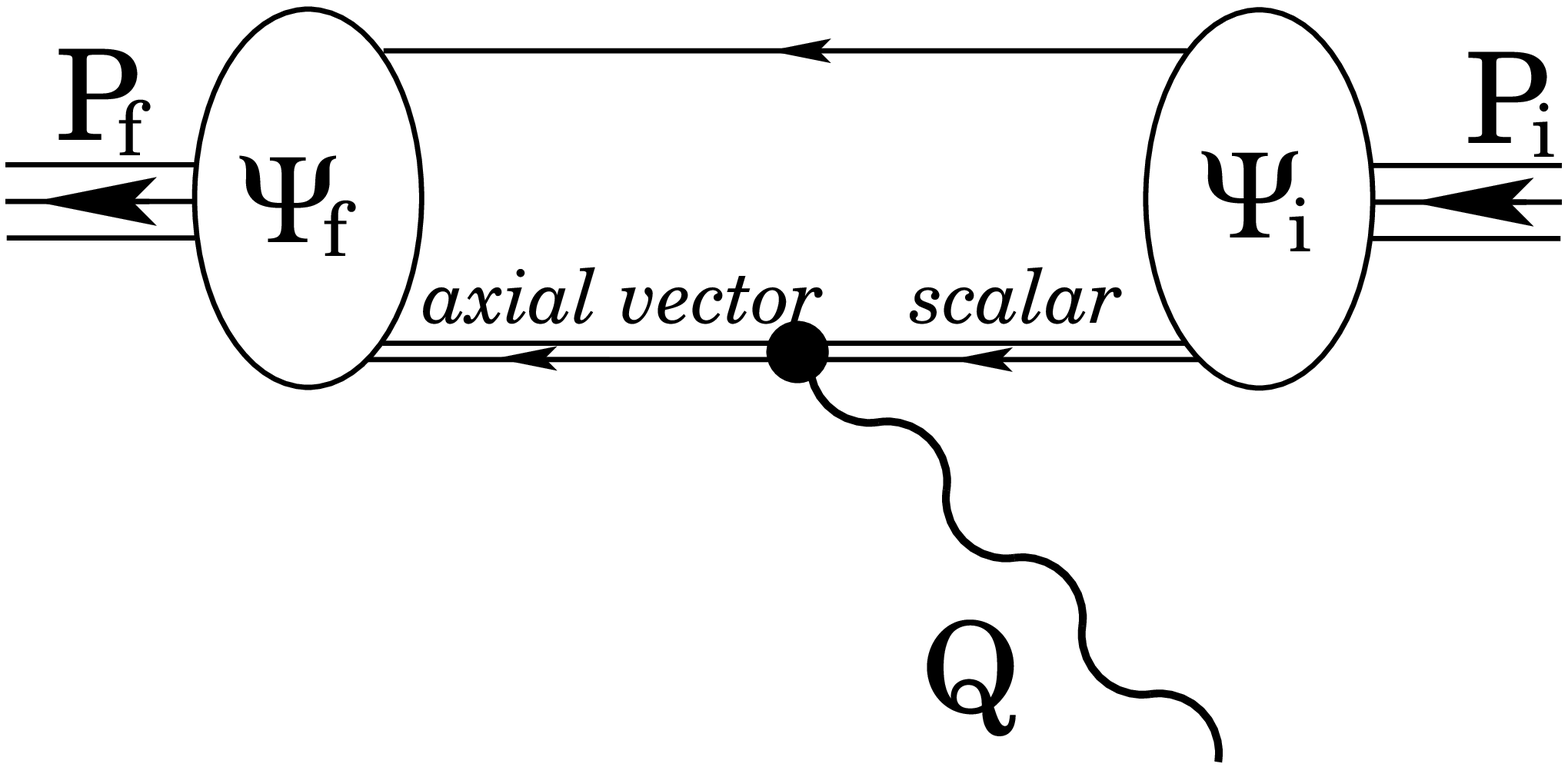}}
\end{minipage}\vspace*{3ex}

\begin{minipage}[t]{0.45\textwidth}
\leftline{\includegraphics[width=0.90\textwidth]{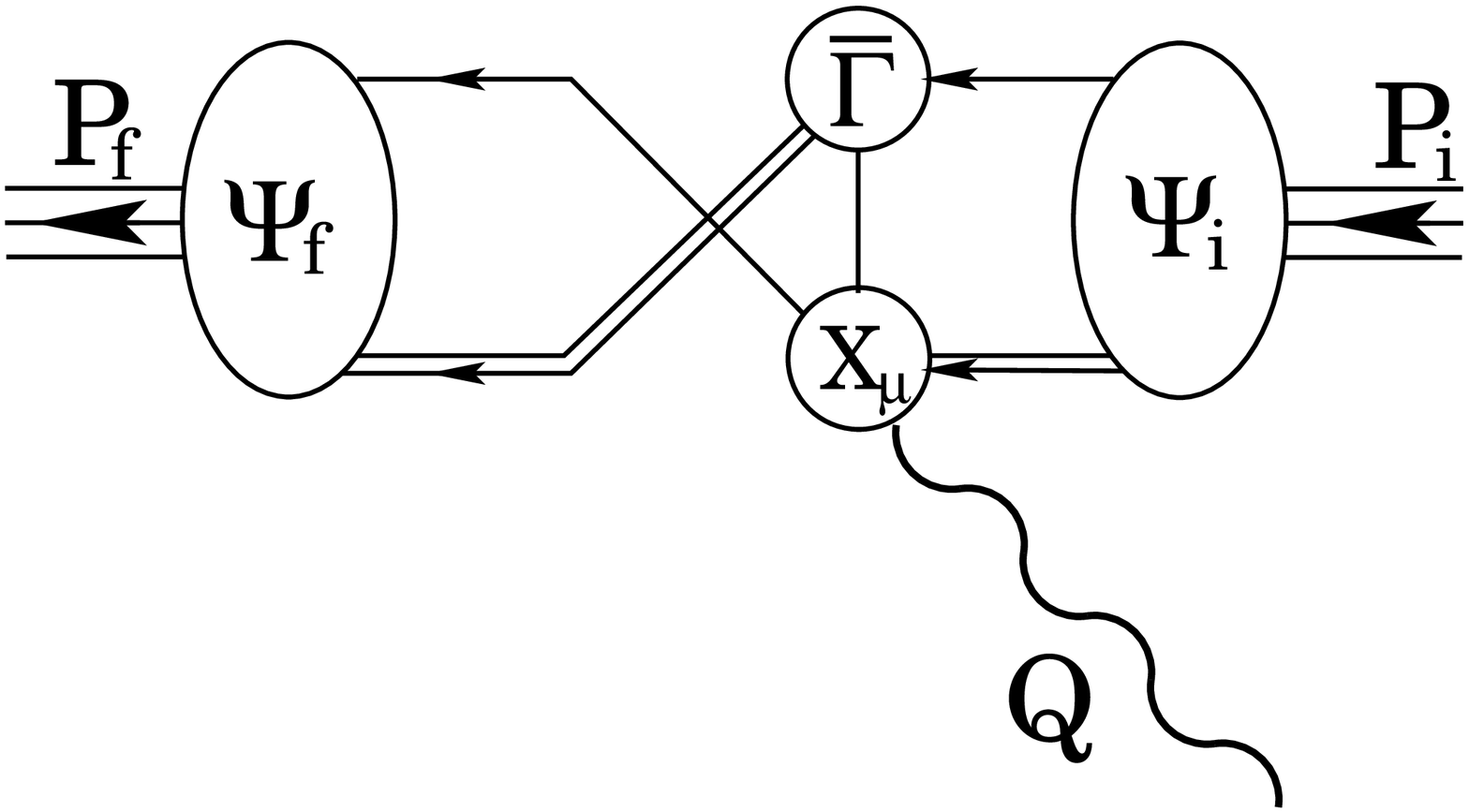}}
\end{minipage}
\begin{minipage}[t]{0.45\textwidth}
\rightline{\includegraphics[width=0.90\textwidth]{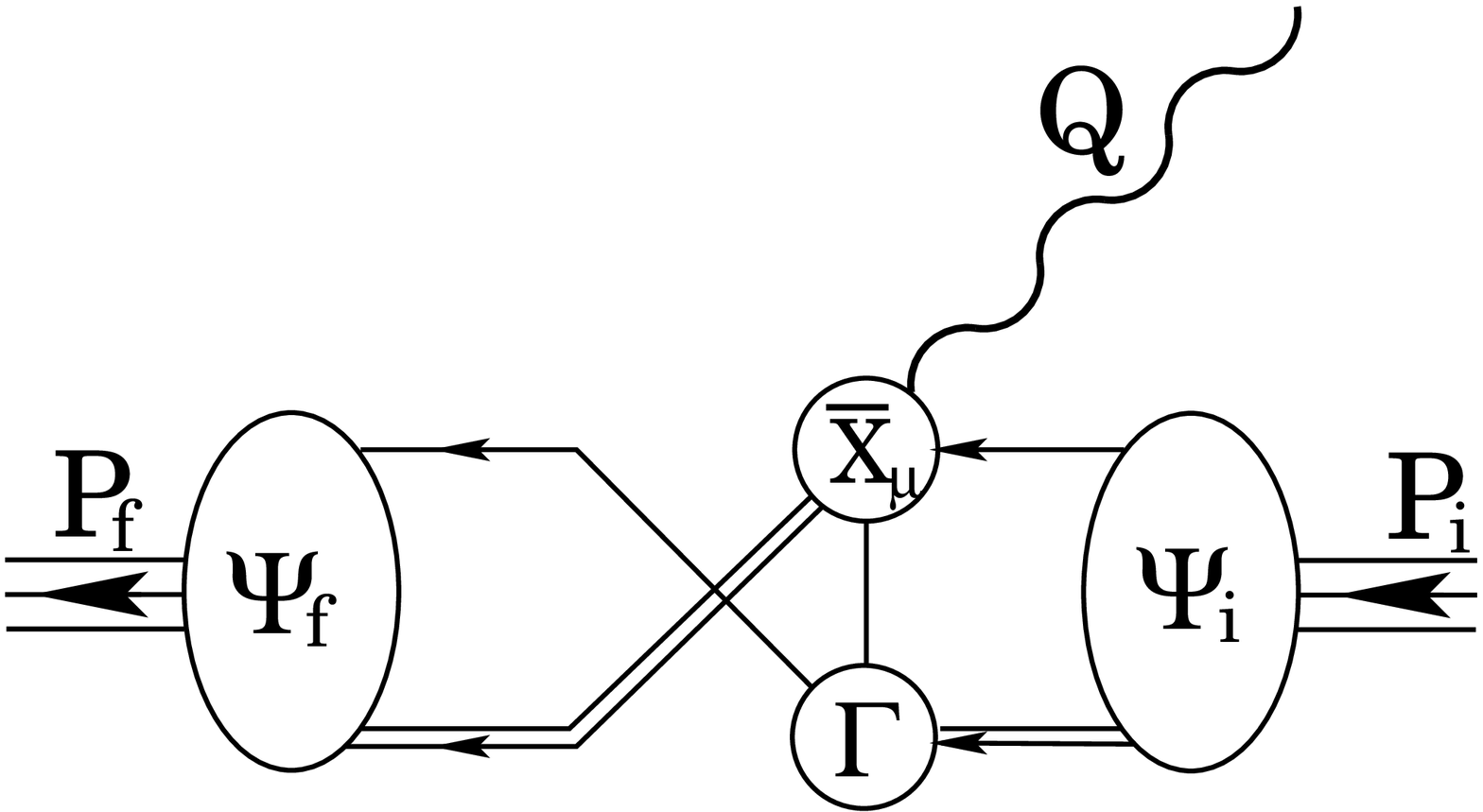}}
\end{minipage}
\end{minipage}
\caption{\label{vertex} Vertex which ensures a conserved current for on-shell nucleons described by the Faddeev amplitudes, $\Psi_{i,f}$, obtained from the Faddeev equation depicted in Fig.\,\protect\ref{faddeevfigure}.  The single line represents the dressed-quark propagator; the double line, the diquark propagator; and $\Gamma$ is the diquark Bethe-Salpeter amplitude.  The remaining vertices are described briefly in the text: the top-left image is Diagram~1; the top-right, Diagram~2; and so on, with the bottom-right image, Diagram~6.  In connection with DIS, the photon line is equated with a zero-momentum insertion.
[A full explanation of the diagrams depicted here is provided in App.\,C of \protect\cite{Cloet:2008re}, from which this figure is adapted.]
}
\end{figure}

In order to calculate nucleon distribution functions one must know the manner in which the nucleon described by the Faddeev equation couples to a photon.  That is derived in \cite{Oettel:1999gc} and illustrated in Fig.\,\ref{vertex}.  Naturally, the current depends on the electromagnetic properties of the diquark correlations.  A detailed explanation of the diagrams in Fig.\,\ref{vertex} is presented in App.\,C of  \protect\cite{Cloet:2008re}.  Here we only provide a brief explanation.

Diagram~1 represents the photon coupling directly to the bystander quark. It is a necessary condition for current conservation that the quark-photon vertex satisfy the Ward-Takahashi identity.  Since the quark is dressed, the vertex is not bare.  It can be obtained by solving an inhomogeneous Bethe-Salpeter equation.

Diagram~2 depicts the photon coupling directly to a diquark correlation.  Naturally, the diquark propagators match the line to which they are attached.  Moreover, the interaction vertices satisfy Ward-Takahashi identities and obey QCD constraints.

Diagram~3 shows a photon coupling to the quark that is exchanged as one diquark breaks up and another is formed.  It is a two-loop diagram.  It is noteworthy that the process of quark exchange provides the attraction necessary in the Faddeev equation to bind the nucleon.  It also guarantees that the Faddeev amplitude has the correct antisymmetry under the exchange of any two dressed-quarks.  This key feature is absent in models with elementary (noncomposite) diquarks.

Diagram~4 differs from Diagram~2 in expressing the contribution to the nucleons' form factors owing to an electromagnetically induced transition between scalar and axial-vector diquarks.  This transition vertex is a rank-2 pseudotensor, kindred, e.g., to the matrix element describing the $\rho\, \gamma^\ast \pi^0$  transition \cite{Maris:2002mz}.

Diagrams~5 \& 6 are the so-called ``seagull'' terms, which appear as partners to Diagram~3 and arise because binding in the Faddeev equations is effected by the exchange of a dressed-quark between \textit{nonpointlike} diquark correlations \cite{Oettel:1999gc}.  The new elements in these diagrams are the couplings of a photon to two dressed-quarks as they either separate from (Diagram~5) or combine to form (Diagram~6) a diquark correlation.  As such they are components of the five point Schwinger function which describes the coupling of a photon to the quark-quark scattering kernel.  These terms vanish if the diquark correlation is represented by a momentum-independent Bethe-Salpeter-like amplitude; i.e., the diquark is pointlike.

%\pagebreak 
\newpage

%\input references
%\bibliographystyle{apsrmp}
%\bibliography{struc_rmp}

\end{document}